\documentstyle[12pt]{article}
\newcommand{\be}{\begin{equation}}
\newcommand{\ee}{\end{equation}}
\renewcommand{\theequation}{\arabic{section}.\arabic{equation}}
\newcommand{\bear}{\begin{eqnarray}}
\newcommand{\ear}{\end{eqnarray}}

\newcommand{\Tr}{\rm Tr}
\newcommand{\Li}{\cal L}
\newcommand{\G}{{\cal G}}
\newcommand{\K}{{\cal K}}
\newcommand{\B}{{\cal B}}
\newcommand{\F}{{\cal F}}
\newcommand{\M}{{\cal M}}
\newcommand{\W}{{\cal W}}
\newcommand{\T}{{\cal T}}
\newcommand{\di}{{\rm d}}

\newcommand{\dA}{\sqcup\!\!\!\!\sqcap}

\def\eins{  1\!{\rm l}  }
\newcommand{\QCD}{\rm QCD}
\newcommand{\GeV}{\rm GeV}
\newcommand{\slp}{\raise.15ex\hbox{$/$}\kern-.57em\hbox{$\partial$}}
\newcommand{\slG}{\raise.15ex\hbox{$/$}\kern-.57em\hbox{$G$}}
\newcommand{\grgl}{\:\hbox to -0.2pt{\lower2.5pt\hbox{$\sim$}\hss}
{\raise3pt\hbox{$>$}}\:}
\newcommand{\klgl}{\:\hbox to -0.2pt{\lower2.5pt\hbox{$\sim$}\hss}
{\raise3pt\hbox{$<$}}\:}
\newcommand{\lvec}{{\buildrel\leftarrow\over\slp}}
\newcommand{\rvec}{{\buildrel\rightarrow\over\slp}}

\def\vec#1{\mathchoice{\mbox{\boldmath$\displaystyle\bf#1$}}
{\mbox{\boldmath$\textstyle\bf#1$}}
{\mbox{\boldmath$\scriptstyle\bf#1$}}
{\mbox{\boldmath$\scriptscriptstyle\bf#1$}}}

\begin{document}
\pagenumbering{arabic}
\begin{titlepage}
\begin{flushright}
HD--THEP--96--38
\end{flushright}
\vspace{.5cm}
\begin{center}
{\bf\LARGE High Energy Collisions} \\
\bigskip
{\bf\LARGE and Nonperturbative QCD\footnote{
Based on lectures given at the workshop ``Topics in
Field Theory'', organized by the Graduiertenkolleg
Erlangen-Regensburg 1993 in Banz, Germany, and at the
``35. Internationale Universit\"atswochen f\"ur Kern- und
Teilchenphysik 1996'' in Schladming, Austria.}}\\
\vspace{.5cm}
O. Nachtmann\\
\bigskip
Institut  f\"ur Theoretische Physik, Universit\"at Heidelberg,\\
Philosophenweg 16,
D-69120 Heidelberg, FRG\\
\end{center}
\vspace{1cm}
\noindent{\bf Abstract:}
We discuss various ideas on the nonperturbative vacuum structure
in QCD. The stochastic vacuum model of Dosch and
Simonov is presented in some detail. We show how this model produces
confinement. The model incorporates the idea of the QCD vacuum
acting like a dual superconductor due to
an effective chromomagnetic monopole
condensate. We turn then to high energy, small momentum transfer
hadron-hadron scattering. A field-theoretic formalism to treat
these reactions is developed, where the basic quantities governing
the scattering amplitudes are correlation functions of light-like
Wegner-Wilson lines and loops. The evaluation of these correlation
functions with the help of the
Minkowskian version of the stochastic vacuum model is discussed.
A further surprising manifestation of the nontrivial vacuum structure
in QCD may be
the production of anomalous soft photons in hadron-hadron
collisions. We interpret these photons as being due to
``synchrotron radiation from the vacuum''. A duality argument
leads us from there to the expectation of anomalous pieces proportional
to $(Q^2)^{1/6}$ in the electric form factors of the nucleons for
small $Q^2$. Finally we sketch the idea that in the Drell-Yan
reaction, where a quark-antiquark pair annihilates with
the production of a lepton pair, a ``chromodynamic Sokolov-Ternov
effect'' may be at work. This leads to a spin correlation of the
$q\bar q$ pair, observable through the angular distribution of the
lepton pair.
\end{titlepage}

\newpage
\section*{Table of Contents}
\begin{enumerate}
\item Introduction
\item The QCD Vacuum
\begin{itemize}
\item[2.1] Connectors
\item[2.2] The non-abelian Stokes theorem
\item[2.3] The cumulant expansion
\item[2.4] The basic assumptions of the
stochastic vacuum model
\item[2.5] The Wegner-Wilson loop in the stochastic vacuum model
\end{itemize}
\item Soft Hadronic Reactions
\begin{itemize}
\item[3.1] General considerations
\item[3.2] The functional integral approach to parton-parton scattering
\item[3.3] The Eikonal expansion
\item[3.4] The quark-quark scattering amplitude
\item[3.5] The scattering of systems of quarks, antiquarks, and gluons
\item[3.6] The scattering of wave packets of
partons representing mesons
\item[3.7] The evaluation of scattering amplitudes in the
Minkowskian version of the stochastic vacuum model
\end{itemize}
\item ``Synchrotron Radiation'' from the vacuum, electromagnetic
form factors of hadrons, and spin correlations in the Drell-Yan reaction
\item Conclusions
\end{enumerate}

\noindent Appendices
\begin{itemize}
\item[A.] Higher cumulant terms and dynamic fermions in the
calculation of the Wegner-Wilson loop in the stochastic vacuum
model
\item[B.] The scattering of gluons
\item[C.] The scattering of baryons
\end{itemize}

\newpage
\section{Introduction}
In these lectures I would like to review some ideas on the way
nonperturbative QCD may manifest itself in high energy collisions.
Thus we will be concerned with strong interactions where we claim
to know the fundamental Lagrangian for a long time now
\cite{1}:
\be\label{1.1}
{\Li}_{\QCD}(x)=-\frac{1}{4}G^a_{\lambda\rho}(x)
G^{\lambda\rho a}(x)+\sum_q\bar q(x)(i\gamma^\lambda D_\lambda
-m_q)q(x).\ee
Here $q(x)$ are the quark fields for the various quark flavours
$(q=u, d, s, c, b, t)$ with masses $m_q$. We denote the gluon
potentials by $G_\lambda^a(x)\ (a=1,...,8)$ and the gluon
field strengths by
\be\label{1.2}
G^a_{\lambda\rho}(x)=\partial_\lambda G_\rho^a(x)-\partial_\rho
G_\lambda^a(x)-gf_{abc}G^b_\lambda(x)G_\rho^c(x),\ee
where $g$ is the strong coupling constant
and $f_{abc}$ are the structure constants of SU(3). The covariant
derivative of the quark fields is
\be\label{1.200}
D_\lambda q(x)=(\partial_\lambda+ig G_\lambda^a\frac{\lambda_a}{2})
q(x),\ee
with $\lambda_a$ the Gell-Mann matrizes of the SU(3) group. The
gluon potential and field strength matrizes are defined as
\bear\label{1.201}
G_\lambda(x):&=&G_\lambda^a(x)\frac{\lambda_a}{2},\nonumber\\
G_{\lambda\rho}(x):&=&G_{\lambda\rho}^a(x)\frac{\lambda_a}{2}.
\ear
The Lagrangian (\ref{1.1}) is invariant under SU(3) gauge
transformations.
Let $x\to U(x)$ be an arbitrary matrix function, where for fixed
$x$ the $U(x)$ are SU(3) matrices:
\bear\label{1.202}
U(x)U^\dagger(x)&=&1,\nonumber\\
{\rm det} U(x)&=&1.\ear
With the transformation laws:
\be\label{1.203}
q(x)\to U(x)q(x),\ee
\be\label{1.204}
G_\lambda(x)\to U(x)G_\lambda(x)U^\dagger(x)-\frac{i}{g}
U(x)\partial_\lambda U^\dagger(x),\ee
we find
\[G_{\lambda\rho}(x)\to U(x)G_{\lambda\rho}(x)U^\dagger(x),
\]
and invariance of ${\Li}_{\QCD}$:
\be\label{1.205}
{\Li}_{\QCD}(x)\to{\Li}_{\QCD}(x).\ee

If we want to derive results from the Lagrangian (\ref{1.1}),
we face problems, the most notable being that ${\Li}_{\QCD}$
is expressed in terms of quark and gluon fields whose quanta
have not been observed as free particles. In the real world
we observe only hadrons, colourless objects, quark and
gluons are permanently confined. Nevertheless it has been
possible in some cases to derive first principle results which
can be compared with experiment, starting from ${\Li}_{\QCD}$
(\ref{1.1}). These are in essence the following:

(1) Pure short-distance phenomena: Due to asymptotic freedom \cite{2}
the QCD coupling constant becomes small in this regime and one
can make reliable perturbative calculations. Examples of pure
short distance processes are for instance the total cross section
for electron-positron annihilation into hadrons and the
total hadronic decay rate of the $Z$-boson.

(2) Pure long-distance phenomena: Here one is in the nonperturbative
regime of QCD and one has to use numerical methods to obtain first
principle results from ${\Li}_{\QCD}$, or rather from
the lattice version of ${\Li}_{\QCD}$ introduced by Wilson \cite{3}.
Today, Monte Carlo simulations of lattice QCD are a big industry
among theorists. Typical quantities one can calculate in this
way are hadron masses and other low energy hadron properties.
(For an up-to-date account of these methods cf. \cite{4}).

There is a third regime of hadronic phenomena, hadron-hadron collisions,
which are -- apart from very low energy collisions -- neither
pure long-distance nor pure short-distance phenomena. Thus, none
of the above-mentioned theoretical methods apply directly. Traditionally
one classifies high-energy had\-ron-hadron collisions as
``hard'' and ``soft'' ones:

(3) High energy hadron-hadron collisions:

\qquad (a) hard collisions,

\qquad (b) soft collisions.

A typical hard reaction is the Drell-Yan process, e.g.

\newpage

\bear\label{1.3}
\pi^-+N\to&&\gamma^*+X\nonumber\\
&&\hookrightarrow \ell^+\ell^-\ear
where $\ell=e,\mu$. All energies and momentum
transfers are assumed to be large.
However, the masses of the $\pi^-$ and $N$ in the initial state
stay fixed and
thus we are not dealing with a pure short distance phenomenon.

In the reaction (\ref{1.3}) we claim to see directly the
fundamental quanta of the
theory, the partons, i.e. the quarks and gluons,
in action (cf. Fig. 1). In the
usual theoretical  framework for hard reactions,
the QCD improved parton model (cf.
e.g. \cite{5}), one describes the reaction of the partons,
in the Drell-Yan case the
$q\bar q$ annihilation into a virtual  photon,
by perturbation theory. This should be
reliable, since the parton process involves only
high energies and high momentum
transfers. All the long distance physics due to the
bound state nature of the
hadrons is then lumped into parton distribution functions
of the participating hadrons. This is called
the \underbar{factorization hypothesis},
which after early investigations of soft initial and
 final state interactions \cite{6} was formulated
and studied in low orders of QCD perturbation theory in \cite{7}.
Subsequently, great theoretical effort has gone into
proving factorization in the framework of QCD
perturbation theory \cite{8}-\cite{10}. The result seems to be that  
factorization is most probably correct there (cf. the
discussion in \cite{11}).
However, it is legitimate to ask if factorization is
respected also by nonperturbative effects.
To my knowledge this question was first asked in \cite{12}
-\cite{14}.
In Sect. 4 of these lectures I will come back to this question
and will argue that there may be evidence for a breakdown
of factorization in the Drell-Yan reaction due to QCD vacuum
effects.

Let us consider now soft high energy
collisions. A typical reaction in this class is proton-proton elastic
scattering:
\be\label{1.4}
p+p\to p+p\ee
at c.m. energies $E_{\rm cm}=\sqrt{s}\grgl
5{\GeV}$ say and small momentum
transfers $\sqrt{|t|}=|q|\klgl1{\GeV}$.
Here we have two scales, one staying
finite, one going to infinity:
\bear\label{1.5}
E_{\rm cm}&&\to \infty,\nonumber\\
|q|&&\klgl 1{\GeV}.\ear
Thus, none of the above mentioned calculational  methods is directly applicable.
Indeed, most theoretical papers dealing with reactions in this class  develop and
apply \underbar{models} which are partly older than QCD, partly QCD ``motivated''.
Let me list \underbar{some models for hadron-hadron elastic scattering} at high
energies:

geometric \cite{15},

eikonal \cite{16},

additive quark model \cite{17},

Regge poles \cite{18},

topological expansions and strings \cite{19},

valons \cite{20},

leading log summations \cite{21},

two-gluon exchange \cite{22},

the Donnachie-Landshoff model for the
``soft Pomeron'' \cite{22a}.

It would be a forbidding  task to collect all
references  in this field. The
references given above should thus only be
considered as representative ones. In
addition I would like to mention the
inspiring general field theoretic
considerations for high energy scattering and particle
production by Heisenberg
\cite{23} and the impressive work by Cheng and Wu on
high energy behaviour in field
theories in the framework of perturbative calculations \cite{24}.

I will now argue that the theoretical description of measurable
quantities of soft
high energy reactions like the total cross sections should
involve in an essential
way \underbar{nonperturbative} QCD.
To see this,  consider massless pure gluon
theory where all ``hadrons'' are
massive glueballs. Then we know from the
renormalization group analysis that the
glueball masses must behave as
\be\label{1.6}
m_{\rm glueball}\propto M e^{-c/g^2(M)}\ee
for $M\to\infty$, i.e. for $g(M)\to 0$, due to asymptotic freedom.
Here $M$  is the renormalization scale, $g(M)$ is the
QCD coupling strength at this scale and $c$ is
a constant:
\bear\label{1.7}
&&\frac{g^2(M)}{4\pi^2}\longrightarrow\frac{12}{33\ln (M^2/\Lambda^2)}
\quad{\rm for}\quad M\to\infty,\nonumber\\
&&c=\frac{8\pi^2}{11},\nonumber\\
&&\Lambda:\quad {\rm QCD\ scale\ parameter}.\ear

Masses in massless Yang-Mills theory are a purely nonperturbative
phenomenon, due to ``dimensional transmutation''. Scattering of  
glueball-had\-rons in
massless pure gluon theory should look very similar to
scattering of hadrons in the
real world,  with finite total cross sections,
amplitudes with analytic $t$
dependence etc. At least, this would be my expectation. If the total cross section
$\sigma_{\rm tot}$ has a finite limit as $s\to\infty$ we must have from the same
renormalization group arguments:
\be\label{1.8}
\lim_{s\to\infty}\sigma_{\rm tot}(s)\propto M^{-2} e^{2c/g^2(M)}\ee
for $g(M)\to 0$.
In this case, the total cross sections in pure gluon theory are also nonperturbative
objects! It is easy to see that this conclusion is not changed
if $\sigma_{\rm
tot}(s)$ has a logarithmic behaviour with $s$ for $s\to\infty$, e.g.
\be\label{1.9}
\sigma_{\rm tot} (s)\to {\rm const} \times (\log s)^2.\ee
I would then expect that also in full QCD  total cross sections are nonperturbative
objects, at least as far as hadrons made
out of light quarks are concerned.

Some time ago P.V. Landshoff and myself
started to think about a possible
connection between the nontrivial vacuum structure of QCD -- a typical
nonperturbative phenomenon -- and soft high energy reactions \cite{25}.
In the following  I will first review  some common folklore
on the QCD vacuum and discuss in more detail the so-called
``stochastic vacuum model''. I will then
sketch possible consequences of these ideas
for high energy collisions.

\section{The QCD Vacuum}
\setcounter{equation}{0}

According to current theoretical  prejudice  the
vacuum state in QCD has a very
complicated  structure \cite{26}-\cite{36}.
It was first noted by Savvidy \cite{26} that by
introducing a constant chromomagnetic  field
\be\label{2.1}
\vec B^a=\vec n\eta^a B,\ (a=1, \ldots, 8),\ee
into the \underbar{perturbative} vacuum one can
 lower the vacuum-energy density
$\varepsilon(B)$. Here $\vec n$ and $\eta^a$
are constant unit vectors in ordinary
and colour  space.  The result of his one-loop calculation was
\be\label{2.2}
\varepsilon(B)=\frac{1}{2}B^2+\frac{\beta_0g^2}{32\pi^2}
B^2\left[\ln\frac{B}{M^2}-\frac{1}{2}\right]\ee
where $g$ is the strong coupling constant, $M$
is again the renormalization scale,
and $\beta_0$ is given by the lowest order
term in the Callan-Symanzik $\beta$-function:
\be\label{2.3}
M\frac{{\rm d}g(M)}{{\rm d}M}=:\beta(g)=-\frac{\beta_0}
{16\pi^2}g^3+\ldots\ee
For 3 colours and $f$ flavours:
\be\label{2.4}
\beta_0=11-\frac{2}{3} f.\ee
Thus, as long as we have asymptotic  freedom, i.e. for $f\leq 16$,
the energy
density $\varepsilon(B)$ looks as indicated schematically
in Fig. 2 and has its
minimum for $B=B_{\rm vac}\not=0$. Therefore, we should
expect the QCD-vacuum to
develop spontaneously a chromomagnetic field,
the situation being similar to that in
a ferromagnet below the Curie temperature where we have spontaneous magnetization.

Of course, the vacuum state in QCD has to be
relativistically invariant and cannot
have a preferred direction in ordinary space and colour space. What has been
considered \cite{32}  are states composed of
domains with random orientation of the
gluon-field strength (Fig. 3). This is analogous  to Weiss domains in a
ferromagnet.  The vacuum state should then be a
suitable linear superposition of
states with various domains and orientation of
the fields inside the domains. This
implies that the orientation of the fields in the domains as well as the  
boundaries of the domains will fluctuate.

A very detailed picture for the QCD vacuum along
these lines has been developed in
ref. \cite{32}. I cannot refrain from comparing
this modern picture of the QCD
vacuum (Fig. 4a) with the ``modern picture'' of the ether developed by
\underbar{Maxwell} more than 100 years ago
\cite{37} (Fig. 4b). The analogy is quite striking
and suggests to me that with time passing
on we may also be able to find simpler
views on the QCD vacuum. Remember that  Einstein made great
progress by eliminating the ether from electrodynamics.
In the following we will
adopt the picture of the QCD vacuum as developed in refs.
\cite{26}-\cite{33},\cite{35} and
outlined above as a \underbar{working hypothesis}.

Let me now come to the values for the field strengths $\vec E^a$
and $\vec B^a$ in
the vacuum. These must also be determined by $\Lambda$, the QCD scale parameter, the
only dimensional parameter in QCD if we disregard
the quark masses. Therefore, we
must have on dimensional grounds for the
renormalization group invariant quantity $(gB)^2$
\be\label{2.5}
(gB)^2\propto\Lambda^4.\ee
But we have much more detailed information
on the values of these field strengths
due to the work of Shifman, Vainshtein, and Zakharov (SVZ),
who introduced the
\underbar{gluon condensate} and first estimated its value
using sum rules for
charmonium states \cite{29}:
\bear\label{2.6}
<0|\frac{g^2}{4\pi^2} G^a_{\mu\nu}(x)G^{\mu\nu a}(x)|0>&\equiv
&<0|\frac{g^2}{2\pi^2}\left(\vec B^a(x)\vec B^a(x)-\vec E^a(x)\vec
E^a(x)\right)|0>\nonumber\\
&\equiv &G_2= (2.4\pm 1.1)\cdot 10^{-2} {\rm GeV}^4\nonumber\\
&=&(335-430 {\rm MeV})^4.\ear
Here we quote numerical values as given in the review \cite{38}.
A simple analysis shows that this implies
\be\label{2.7}
<0|g^2\vec B^a(x)\vec B^a(x)|0>=-<0|g^2\vec E^a(x)\vec E^a(x)|0>=\pi^2  
G_2\simeq(700{\rm MeV})^4.\ee
To prove eq. (\ref{2.7}) we note that Lorentz- and
parity-invariance require the
vacuum expectation value  of the uncontracted product of two gluon field strengths
to be of the form
\be\label{2.8}
<0|\frac{g^2}{4\pi^2} G^a_{\mu\nu}(x) G^b_{\rho\sigma}
(x)|0>=(g_{\mu\rho}
g_{\nu\sigma}-g_{\mu\sigma} g_{\nu\rho})\delta^{ab}
\frac{G_2}{96}\ee
where $G_2$ is the same constant as in (\ref{2.6}).
Taking appropriate contractions leads to (\ref{2.6}) and (\ref{2.7}).

We find that $<0|\vec B^a(x)\ \vec B^a(x)|0>$ is positive,
$<0|\vec E^a(x)\ \vec E^a(x)|0>$ negative! This can happen
because we are really considering
products of field operators, normal-ordered with respect
to the perturbative
vacuum. The interpretation of eq. (\ref{2.7}) is, therefore,
that the B-field
fluctuates with bigger amplitude, the E-field with
smaller amplitude than in the perturbative vacuum state.

What about the size $a$ of the colour domains and the
fluctuation times $\tau$
of the colour fields? On dimensional grounds we must have
\be\label{2.9} a\sim \tau\sim\Lambda^{-1}.\ee

A detailed model for the QCD vacuum incorporating the
gluon condensate idea
and a fall-off of the correlation of two field
strengths with distance
was proposed in \cite{39}: the ``stochastic vacuum model'' (SVM).
In the following we will discuss the basic assumptions of
the model and then apply it to derive the area law for the
Wegner-Wilson loop, i.e. confinement of static quarks. For
the rest of this section we will work in Euclidean space-time.
To accomplish the analytic continuation from Minkowski to
Euclidean space-time we have to make the following
replacements for $x^\lambda$ and $G^\lambda$ (cf. (1.4)):
\bear\label{2.10}
&&x^0\to -iX_4,\nonumber\\
&&\vec x\to\vec X,\nonumber\\
&&G^0\to i{\G}_4,\nonumber\\
&&\vec G\to-\vec{\G}.\ear
Here $X=(\vec X,X_4)$ denotes an Euclidean space-time point
and ${\G}_\alpha$ $(\alpha=1,...,4)$ the Euclidean gluon
potential. With (\ref{2.10}) we get
\bear\label{2.11}
&&(x\cdot y)\to-(X\cdot Y)=-X_\alpha Y_\alpha,\nonumber\\
&&-ig\int {\di}x^\mu G_\mu(x)\to -ig\int {\di}X_\alpha{\G}_\alpha(X),
\nonumber\\
&&G^{0j}\to -i{\G}_{4j},\nonumber\\
&&G^{jk}\to {\G}_{jk},\ear
where $1\leq j,k\leq 3,\quad 1\leq \alpha,\beta\leq 4$ and
\be\label{2.12}
{\G}_{\alpha\beta}=\partial_\alpha{\G}_\beta-\partial_
\beta{\G}_\alpha+ig[{\G}_\alpha,{\G}_\beta]\ee
is the Euclidean gluon field strength tensor.

\subsection{Connectors}
Consider classical gluon fields in Euclidean space-time.
Let $X, Y$ be two points there and $C_X$ a curve
from $X$ to $Y$ (Fig. 5).
We define the connector, the non-abelian generalization of the
``Schwinger string'' \cite{40} of QED as
\be\label{2.13}
V(Y,X;C_X)={\rm P}\{\exp[-ig\int_{C_X}
{\di}Z_\alpha {\G}_\alpha(Z)]\}.
\ee
where ${\rm P}$ means path ordering. The connector can be obtained
as solution of a differential equation. Let
\bear\label{2.14}
&&\tau\to Z(\tau),\nonumber\\
&&\tau_1\leq\tau\leq\tau_2,\nonumber\\
&&Z(\tau_1)=X,\quad Z(\tau_2)=Y,\ear
be a parametrization of $C_X$. Consider the differential
equation for a $3\times 3$ matrix function $V(\tau)$:
\be\label{2.15}
\frac{\di}{{\di}\tau}V(\tau)=-ig\frac{{\di}Z_\alpha(\tau)}
{{\di}\tau}
{\G}_\alpha(Z(\tau))V(\tau),\ee
with the boundary condition
\be\label{2.16}
V(\tau_1)={\rm 1\!l}.\ee
The solution of (\ref{2.15}), (\ref{2.16}) gives for
$\tau=\tau_2$ just the connector (\ref{2.13}).

Under a gauge transformation
\be\label{2.17}
{\G}_\alpha(X)\to U(X){\G}_\alpha(X)U^\dagger(X)-
\frac{i}{g}U(X)\partial_\alpha U^\dagger(X)\ee
where $U(X)\in SU(3)$, we have
\be\label{2.18}
V(Y,X;C_X)\to U(Y)V(Y,X;C_X)U^{-1}(X).\ee

The connector can be used to ``shift'' various objects from
one space-time point to another in a gauge-covariant way.
We define for instance the field strength tensor shifted
from $X$ to $Y$ along $C_X$ as
\be\label{2.19}
\hat{\G}_{\alpha\beta}(Y,X;C_X):=V(Y,X;C_X){\G}_{\alpha\beta}(X)
V^{-1}(Y,X;C_X).\ee
Under a gauge transformation $\hat{\G}_{\alpha\beta}(Y,X;C_X)$
transforms like a field strength tensor at $Y$:
\be\label{2.20}
\hat{\G}_{\alpha\beta}(Y,X;C_X)\to U(Y)
\hat{\G}_{\alpha\beta}(Y,X;C_X)U^{-1}(Y).\ee

Connectors can, of course, be defined for arbitrary SU(3)
representations, not only for the fundamental one used in
(\ref{2.13}). Let $T_a(a=1,...8)$ be the generators of
SU(3) in some arbitrary unitary representation $R$ where
\be\label{2.21}
[T_a,T_b]=if_{abc}T_c.\ee
We define the connector for this representation by:
\be\label{2.22}
V_R(Y,X;C_X):={\rm P}\ \exp[-ig\int_{C_X}
{\di}Z_\alpha{\G}_\alpha^a(Z)
T_a].\ee

We list some basic properties of connectors:

(i) For 2 adjoining curves $C_1,C_2$ (Fig. 6), the connectors are
multiplied:
\be\label{2.23}
V_R(X_3,X_1;C_2+C_1)=V_R(X_3,X_2;C_2).V_R(X_2,X_1;
C_1).\ee

(ii) Let $C$ be a curve from $X$ to $Y$ and $\bar C$ the
same curve but oriented in inverse direction, running from
$Y$ to $X$ (Fig. 6).
Then
\be\label{2.24}
V_R(X,Y;\bar C)V_R(Y,X;C)={\rm 1\!l},\ee
\be\label{2.25}
V_R^\dagger(Y,X;C)=V_R(X,Y;\bar C).\ee
The product of the connectors for the path and the reverse
path is equal to the unit matrix. The reversal of the path is
equivalent to hermitian conjugation.

We leave the proof of (\ref{2.23})-(\ref{2.25})
as an exercise for the reader.

\subsection{The non-abelian Stokes theorem}
In this subsection we will derive the non-abelian generalization
of the Stokes theorem \cite{41}. Consider a surface $S$
in Euclidean space time with boundary $C=\partial S$. Let
$X$ be some point on $C$ as indicated in Fig. 7 and consider
the connector (\ref{2.22}) from $X$ back to $X$ along $C$:
\be\label{2.26}
V_R(X,X;C)={\rm P}\ \exp[-ig\int_C
{\rm d}Z_\alpha{\G}^a_\alpha(Z)T_a].\ee
The problem is to transform this line integral into a
surface integral.

We start by considering a point $Z_1$ in $S$ and a small
plaquette formed by curves
$C_1,..,C_4$
where one corner point is $Z_1$  (Fig. 7).
We choose a coordinate system on $S$
in the neighbourhood of $Z_1$:
\be\label{2.27}
(u,v)\to Z(u,v)\ee
such that
\bear\label{2.28}
&&Z_1=Z(0,0),\nonumber\\
&&Z_2=Z(\Delta u,0),\nonumber\\
&&Z_3=Z(\Delta u,\Delta v),\nonumber\\
&&Z_4=Z(0,\Delta v)\ear
and the curves $C_1,C_3(C_2,C_4)$ correspond to lines
$v$=const. ($u$=const.). The matrix representing the
line integral around the small plaquette is
\be\label{2.29}
U_R(\Delta u,\Delta v):=V_R(C_4)V_R(C_3)V_R(C_2)V_R(C_1).\ee
We want to make a Taylor expansion of $U_R(\Delta u,\Delta v)$
in $\Delta u$ and $\Delta v$. From (\ref{2.24}) we find
immediately that
\be\label{2.30}
U_R(0,\Delta v)=U_R(\Delta u,0)=\eins.\ee
Thus the lowest order term in the expansion after the
zeroth order is proportional to $\Delta u\cdot\Delta v$
and we get easily:
\bear\label{2.31}
&&U_R(\Delta u,\Delta v)={\rm 1\!l}-ig\frac{1}{2}
\Delta\sigma_{\alpha\beta}G^a_{\alpha\beta}(Z_1)T_a\nonumber\\
&&+O(\Delta u^2\Delta v,\Delta u\Delta v^2),\ear
where
\be\label{2.32}
\Delta\sigma_{\alpha\beta}=\Delta u\Delta v\frac{\partial
(Z_\alpha,Z_\beta)}{\partial(u,v)}.\ee
In the limit $\Delta u,\Delta v\to 0$ $\Delta\sigma_{\alpha
\beta}$ becomes the surface element $d\sigma_{\alpha\beta}$ of the
plaquette.

The next step is to choose an arbitrary point $Y$, the reference
point, on the surface $S$ and to draw a fan-type net on $S$
as a spider would do (Fig. 8). The
system of curves of the net consists
of the curve $C_X$ running from $X$ to $Y$, then $\bar C_{Z_1}$ from
$Y$ to $Z_1$, then around a small plaquette at $Z_1$, back
to $Y$ along $C_{Z_1}$ and so on. The final curve is $\bar C_X$
from $Y$ to $X$. Apart from the initial and final curves
$C_X$ and $\bar C_X$ we have a system of plaquettes with
``handles'' connecting them to $Y$. With the help of (\ref{2.23})
-(\ref{2.25}) we see that the connector along the whole net
is equivalent to the original connector (\ref{2.26}).
\bear\label{2.33}
&&V_R(X,X;C)=V_R(X,Y;\bar C_X)\cdot\ {\rm product\
of\ connectors}\nonumber\\
&&{\rm for\ the\ plaquettes\ with\ handles\ }\cdot V_R(Y,X;C_X).
\ear
Let us consider one plaquette with handle, for instance
the one at $Z_n$ in Fig. 8. For this contribution to (\ref{2.33})
we get
\bear\label{2.34}
&&V_R(Y,Z_n ;C_{Z_n})V_R({\rm plaquette\ at}\ Z_n)
V_R(Z_n,Y;\bar C_{Z_n})
\nonumber\\
&&= V_R(Y,Z_n; C_{Z_n})\left[\eins-ig\frac{1}{2}\Delta
\sigma_{\alpha\beta}\G^a_{\alpha\beta}(Z_n)T_a+\cdots\right]
V_R(Z_n,Y;\bar C_{Z_n})\nonumber\\
&&=\eins-ig\frac{1}{2}\Delta\sigma_{\alpha\beta}
\hat\G^a_{\alpha\beta}(Y,Z_n; C_{Z_n})T_a+\cdots.\ear
Here we use (\ref{2.31}) and the shifted field strengths as defined in
(\ref{2.19}). We leave it as an excercise to the reader to show that
(\ref{2.19}) implies
\bear\label{2.35}
&&V_R(Y,Z_n;C_{Z_n})\G^a_{\alpha\beta}(Z_n)T_aV_R(Z_n,Y;\bar C_{Z_n})
\nonumber\\
&&=\hat{\G}^a(Y,Z_n;C_{Z_n})T_a\ear
for arbitrary representation $R$ of $SU(3)$.

Inserting (\ref{2.34}) in (\ref{2.33}) and
summing up the contribution of
all pla\-quettes with handles, where we
have of course to respect the
ordering, we get in the limit that the net is infinitesimally fine:
\bear\label{2.36}
&&V_R(X,X;C)= V_R(X,Y;\bar C_X)\cdot\nonumber\\
&&{\rm P}\exp \left[-i\frac{g}{2}\int_S {\di}\sigma_{\alpha\beta}
\hat{\G}^a_{\alpha\beta}(Y,Z;C_Z)T_a\right]
V_R(Y,X;C_X).\ear
Here $P$ denotes the ordering on the whole surface as implied by the
net. Usually one takes the trace in (\ref{2.36}) which leads with
(\ref{2.24}) to
\be\label{2.37}
{\Tr}\ V_R(X,X;C)={\Tr}\ {\rm P} \exp\left[-i\frac{g}
{2}\int_S {\rm d}\sigma_{\alpha
\beta}\hat{\G}^a_{\alpha\beta}(Y,Z;C_Z)T_a\right].\ee
This is the desired non-abelian version of  Stokes' theorem. We leave it
to the reader as an exercise to show that for the abelian case
(\ref{2.37}) reduces to the conventional Stokes theorem.

\subsection{The cumulant expansion}
As a last mathematical tool for making calculations with
 the SVM we discuss the cumulant expansion \cite{42}.
Consider functions
\be\label{2.38}
\tau\to B(\tau)\ee
on the interval $O\leq \tau\leq 1$ where $B(\tau)$ are
quadratic matrices. We assume that an averaging procedure over
products of
the  functions  $B(\cdot)$ is defined:
\[ E( B(\tau_1)),E(B(\tau_1)B(\tau_2)),\ldots.\]
We consider first the case that all averages $E(\cdot)$
are $c$ numbers and that
\be\label{2.39}
E(1)=1.\ee
Let us consider the expectation value of
the $\tau$-ordered exponential:
\be\label{2.40}
f(t):=E({\rm P}\exp\left[ t\int^1_0 {\di}\tau B(\tau)\right]),\ee
where $t\in{\cal C}$.  The cumulant expansion asserts that
$\ln f(t)$ can be expanded as
\be\label{2.41}
\ln f(t)=\sum^\infty_{n=1}\frac{t^n}{n!}\int^1_0 {\rm d}\tau_1
\ldots\int^1_0 {\rm d}\tau_n K_n(\tau_1,\ldots,\tau_n),\ee
where the $n$-th cumulant $K_n(\tau_1,\ldots,\tau_n)$
is a symmetric function of its arguments. A frequently
used notation for the $K_n$ is:
\be\label{2.42}
K_n(\tau_1,\ldots,\tau_n)
\equiv\langle\langle B(\tau_1)\ldots B(\tau_n)\rangle\rangle.\ee

To prove  (\ref{2.41}) we proceed as follows.  Expanding in powers
of $t$ on the r.h.s. of (\ref{2.40}) we get
\be\label{2.43}
f(t)=1+\sum^\infty_{n=1}\frac{t^n}{n!}{\cal B}_n\ee
where
\be\label{2.44}
{\cal B}_n=\int^1_0 {\di}\tau_1\ldots
\int^1_0 {\di}\tau_n E({\rm P}(B(\tau_1)\ldots B(\tau_n))).\ee
Now we expand $\ln f(t)$:
\be\label{2.45}
\ln f(t)=\sum^\infty_{n=1}\frac{t^n}{n!}{\cal K}_n,\ee
where the expansion coefficients   ${\cal K}_n$ are obtained from:
\bear\label{2.46}
f(t)&=&\exp(\ln f(t)),\nonumber\\
1+\sum^\infty_{n=1}\frac{t^n}{n!}{\cal B}_n&=&\exp
\left[\sum^\infty_{n=1}\frac{t^n}{n!}{\cal K}_n\right].
\ear  From this we obtain the ${\cal K}_n$ as solution of the
following system of equations:
\bear\label{2.47}
{\cal B}_1&=&{\cal K}_1,\nonumber\\
{\cal B}_2&=&{\cal K}_2+{\cal K}_1^2,\nonumber\\
{\cal B}_3&=&{\cal K}_3+\frac{3}{2}({\cal K}_2{\cal K}_1+
{\cal K}_1{\cal K}_2)+{\K}^3_1,
\nonumber\\
{\cal B}_4&=&{\K}_4+2({\K}_3{\K}_1+{\K}_1\K_3)+
3{\K}^2_2+2({\K}_2{}\K^2_1+{\K}_1
{\K}_2{\K}_1+{\K}_1^2{\K}_2)+{\K}^4_1,\nonumber\\
&&\ldots.\ear
Clearly, this system can be inverted and we get ${\K}_n$ as sum of
monomials of the form
\be\label{2.48}
{\B}_{i_1}\cdot {\B}_{i_2}\ldots {\B}_{i_k},\ee
where
\be\label{2.49}
\sum^k_{j=1}i_j=n.\ee
Using (\ref{2.44}), every monomial (\ref{2.48}) can be written as
$n$-fold  integral over $\tau_1,\ldots,\tau_n$ with $0\leq \tau_j
\leq 1\ (j=1,\ldots,n)$. Since the integration domain is symmetric
under arbitrary permutations of $\tau_1,\ldots,\tau_n$ we can symmetrize
the integrand completely. In this way we get
\be\label{2.50}
\K_n=\int^1_0 {\di}\tau_1\ldots\int^1_0 {\di}
\tau_nK_n(\tau_1,\ldots,\tau_n),
\ee
where $K_n$ is a totally symmetric function of its arguments.
Inserting (\ref{2.50}) in (\ref{2.45}) we get the cumulant expansion
(\ref{2.41}), q.e.d. Explicitly we find for the first few cumulants:
\bear\label{2.51}
K_1(1)&=&E(B(1)),\nonumber\\
K_2(1,2)&=&E({\rm P}(B(1)B(2)))\nonumber\\
&&-\frac{1}{2}E(B(1))E(B(2))-\frac{1}{2}E(B(2))E(B(1)),\nonumber\\
K_3(1,2,3)&=&E({\rm P}(B(1)B(2)B(3)))\nonumber\\
&&-\frac{1}{2}[E({\rm P}(B(1)B(2)))E(B(3))\nonumber\\
&&+E(B(1))E({\rm P}(B(2)B(3)))\nonumber\\
&&+{\rm cycl.\ perm.\ }]\nonumber\\
&&+\frac{1}{3}[E(B(1))E(B(2))E(B(3))+{\rm perm.\ }],\nonumber\\
&&\ldots.\ear
Here we write as a shorthand notation
$K_1(1)\equiv K_1(\tau_1),\ B(1)\equiv B(\tau_1)$ etc.

The cumulant expansion (\ref{2.41}) has the so-called ``cluster''
property: Let us assume that the expectation values
of the P-ordered
products factorize
\be\label{2.52}
E({\rm P}(B(1)\ldots B(n)))=\frac{1}{n!}\left\{E(B(1))\ldots E(B(n))
+{\rm perm.}\right\}\ee
for all $n\geq 2$ and all
\be\label{2.53}
|\tau_i-\tau_j|\geq\tau_{\rm min}\quad (i\not= j).\ee
We can then show that the cumulants for $n\geq 2$ vanish:
\be\label{2.54}
K_n(1,\ldots, n)=0,\quad(n\geq 2)\ee
if the $\tau_i$ satisfy (\ref{2.53}).

To prove (\ref{2.54}) we show first that it is true for
$n=2$. Indeed, from (\ref{2.51}) and (\ref{2.52}) we get for
$|\tau_1-\tau_2|\geq \tau_{\rm min}$
\bear\label{2.55}
K_2(1,2)&=&\frac{1}{2!}\{E(B(1))E(B(2))+{\rm perm.}\}\nonumber\\
& &-\frac{1}{2} E(B(1))E(B(2))-\frac{1}{2}E(B(2))E(B(1))\nonumber\\
&=&0.\ear
Now we proceed by mathematical induction. Assume that (\ref{2.54})
has been shown for all $k$ with $2\leq k\leq n-1$.
We have from (\ref{2.47}):
\bear\label{2.56}
K_n(1,\ldots,n)&=&E({\rm P}(B(1)\ldots B(n)))\nonumber\\
&&-\frac{1}{n!}[K_1(1)\ldots K_1(n)+{\rm perm.}]\nonumber\\
&&+\ [{\rm symmetrized\ products\  of\ cumulants\ } K_k(1,\ldots,k)
\nonumber\\
&&
\quad {\rm with}\ 1\leq k\leq n-1]\ear
but at least one factor with $k\geq 2$. With (\ref{2.52}) we get
now immediately  $K_n(1,\ldots,n)=0$ in the region defined by
 (\ref{2.53}), q.e.d.

Up to now we have assumed the expectation values $E(\cdot)$ to be
$c$-numbers. In this case many of the formulae (\ref{2.47})
- (\ref{2.56}) can be simplified by using
\bear\label{2.57}
&&E(B(1))E(B(2))=E(B(2))E(B(1)),\nonumber\\
&&{\rm etc.}\ear
We have, on purpose, not used such commutativity relations, since
we are now going to generalize the cumulant expansion to the case
where
\be\label{2.58}
E(B(1)),E(B(1)B(2)),\ldots\ee
are themselves \underbar{quadratic matrix valued} expectation
values with
\be\label{2.59}
E(\eins)=\eins.\ee
Then we have, of course, in general, no more the commutativity
relations (\ref{2.57}). But all formulae (\ref{2.40}) - (\ref{2.56})
are written in such a way that they remain true also for the case
of \underbar{matrix valued} expectation values $E(\cdot)$.

\subsection{The basic assumptions of the stochastic vacuum model}

The basic object of the SVM is the correlator of two field
strengths shifted to a common reference point. Let $X,X'$
be two points in Euclidean space-time, $Y$ a reference point
and $C_X$, $C_{X'}$ curves from $X$ to $Y$ and
$X'$ to $Y$, respectively (Fig. 5). We consider the shifted field
strengths as defined in (\ref{2.19}) and the
vacuum expectation value of their product in the sense of
Euclidean QFT:
\be\label{2.60}
\langle\frac{g^2}{4\pi^2}\left[\hat{\G}^a_{\mu\nu}(Y,X;C_X)\hat
{\G}^b_{\rho\sigma}(Y,X';C_{X'})\right]
\rangle=:\frac{1}{4}\delta^{ab}
F_{\mu\nu\rho\sigma}(X,X',Y;C_X,C_{X'}).
\ee
Here (\ref{2.20}) and colour conservation allow us to
write the r.h.s. of (\ref{2.60}) proportional to
$\delta^{ab}$. It is easy to see that $F_{\mu\nu\rho\sigma}$
depends only on $X,X'$ and the curve $C_X+\bar C_{X'}$ connecting
them; i.e. the reference point $Y$ can be freely shifted on
the connecting curve. In the SVM one makes now the strong
assumption that the correlator (\ref{2.60}) even does not
depend on the connecting curve at all:

\begin{itemize}
\item{\bf Ass. 1:} $F_{\mu\nu\rho\sigma}$
is independent of the
reference point $Y$ and of the curves $C_X$ and $C_{X'}$.
\end{itemize}

Translational, $O(4)$- and parity invariance
require then the correlator
(\ref{2.60}) to be of the following form:
\bear\label{2.61}
&&F_{\mu\nu\rho\sigma}=F_{\mu\nu\rho\sigma}(Z)
=\frac{1}{24}G_2\Bigl\{
\left(\delta_{\mu\rho}\delta_{\nu\sigma}-\delta_{\mu\sigma}\delta_{\nu
\rho}\right)\kappa D(-Z^2)\nonumber\\
&&+\frac{1}{2}\Bigl[\frac{\partial}{\partial  
Z_\nu}\left(Z_\sigma\delta_{\mu\rho}-Z_\rho\delta_{\mu\sigma}\right)
+\frac{\partial}{\partial  
Z_\mu}\left(Z_\rho\delta_{\nu\sigma}-Z_\sigma\delta_{\nu\rho}\right)
\Bigr](1-\kappa)D_1(-Z^2)\Bigr\}.\nonumber\\
&&\ear
Here $Z=X-X'$, $G_2$ is the gluon condensate, $D,D_1$ are invariant
functions normalized to
\be\label{2.62}
D(0)=D_1(0)=1\ee
and $\kappa$ is a parameter measuring the non-abelian
character of the correlator.

Indeed, if we consider an abelian theory we have to replace
the gluon field strengths ${\cal G}_{\mu\nu}$ by abelian
field strengths ${\cal F}_{\mu\nu}$, which satisfy the homogenous
Maxwell equation
\be\label{2.63}
\epsilon_{\mu\nu\rho\sigma}\partial_\nu{\cal F}_{\rho\sigma}(X)=0,\ee
if there are no magnetic monopoles present.
It is easy to see that this implies $\kappa=0$ in (\ref{2.61}). Thus, in
an abelian theory, the $D$-term in (\ref{2.61}) is absent without
magnetic monopoles but would be non-zero if the vacuum contained
a magnetic monopole condensate. In the non-abelian theory the
$D$-term has no reason to vanish. In fact, we will see that
it dominates over the $D_1$-term. The abelian analogy suggests
an interpretation of the $D$-term
as being due to an effective chromomagnetic monopole condensate
in the QCD vacuum.

Two further assumptions are made in the SVM:

\begin{itemize}
\item{\bf Ass. 2:} The correlation functions
$D(-Z^2)$ and $D_1(-Z^2)$ fall off rapidly for $Z^2\to\infty$.
There exists a characteristic finite correlation length $a$,
which we define as
\be\label{2.63a}
a:=\int^\infty_0{\rm d}Z\ D(-Z^2).\ee
\end{itemize}

A typical ansatz for the function $D$, incorporating assumption 2,
is as follows:
\be\label{2.64}
D(-Z^2)=\frac{27}{64}a^{-2}\int {\rm d}^4Ke^{iKZ}
K^2\left[K^2+\left(\frac{3\pi}{8a}\right)^2\right]^{-4},\ee
which leads to
\be\label{2.65}
D(-Z^2)\propto\exp\left(-\frac{3\pi|Z|}{8a}\right)
\qquad {\rm for}\ Z^2\to\infty.\ee
The function $D_1$ is chosen such that
\be\label{2.66}
\left(4+Z_\mu\frac{\partial}{\partial Z_\mu}\right)
D_1(-Z^2)=4D(-Z^2)\ee
which leads to
\be\label{2.66a}
D_1(-Z^2)=(Z^2)^{-2}\int^{Z^2}_0{\rm d}v2vD(-v).\ee
With (\ref{2.66}) the contracted field strength correlator
has the form (cf. (\ref{2.61}):
\be\label{2.67}
F_{\mu\nu\mu\nu}=\frac{1}{2}G_2D(-Z^2).\ee

The ansatz  (\ref{2.64}), (\ref{2.66})
can be compared to a lattice gauge theory calculation
of the gluon field strength correlator \cite{43} in order to fix
the parameters. One finds (cf. \cite{43}, \cite{44}
and Fig. 9):
\bear\label{2.68}
a&=&0.35\ {\rm fm},\nonumber\\
\kappa&=&0.74,\nonumber\\
G_2&=&(496\ {\rm MeV})^4,\ear
with an educated guess for the error of $\approx 10 \%$.
Of course, from Fig. 9 we get only the product $\kappa\cdot G_2$.
These quantities are obtained separately by measuring on the lattice
several components of the correlator (\ref{2.60}). The value for
$G_2$ in (\ref{2.68}) from the lattice calculations is somewhat
larger than from phenomenology. This can be understood as follows.
The lattice calculations of \cite{43}
are for the pure gluon theory. In
the real world light quarks are present. Their effect is estimated
to reduce the value of $G_2$ substantially \cite{45}.

We note that the correlation length is smaller,
albeit not much smaller
than a typical radius $R$ of a light hadron
(cf. e.g. \cite{46}, \cite{47}, \cite{48},\cite{49}):
\be\label{2.69}R\sim 0.7-1\ {\rm fm}.\ee
Still
\be\label{2.70}
a^2/R^2\approx 0.2-0.3\ee
is a reasonably small number and this will be important for us in the
following.

We come now to the third assumption made in the SVM:

\begin{itemize}
\item{\bf Ass. 3:} Factorization of higher point gluon
field strength correlators.
\end{itemize}

In detail assumption 3 reads as follows (cf. \cite{44}):

All expectation values of an odd number of products
of shifted field strengths vanish:
\be\label{2.71}
\langle\hat{\G}(1)...\hat{\G}(2n+1)\rangle=0
\quad{\rm for}\quad n=0,1,2,... .\ee
Here we set as shorthand
\be\label{2.72}
\hat{\G}(i)\equiv\hat{\G}^{a_i}_{\alpha_i
\beta_i}(Y,X_i;C_{X_i}).\ee
For even number of shifted field strengths we set in the
SVM:
\bear\label{2.73}
\langle\hat{\G}(1)...\hat{\G}(2n)\rangle&=&
\sum_{all\ pairings}\langle\hat{\G}(i_1)\hat G(j_1)\rangle
...\langle\hat{\G}(i_n)\hat{\G}(j_n)\rangle,\nonumber\\
&&(i_1,j_1)...(i_n,j_n)\ear
where $n=1,2,...$ .

We note that $\langle\hat{\G}(1)\rangle$ must vanish
due to colour conservation since the QCD vacuum has no
preferred direction in colour space. The vanishing
of the other correlators of odd numbers of field
strengths, postulated in (\ref{2.71}), as well as the
factorization property (\ref{2.73}) are strong dynamical
assumptions. They mean that the vacuum fluctuations are assumed
to be of the simplest type: a Gaussian random process.

For some applications of the SVM, for instance the
calculation of the Wegner-Wilson loop described below,
assumption 3 is not crucial and can be relaxed. But for the
applications of the SVM to high energy scattering (cf. sect. 3)
assumption 3 is crucial. In any case we prefer to specify
the model completely, thus giving it maximal predictive power.
On the other hand, of course, the model can then more easily
run into difficulties in comparison with
experiments.

\subsection{The Wegner-Wilson loop in the
stochastic \protect\\vacuum model}

We have now specified the SVM completely and can proceed to show
how this model produces \underbar{confinement}. We consider a static
quark-antiquark pair at distance $R$ from each other and ask for the
potential $V(R)$. To calculate $V(R)$ we start with a rectangular
Wegner-Wilson loop in the $X_1-X_4$ plane (Fig. 10): Let $C$ be
the loop and $S$ the rectangle, $C=\partial S$. Then
\be\label{2.74}
W(C)=\frac{1}{3}\langle {\Tr}\ {\rm P}\ {\exp}
[-ig\int_C {\rm d} Z_\mu {\cal G}_\mu(Z)]\rangle
\ee
and
\be\label{2.75}
V(R)=-\lim_{T\to\infty}\frac{1}{T}\ln W(C).\ee
To evaluate $W(C)$ in the SVM we first transform the line integral
of the potentials in (\ref{2.74}) into a surface integral of field
strengths, using the non-abelian version of Stokes theorem
(cf. sect. 2.2). For this we choose a reference point $Y$ in $S$.
We get then
\be\label{2.76}
W(C)=\frac{1}{3}\langle {\Tr}\ {\rm P}\ {\exp}[-ig\int_S {\rm d}
X_1{\rm d}X_4
\hat{\cal G}_{14}(Y,X,C_X)]\rangle\ee
where $\hat{\cal G}_{\mu\nu}$ are the field
strengths parallel-transported from $X$ to $Y$ along a straight line
$C_X$. The path-ordered way to integrate over $S$ in a fan-type
net (cf. Fig. 8) is indicated by ${\rm P}$. To evaluate the expectation
value of the path-ordered exponential in (\ref{2.76}), we
will use the technique of the cumulant expansion and the
assumptions 1-3 of the SVM. We make the following replacements
in the formulae (\ref{2.38}) ff. of sect. 2.3:
\bear\label{2.77}
&&\tau\to(X_1,X_4),\nonumber\\
&&B(\tau_i)\to\hat{\G}_{14}(Y,X^{(i)},C_{X^{(i)}}),
\nonumber\\
&&E(B(\tau_1)..B(\tau_n))\to\frac{1}{3}{\Tr}\langle\hat{\G}
_{14}(Y,X^{(1)};C_{X^{(1)}})...\hat{\G}_{14}(Y,X^{(n)};
C_{X^{(n)}})\rangle,\nonumber\\
&&t\to -ig,\nonumber\\
&&f(t)\to W(C).\ear
With this we can express $W(C)$ as an exponential of cumulants
as given in (\ref{2.41}). From Ass. 3 of the SVM we get that
all cumulants for odd numbers of gluon field strengths vanish.
The lowest nontrivial cumulant is the second one, $K_2$, and from
(\ref{2.51}) we find
\be\label{2.78}
K_2(1,2)\to\frac{1}{3}{\Tr}\langle {\rm P}(\hat{\G}
_{14}(Y,X^{(1)};C_{X^{(1)}})\hat{\G}_{14}(Y,X^{(2)};C_{X^{(2)}}
)\rangle.\ee
If we cut off the cumulant expansion (\ref{2.41}) at $n=2$ we get
then for the Wegner-Wilson loop
\bear\label{2.79}
&&W(C)=\exp\Bigl\{-\frac{g^2}{2}\int_S {\rm d}
X_1{\rm d}X_4\int_S {\rm d}X_1'{\rm d}X_4'
\nonumber\\
&&\langle\frac{1}{3}{\Tr}\ {\rm P}\ [\hat{\cal G}_{14}(Y,X,C_X)\hat{\cal  
G}_{14}(Y,X',C_{X'})]
\rangle\Bigr\}\nonumber\\
&&=\exp\left\{-\frac{2\pi^2}{3}\int_S{\di}X_1{\di}X_4\int_S
{\di}X_1'{\di}X_4'F_{1414}(X-X')\right\},\ear
where in the last step we used (\ref{2.61}), i.e. Ass. 1
of the SVM. Next
the assumption 2 of short-range correlation for the field
strengths enters in a crucial way. This implies that for large
Wegner-Wilson loops the integration over $X_1',X_4'$ keeping
$X_1,X_4$ fixed in (\ref{2.79}) gives essentially a factor $a^2$.
The remaining $X_1,X_4$ integration gives then the area of $S=RT$.
Thus we arrive at an area law for the Wegner-Wilson loop for $R,T\gg a$:
\be\label{2.80}
W(C)=e^{-\sigma\cdot R\cdot T},\ee
where the constant $\sigma$ is obtained as
\be\label{2.81}
\sigma=\frac{\pi^3\kappa G_2}{36}\int^\infty_0
{\rm d}Z^2D(-Z^2)=\frac{32\pi\kappa G_2a^2}{81}.\ee
We leave the proof of (\ref{2.81}) as an exercise for the reader.
Comparing (\ref{2.75}) and (\ref{2.80}) we find that the
SVM produces a linearly rising potential
\be\label{2.82}
V(R)=\sigma\cdot R\quad{\rm for}\quad R\gg a,\ee
where $\sigma$ is the string tension.

The results (2.82)-(2.84)  were derived in the framework
of the SVM in \cite{39}, where the model was introduced,
and they are
interesting in quite a number of respects:

Only for $\kappa\not=0$ does one get an area law and thus confinement.
The $D_1$ term which is the only one present in the abelian theory
produces no confinement. In the SVM confinement
is related to an effective chromomagnetic monopole condensate
in the vacuum.

The \underbar{short range} correlation for the field strengths
produces a \underbar{long range} correlation for the potentials
if the $D$ term is present in (\ref{2.61}), i.e. if $\kappa\not=0$.

The string tension $\sigma$ is obtained numerically with the
input (\ref{2.68}) as
\be\label{2.83}
\sigma=(420\ {\rm MeV})^2\ee
This is very consistent with the phenomenological determination
of $\sigma\simeq (430\ {\rm MeV})^2$ from the charmonium spectrum
\cite{50}.

The SVM has been applied in many other studies of low energy
hadronic phenomena (cf. \cite{38} for a review).
It was, for instance, possible to calculate flux distributions
around a static quark-antiquark pair \cite{51}.
The results compare well with lattice gauge theory calculations
wherever the latter are available.

Let us come back to the calculation of the Wegner-Wilson loop
above. It is legitimate to ask about the contribution of higher
cumulants to $W(C)$. How do they modify (\ref{2.79})?
It turns out that higher cumulants
may cause some problems, which
we discuss in Appendix A together with a proposal for their
remedy.

\section{Soft Hadronic Reactions}
\setcounter{equation}{0}
\subsection{General considerations}
In this section we will present a microscopic approach towards
hadron-hadron diffractive scattering (cf. \cite{25}, \cite{52}).
Consider as an example elastic scattering of two hadrons $h_1,h_2$
\be\label{3.1}
h_1+h_2\to h_1+h_2\ee
at high energies and small momentum transfer. We will look at
reaction (\ref{3.1}) from the point of view of an
observer living in the ``femto-universe'', i.e. we
imagine having a microscope with resolution much better
than 1 fm for observing what happens during the collision. Of course,
we should choose an \underbar{appropriate resolution}
for our microscope. If we choose the resolution much too good,
we will see too many details of the internal structure of the
hadrons which are irrelevant for the reaction considered and
we will miss the essential features. The same is true if the
resolution is too poor. In \cite{52} we used a series of simple
arguments based on the uncertainty relation to estimate this
appropriate resolution.

Let $t=0$ be the nominal collision time of the hadrons in (\ref{3.1})
in the c.m. system. This is the time when the hadrons $h_{1,2}$
have maximal spatial overlap. Let furthermore be $t_0/2$ the
time when, in an inelastic collision, the first produced hadrons
appear. We estimate $t_0\approx 2$ fm from the LUND model of
particle production \cite{53}. Then the appropriate resolution, i.e.
the cutoff in transverse parton momenta $k_T$ of the hadronic wave
functions to be chosen for describing reaction (\ref{3.1}) in
an economical way is
\be\label{3.2}
k^2_T\leq\sqrt s/(2t_0)\ee
where $\sqrt s$ is the c.m. energy. Modes with higher $k_T$ can be
assumed to be integrated out. With this we could argue that over the
time interval
\be\label{3.3}
-\frac{1}{2}t_0\leq t\leq \frac{1}{2}t_0\ee
the following should hold or better: could be assumed:

(a) The parton state of the hadrons does not change qualitatively,
i.e. parton annihilation and production processes can be neglected
for this time.

(b) Partons travel in essence on straight lightlike world lines.

(c) The partons undergo ``soft'' elastic scattering.

The strategy is now to study first soft parton-parton
scattering in the femto-universe. There, the relevant interaction
will turn out to be mediated by the gluonic vacuum fluctuations.
We have argued at length in section 2 that these have a highly
nonperturbative character. In this way the nonperturbative
\underbar{QCD vacuum structure} will enter the picture for
high energy soft hadronic reactions. Once we have solved the
problem of parton-parton scattering we have to fold the
partonic $S$-matrix with the hadronic wave functions of the
appropriate resolution (\ref{3.2}) to get the hadronic
$S$-matrix elements.

We will now give an outline of the various steps in this program.

\subsection{The functional integral approach to \protect\\
parton-parton scattering}

Consider first quark-quark scattering:
\be\label{3.4}
q(p_1)+q(p_2)\to q(p_3)+q(p_4),\ee
where we set

\newpage

\bear\label{3.5}
&&s=(p_1+p_2)^2\nonumber\\
&&t=(p_1-p_3)^2\nonumber\\
&&u=(p_1-p_4)^2.\ear
Of course, free quarks do not exist in QCD, but let us close our
eyes to this at the moment.
Now we should calculate the scattering of the quarks
over the finite time interval (\ref{3.3}) of length
$t_0\approx$ 2 fm. Let us assume that 2 fm is nearly infinitely
long on the scale of the femto universe and use the
standard reduction formula, due to Lehmann,
Symanzik, and Zimmermann, to relate the $S$-matrix element
for (\ref{3.4}) to an integral over the 4-point function
of the quark fields. We use the following normalization for
our quark states
\bear\label{3.5a}
\langle q(p_j,s_j,A_j)&|& q(p_k,s_k,A_k)\rangle\nonumber\\
&&=\delta_{s_js_k}\delta_{A_jA_k}(2\pi)^3\sqrt{2p^0_j
2p^0_k}\delta^3(\vec p_j-\vec p_k)\nonumber\\
&&\equiv \delta(j,k),\ear
where $s_j,s_k$ are the spin and $A_j,A_k$ the
colour indices. With this we get
\bear\label{3.6}
&&\langle q(p_3,s_3,A_3)q(p_4,s_4,A_4)|S|q(p_1,
s_1,A_1)q(p_2, s_2,A_2)\rangle\nonumber\\
&&\equiv
\langle3,4|S|1,2\rangle\nonumber\\
&&=\langle3,4|1,2\rangle+Z^{-2}_\psi
\left\{(4|(i\rvec-m_q')\otimes(3|(i\rvec-m_q')
\right.\nonumber\\
&&\hphantom{=\langle3,4|1,2\rangle+Z^{-2}_\psi
\left\{\right.}
\langle 0|{\rm T}(q(4)q(3)\bar q(1)\bar q(2))|0\rangle
\nonumber\\
&&\left.\hphantom{=\langle3,4|1,2\rangle+Z^{-2}_\psi
\left\{\right.}
(i\lvec+m_q')|1)\otimes
(i\lvec+m_q')|2)\right\}.\ear
Here $Z_\psi$ is the quark wave function renormalization constant and
$m_q'$ the renormalized quark mass. We use a shorthand
notation
\bear\label{3.7}
&&|j)\equiv {\it u}_{s_j,A_j}(p_j)e^{-ip_jx_j},\nonumber\\
&&(j|=e^{ip_jx_j}\bar {\it u}_{s_j,A_j}(p_j),\nonumber\\
&&q(j)\equiv q(x_j),\nonumber\\
&&(j=1,..,4),\ear
where ${\it u}$ is the spinor in Dirac and colour space.
Two repeated arguments $j,k,...$ imply a space-time
integration, for instance
\be\label{3.8}
\bar q(1)(i\lvec+m_q')|1)\equiv
\int {\rm d}x_1\bar q(x_1)(i\lvec+m_q')
e^{-ip_1x_1}{\it u}_{s_1,A_1}(p_1).\ee
Thus in (\ref{3.6}) we have four integrations over
$x_1,...,x_4$.

We can represent the 4-point function of the quark
fields as a functional integral:
\bear\label{3.9}
&&\langle 0|{\rm T}(q(4)q(3)\bar q(1)\bar q(2))|0\rangle\nonumber\\
&&=Z^{-1}\int{\cal D}(G,q,\bar q)\exp\big\{i\int
{\rm d}x{\cal L}_{\QCD}
(x)\big\}q(4)q(3)\bar q(1)\bar q(2),\ear
where $Z$ is the partition function:
\be\label{3.10}
Z=\int{\cal D}(G,q,\bar q)\exp\big\{i\int {\rm d}x{\Li}_{\QCD}(x)
\big\}.\ee

The QCD Lagrangian (\ref{1.1}) is bilinear in the quark and
antiquark fields. Thus -- as is well known -- the functional
integration over $q$ and $\bar q$ can be carried out immediately.
After some standard manipulations we arrive at the
following expression:
\bear\label{3.11}
&&\langle 0|{\rm T}(q(4)q(3)\bar q(1)\bar q(2))|0\rangle\nonumber\\
&&=\frac{1}{Z}\int{\cal D}(G)\exp\left\{-i\int {\rm d}
x\frac{1}{2}{\rm Tr}
(G_{\lambda\rho}(x)G^{\lambda\rho}(x))\right\}\nonumber\\
&&\det[-i(i\gamma^\lambda D_\lambda-m_q+i\epsilon)]\nonumber\\
&&\left\{\frac{1}{i}S_F(4,2;G)\frac{1}{i}S_F(3,1;G)
-(3\leftrightarrow4)\right\}.\ear
Here $S_F(j,k;G)\equiv S_F(x_j,x_k;G)$
is the unrenormalized quark propagator in the given
gluon potential $G_\lambda(x)$. We have
\be\label{3.12}
(i\gamma^\mu D_\mu-m_q)S_F(x,y;G)=-\delta(x-y).\ee
Functional integrals as in (\ref{3.11}) will occur frequently
further on. Let $F(G)$ be some functional of the gluon
potentials. We will denote the functional integral over
$F(G)$ by brackets $\langle F(G)\rangle_G$:
\bear\label{3.13}
&&\langle F(G)\rangle_G:=\frac{1}{Z}\int {\cal D}(G)
\exp\left\{-i\int {\di}x\frac{1}{2}{\Tr}
(G_{\lambda\rho}G^{\lambda\rho})
\right\}\cdot\nonumber\\
&&{\det}[-i(i\gamma^\lambda D_\lambda-m_q+ i\epsilon)]
F(G).\ear

Now we insert (\ref{3.11}) in (\ref{3.6}) and get
\be\label{3.14}
\langle 3,4|S|1,2\rangle=\langle3,4|1,2\rangle-Z_\psi^{-2}
\langle{\M}^F_{31}(G){\M}^F_{42}(G)-(3\leftrightarrow 4)
\rangle_G,\ee
where
\bear\label{3.15}
&&{\M}^F_{kj}(G)=(k|(i\rvec-m_q')S_F(i\lvec+m_q')
|j),\nonumber\\
&&(k=3,4;\ j=1,2).
\ear

The term ${\M}^F_{31}\cdot{\M}^F_{42}$ on the r.h.s. of
(\ref{3.14}) corresponds to the $t$-channel exchange diagrams,
the second term, where the role of the quarks 3 and 4 is
interchanged, to the $u$-channel exchange diagrams (Fig. 11).
The latter term should be unimportant for high energy, small
$|t|$ scattering. Thus we neglect it in the following.
For the scattering of different quark flavours it
is absent anyway. We set, therefore:
\be\label{3.15a}
\langle 3,4|S|1,2\rangle\cong\langle 3,4|1,2\rangle-
Z_\psi^{-2}\langle{\M}^F_{31}(G){\M}^F_{42}(G)\rangle_G.\ee

We can interpret ${\M}^F_{kj}(G)$ as scattering amplitude for
quark $j$ going to $k$ in the fixed gluon potential
$G_\lambda(x)$. To see this, let us define the wave function
\be\label{3.16}
|\psi^F_{p_j})=S_F(i\lvec+m_q')|j)\ee
which satisfies the Dirac equation with the gluon potential
$G_\lambda(x)$:
\be\label{3.17}
(i\gamma^\lambda D_\lambda-m_q)|\psi^F_{p_j})=0.\ee
Furthermore we use the Lippmann-Schwinger equation for $S_F$:
\be\label{3.18}
S_F=S_F^{(0)}-S_F^{(0)}(g{\slG}-\delta m_q)S_F\ee
where $S_F^{(0)}$ is the free quark propagator
for mass $m_q'$ and $\delta m_q=
m_q'-m_q$ is the quark mass shift. Inserting (\ref{3.18})
and (\ref{3.16}) in (\ref{3.15}) gives after some simple
algebra
\be\label{3.19}
{\M}^F_{kj}(G)=(p_k|(g{\slG}-\delta m)|\psi^F_{p_j}).\ee
This represents ${\M}^F_{kj}$ in the form a scattering amplitude
should have: a complete incoming wave is folded with the potential
and the free outgoing wave. However, there is a small problem.
The wave function $|\psi^F_{p_j})$ defined in (\ref{3.16})
does not satisfy the boundary condition which we should have
for using it in the scattering amplitude, i.e. it does not go
to a free incoming wave for time $t\to-\infty$. The wave function
with this boundary condition is obtained by replacing the
Feynman propagator $S_F$ in (\ref{3.16}) by the retarded one, $S_r$.
We have shown in \cite{52} that in the \underbar{high energy
limit} this replacement can indeed be justified for the
calculation of ${\M}^F_{kj}(G)$ if the gluon potential
$G_\lambda(x)$ contains only a limited range of frequencies:
\be\label{3.20}
{\M}_{kj}^F(G)\simeq{\M}^r_{kj}(G)=(p_k|(g{\slG}-\delta m)
|\psi^r_{p_j}),\ee
where
\be\label{3.21}
|\psi^r_{p_j})=S_r(i\lvec+m_q')|j),\ee
which satisfies
\be\label{3.22}
(i\gamma^\mu D_\mu-m_q)|\psi^r_{p_j})=0,\ee
\be\label{3.23}
|\psi_{p_j}^r)\to|j)\quad{\rm for}\quad t\to-\infty.\ee

We summarize the results of this subsection: At high energies and
small $|t|$ the quark-quark scattering amplitude can be obtained
by calculating first the scattering of quark 1 going to 3
and 2 going to 4 in the same fixed gluon potential $G_\lambda(x)$.
Let the corresponding scattering amplitudes be ${\M}^r_{31}(G)$ and
${\M}^r_{42}(G)$ (cf. (\ref{3.20})). Then the product of these
two amplitudes is to be integrated over all gluon potentials with
the measure given by the functional integral (\ref{3.13})
and this gives the quark-quark scattering amplitude (\ref{3.15a}).
The point of our further strategy is to continue making suitable
high energy approximations in the integrand of this functional
integral which will be evaluated finally using the methods of the
stochastic vacuum model.

In the following it will be convenient to choose a coordinate system
for the description of reaction (\ref{3.4}) where the quarks
1,3 move with high velocity in essence in positive $x^3$ direction,
the quarks 2,4 in negative $x^3$ direction. We define the light
cone coordinates
\be\label{3.24}
x_\pm=x^0\pm x^3\ee
and in a similar way the $\pm$ components of any 4-vector. With
this we have for the 4-momenta of our quarks:
\be\label{3.25}
p_j=\left(\begin{array}{ccc}
\frac{1}{2}p_{j+}&+&\frac{{\vec p}_{jT}^2+{m'}_q^2}{2p_{j+}}\\
&{\vec p}_{jT}&\\
\frac{1}{2}p_{j+}&-&\frac{{\vec p}_{jT}^2+
m_{q'}^2}{2p_{j+}}\end{array}\right)\ee
for $j=1,3$ with $p_{j+}\to\infty$ and
\be\label{3.26}
p_k=\left(\begin{array}{ccc}
\frac{1}{2}p_{k-}&+&\frac{{\vec p}_{kT}^2+{m'}_{q'}^2}{2p_{k-}}\\
&{\vec p}_{kT}&\\
-\frac{1}{2}p_{k-}&+&\frac{{\vec p}_{kT}^2+
m_{q'}^2}{2p_{k-}}\end{array}\right)\ee
for $k=2,4$ with $p_{k-}\to\infty$.

\subsection{The Eikonal expansion}

The problem is now to solve the Dirac equation (\ref{3.22})
for arbitrary external gluon potential $G_\lambda(x)$.
Of course, we cannot do this exactly. But we are only
interested in the high energy, small $|t|$ limit. This
suggests to use an eikonal type approach. This works indeed,
but it is not as straightforward as one would think at first,
since the Dirac equation is of first order in the derivatives, whereas
the eikonal expansion is easy to make for for a second-order
differential equation. What we did in \cite{52} was to make an
ansatz for the Dirac field $\psi_{p_j}^r(x)$ in terms
of a ``potential'' $\phi_j(x)$ as follows:
\be\label{3.27}
\psi_{p_j}^r(x)=(i\gamma^\lambda D_\lambda+m_q)\phi_j(x).\ee
A suitable boundary condition for $\phi_j(x)$ which is compatible
with (\ref{3.23}) is
\be\label{3.28}
\phi_j(x)\to\frac{1}{p_j^0+m_q}
\frac{1+\gamma^0}{2}e^{-ip_jx}{\it u}(p_j)\ee
for $t\to-\infty$. Inserting (\ref{3.27}) into the Dirac equation
(\ref{3.22}) gives:
\be\label{3.29}
\{i\gamma^\lambda D_\lambda-m_q\}\{i\gamma^\rho D_\rho
+m_q\}\phi_j(x)=0.\ee
For the case of no gluon field, $G_\lambda(x)=0$, the covariant
derivatives $D_\lambda$ degenerate to the ordinary ones,
$\partial_\lambda$, and (\ref{3.29}) to the Klein-Gordon equation:
\be\label{3.30}
({\dA}+m^2_q)\phi_j(x)=0.\ee
Thus the problem for the Dirac field is in essence
reduced to a scalar field-type problem which is handled
more easily.

Now it is more or less straightforward to turn the theoretical
crank and to obtain the eikonal approximations for $\phi_j(x)$
and $\psi_{p_j}^r(x)$, respectively. Take $j=1$ as an example.
We make the ansatz
\be\label{3.31}
\phi_1(x)=e^{-ip_1x}\tilde\phi_1(x).\ee
Inserting $p_1$ from (\ref{3.25}) we see that on the
r.h.s. of (\ref{3.31}) the fast varying phase factor is
\[\exp(-i\frac{1}{2}p_{1+}x_-),\]
since $p_{1+}\to\infty$. Assuming all the remaining
factors to vary slowly with $x$ we insert (\ref{3.31})
in (\ref{3.29}) and order the resulting
terms in powers of $1/p_{1+}$.
The solution of (\ref{3.29}) to leading order in $1/p_{1+}$
is then easily obtained. The final formula for $\psi^r_{p_1}(x)$
reads:
\be\label{3.32}
\psi^r_{p_1}(x)=V_-(x_+,x_-,\vec x_T).\left\{1
+O\left(\frac{1}{p_{1+}}\right)\right\}e^{-ip_1x}{\it u}(p_1),\ee
where
\be\label{3.33}
V_-(x_+,x_-,\vec x_T)={\rm P}\left\{\exp\left[-\frac{i}{2}
g\int^{x_+}_{-\infty}{\di}x_+'G_-(x_+',x_-,\vec x_T)
\right]\right\}\ee
and P means path ordering. When coming in, the quark
picks up a \underbar{non-abeli-}\\
\underbar{an phase factor}, just the
ordered integral of $G$ along the path.
Of course, $V_-$ is a connector, as studied in section
2.1, but now in Minkowski space and for a straight light-like
line running from $-\infty$ to $x$ (Fig. 12).

In a similar way we obtain for the quark with initial
momentum $p_2$, i.e. the one coming in from
the right, for $p_{2-}\to\infty$:
\be\label{3.34}
\psi_{p_2}^r(x)=V_+(x_+,x_-,\vec x_T)\left\{
1+O\left(\frac{1}{p_{2-}}\right)\right\}
e^{-ip_2x}{\it u}(p_2),\ee
where
\be\label{3.35}
V_+(x_+,x_-,\vec x_T)={\rm P}\left\{\exp\left[-\frac{i}{2}g
\int^{x_-}_{-\infty}{\di}x_-'G_+(x_+,x_-',\vec x_T)
\right]\right\}.\ee
Recall that
\be\label{3.36}
G_\pm(x)=\left(G^{0a}(x)\pm G^{3a}(x)\right)\frac{\lambda
_a}{2}\ee
are matrices in colour space. Thus path ordering in (\ref{3.33})
and (\ref{3.35}) is essential.

A solution for $\psi_{p_1}^r(\psi^r_{p_2})$ as a series
expansion in powers of $1/p_{1+}(1/p_{2-})$
was obtained to all orders in \cite{54}.

\subsection{The quark-quark scattering amplitude}

We can now insert our high energy approximations (\ref{3.32}),
(\ref{3.34}) for $\psi_{p_{1,2}}^r(x)$ in the expression for
${\cal M}_{kj}^{(r)}$ in (\ref{3.20}). The resulting
integrals are easily done and we get for $p_{1+},p_{3+}\to
\infty$:
\bear\label{3.37}
{\cal M}^r_{31}(G)&\to&\int {\di}x e^{i(p_3-p_1)x}\nonumber\\
&&\bar {\it u}(p_3)
(g{\slG}(x)-\delta m)V_-(x)u(p_1)\nonumber\\
&\to& \frac{i}{2}\int {\di}x_+{\di}x_-{\di}^2x_T\ \exp\left[
\frac{i}{2}(p_3-p_1)_+x_--i(\vec p_3-\vec p_1)_T\cdot\vec x_T
\right]\nonumber\\
&&\bar u(p_3)\gamma_+\frac{\partial}{\partial x_+}V_-(x_+,x_-
,\vec x_T){\it u}(p_1)\nonumber\\
&\to&i\sqrt{p_{3+}p_{1+}}\cdot\delta_{s_3,s_1}\int {\di}x_-{\di}
^2x_T\nonumber\\
&&\exp[\frac{i}{2}(p_3-p_1)_+x_--i(\vec p_3-\vec p_1)_T\cdot
\vec x_T]\nonumber\\
&&[V_-(\infty,x_-,\vec x_T)-\eins]_{A_3,A_1}.\ear
In a similar way we obtain for $p_{2-},p_{4-}\to\infty$:
\bear\label{3.38}
{\cal M}^r_{42}(G)&\to&i\sqrt{p_{4-}\cdot p_{2-}}\cdot
\delta_{s_4,s_2}\nonumber\\
&&\int {\di}y_+{\di}^2y_T\exp[\frac{i}{2}(p_4-p_2)_-
y_+-i(\vec p_4-\vec p_2)_T\cdot\vec y_T]\nonumber\\
&&[V_+(y_+,\infty,\vec y_T)-\eins]_{A_4,A_2}.\ear
Here we have written out the spin and colour indices
of the in- and outgoing quarks (cf. (\ref{3.5a}), (\ref{3.6})).
We have used furthermore:
\bear\label{3.39}
&&\bar u_{s_3}(p_3)\gamma^\mu u_{s_1}(p_1)\to
\sqrt{p_{3+}p_{1+}}\cdot\delta_{s_3,s_1}n^\mu_+\quad
{\rm for}\quad p_{1,3+} \to\infty,\nonumber\\
&&\bar u_{s_4}(p_4)\gamma^\mu u_{s_2}(p_2)\to
\sqrt{p_{4-}p_{2-}}\cdot\delta_{s_4,s_2}n^\mu_-\quad
{\rm for}\quad p_{2,4-} \to\infty,\ear
where
\be\label{3.40}
n_\pm^\mu=\left(\begin{array}{c}
1\\0\\0\\ \pm1\end{array}\right).\ee
Finally, the $x_+(x_-)$ integration for ${\M}_{31}^r({\cal M}
^r_{42})$ could be done with the help of (\ref{2.15}) or rather
the analogous equation for connectors in Minkowski space time.

Now we can insert everything in our expression (\ref{3.15a})
for the $S$-matrix element. This gives:
\bear\label{3.41}
&&\langle 3,4|S|1,2\rangle=\langle3,4|1,2\rangle+\nonumber\\
&&\sqrt{p_{3+}p_{1+}p_{4-}p_{2-}}
\cdot\delta_{s_3,s_1}\delta_{s_4,s_2}Z^{-2}_\psi\int {\di}x_-
{\di}^2x_T\int {\di}y_+{\di}^2y_T\nonumber\\
&&\exp\big[\frac{i}{2}(p_3-p_1)_+x_--i(\vec p_3-\vec p_1)_T\cdot\vec
x_T\big]\nonumber\\
&&\exp\big[\frac{i}{2}(p_4-p_2)_-y_+-i(\vec p_4-\vec p_2)_T
\cdot\vec y_T\big]\nonumber\\
&&\left\langle[V_-(\infty,x_-,\vec x_T)-\eins]_{A_3,A_1}
[V_+(y_+,\infty,\vec y_T)-\eins]_{A_4,A_2}\right\rangle_G.
\ear From translational invariance of the functional integral
we have:
\bear\label{3.42}
&&\langle[V_-(\infty,x_-,x_T)-\eins]_{A_3,A_1}
[V_+(y_+,\infty,\vec y_T)-\eins]_{A_4,A_2}
\rangle_G\nonumber\\
&&=\langle[V_-(\infty,0,\vec x_T-\vec y_T)-\eins]_{A_3,A_1}
[V_+(0,\infty,0)-\eins]_{A_4,A_2}\rangle_G.\ear
Inserting (\ref{3.42}) in (\ref{3.41}), we can pull out
the $\delta$-function for the overall energy-momentum
conservation and we get finally:
\bear\label{3.43}
&&\langle3,4|S|1,2\rangle=\langle3,4|1,2\rangle+
+i(2\pi)^4\delta(p_3+p_4-p_1-p_2)\langle3,4|T|1,2\rangle,
\nonumber\\[0.2cm]
&&\langle3,4|T|1,2\rangle=i2\sqrt{p_{3+}p_{1+}p_{4-}p_{2-}}\cdot
\delta_{s_3,s_1}\delta_{s_4,s_2}(-
Z^{-2}_\psi)\int {\di}^2z_Te^{i\vec q_T\cdot\vec z_T}\nonumber\\
&&\left\langle[V_-(\infty,0,\vec z_T)-\eins]_{A_3,A_1}
[V_+(0,\infty,0)-\eins]_{A_4,A_2}\right\rangle_G.\ear
Here $q$ is the momentum transfer:
\bear\label{3.44}
q&=&p_1-p_3=p_4-p_2,\nonumber\\
q^2&=&t.\ear
In the high energy limit $q$ is purely transverse
\be\label{3.45}
q\to\left(\begin{array}{c}
0\\ \vec q_T\\ 0\end{array}\right),
\quad q^2\to-\vec q^2_T.\ee
Using different techniques the type of formula
(\ref{3.43}) was also obtained in \cite{55}.

In (\ref{3.43}) we still have to calculate the wave function
renormalization constant $Z_\psi$. This can be done by
considering a suitable matrix element of the baryon number
current
\[\frac{1}{3}\bar q(x)\gamma^\mu q(x)\]
which is conserved and, therefore, needs no renormalization.
The result is (cf. \cite{52}):
\be\label{3.45a}
Z_\psi=\frac{1}{3}\langle {\Tr}V_-(\infty,0,0)\rangle_G.\ee

Let us summarize the results obtained so far:

The quark-quark scattering amplitude (\ref{3.43}) is
diagonal in the spin indices. Thus we get helicity conservation
in high energy quark-quark scattering. Using (\ref{3.39})
we can write the spin factor in (\ref{3.43}) as
\be\label{3.45b}
2\sqrt{p_{3+}p_{1+}p_{4-}p_{2-}}\delta_{s_3,s_1}\delta_{s_4,s_2}
\cong \bar u_{s_3}(p_3)\gamma^\mu u_{s_1}(p_1)\bar u_{s_4}(p_4)
\gamma_\mu u_{s_2}(p_2)\ee
for $p_{1,3+}\to\infty$ and $p_{2,4-}\to\infty$.
This $\gamma^\mu\otimes\gamma_\mu$ structure was
postulated for high energy quark-quark scattering in the
Donnachie-Landshoff model for the ``Pomeron'' coupling \cite{22a}.
However, a study of quark-antiquark scattering (cf. \cite{52}
and below) reveals that (\ref{3.43}) does \underbar{not} allow
an interpretation in terms of an effective Lorentz vector
exchange between the quarks. The amplitude (\ref{3.43}) has both,
charge conjugation $C$ even and odd contributions. The $C$ even part
corresponds to the ``Pomeron'' and this is \underbar{not}
a Lorentz-vector exchange, but the coherent sum of spin
2,4,6,... exchanges (cf. \cite{56}, \cite{52}).
The $C$-odd part corresponds to the ``Odderon'' introduced in
\cite{57} and this is indeed a Lorentz vector exchange.

The quark-quark scattering amplitude (\ref{3.43})
is governed by the correlation function of two connectors
or string operators $V_{\pm}$ associated with two
light-like Wegner-Wilson lines (Fig. 13).

The first numerical evaluations of (\ref{3.43}) using the
methods of the SVM were done in \cite{58}. However,
it turned out that quark-quark scattering was calculable from
(\ref{3.43}) for \underbar{abelian} gluons only. Indeed, we
are embarked on a program to reproduce in this way the results
obtained in \cite{24} in the framework of perturbation theory
for high energy scattering \cite{59}. For the
\underbar{non-abelian} gluons difficulties arose having to
do with our neglect of quark \underbar{confinement}. This
was really a blessing in disguise and the solution proposed in
\cite{58} was to consider directly hadron-hadron scattering,
representing the hadrons as $q\bar q$ and $qqq$ wave packets
for mesons and baryons, respectively. We will see below how
this is done.

\subsection{The scattering of systems of quarks,
antiquarks, and gluons}

Let us consider now the scattering of systems of partons. As
an example we study the scattering of two $q\bar q$ pairs
on each other:
\be\label{3.46}
q(1)+\bar q(1')+q(2)+\bar q(2')\to q(3)
+\bar q(3')+q(4)+\bar q(4'),\ee
where we set $q(i)\equiv q(p_i,s_i,A_i)$,
$\bar q(i')\equiv \bar q(p_i',s_i',A_i')\ (i=1,...,4)$
with $p_i,s_i,
A_i\ (p_i',s_i',A_i')$ the momentum, spin, and colour
labels for quarks (antiquarks). We assume the particles
with odd (even) indices $i$ to have very large momentum
components in positive (negative) $x^3$ direction (Fig. 14),
i.e. we assume:
\bear\label{3.47}
&&p_{i+},p_{i+}'\to\infty\quad {\rm for}\quad i\ {\rm odd},
\nonumber\\
&&p_{i-},p_{i-}'\to\infty\quad {\rm for}\quad i\ {\rm even}.
\ear
The transverse momenta are assumed to stay limited.

Of course, the reduction formula can also be applied for
the reaction (\ref{3.46}). We have to be
careful in keeping disconnected
pieces. The further strategy is completely analogous to the one
employed in sect. 3.2 for deriving (\ref{3.6})-(\ref{3.15a}).
As we dropped the $u$-channel exchange diagrams in sect. 3.2,
we drop now all terms which are estimated to give a vanishing
contribution to high energy small momentum transfer scattering.
These terms are characterized by large momenta of order
of the c.m. energy $\sqrt s$
flowing through gluon lines, leading to suppression factors
of order $1/s$. Keeping only the $t$-channel exchange terms
and performing all the steps as done in sections 3.2, 3.3
for quark-quark scattering leads
finally to a very simple answer for the $S$-matrix element
correponding to reaction (\ref{3.46}) in
the limit (\ref{3.47}):
\bear\label{3.48}
&&\langle 3,3',4,4'|S|1,1',2,2'\rangle\to\nonumber\\
&&\langle[\delta(3,1)-iZ^{-1}_\psi{\cal M}^r_{31}(G)]
[\delta(3',1')-iZ^{-1}_\psi{\cal M'}^r_{3'1'}(G)]
\nonumber\\
&&\hphantom{\langle}
[\delta(4,2)-iZ^{-1}_\psi{\cal M}^r_{42}(G)]
[\delta(4',2')-iZ^{-1}_\psi{{\cal M}'}^r_{4'2'}(G)]\rangle_G.
\ear
Here $\delta(i,j)$ is as defined in (\ref{3.5a}),
${\cal M}^r_{31}(G)$ and ${\cal M}^r_{42}(G)$ are
as in (\ref{3.20}) and ${{\cal M}'}^r_{3'1'}(G)$
and ${{\cal M}'}^r_{4'2'}(G)$ are the corresponding
amplitudes for the scattering of antiquarks on the
gluon potential $G_\lambda(x)$. We define
\be\label{3.49}
(j'|=\bar{\it v}_{s_j',A_j'} (p_j')e^{-ip_j'x},\ee
where $\it v_{s_j',A_j'}(p_j')$ is the Dirac and
colour spinor for the antiquark $\bar q(j')$. We have
then with $S_r$ the retarded Green's function for quarks
in the gluon potential $G_\lambda(x)$:
\be\label{3.50}
{{\cal M}'}^r_{k'j'}(G)=-(j'|(i\rvec-m_q')S_r(i\lvec
+m_q')|k'),\ee
where $(k',j')=(3',1'),(4',2')$.

In the high energy limit (\ref{3.47}) the
scattering amplitudes (\ref{3.50}) can again be obtained in
the eikonal approximation. Indeed, we can just use
$C$-invari\-ance to get:
\bear\label{3.51}
&&{{\cal M}'}^r_{3'1'}(G)\to i\sqrt{p_{3+}'p_{1+}'}\cdot\delta_{s_3',
s_1'}\nonumber\\
&&\int {\di}x_-{\di}^2x_T\exp\left[
\frac{i}{2}(p_3'-p_1')_+x_--i(\vec p_3'
-\vec p_1')_T\cdot\vec x_T\right]\nonumber\\
&&[V_-^*(\infty,x_-,\vec x_T)-\eins]_{A_3',A_1'},\ear
\bear\label{3.52}
&&{{\cal M}'}^r_{4'2'}(G)\to i\sqrt{p_{4-}'p_{2-}'}\cdot\delta_{s_4',
s_2'}\nonumber\\
&&\int {\di}y_+{\di}^2y_T\exp\left[
\frac{i}{2}(p_4'-p_2')_-y_+-(\vec p_4'
-\vec p_2')_T\cdot\vec y_T\right]\nonumber\\
&&[V_+^*(y_+,\infty,\vec y_T)-\eins]_{A_4',A_2'}.\ear

In section 3.1 we have argued that over the time
interval (\ref{3.3}) we can neglect parton production
and annihilation processes. For the scattering over such
a time interval we should have an effective wave function
renormalization constant
\be\label{3.52a}
Z_\psi=1,\ee
since the deviation of $Z_\psi$ from 1 is just a measure
of the strength of quark splitting processes: $q\to q+G$
etc. In the calculation of $Z_\psi$ in the framework of the SVM
to be described below, one finds indeed $Z_\psi=1$, showing
the consistency of this approach with the simple physical picture
of sect. 3.1. Anticipating this result we see from (\ref{3.5a}),
(\ref{3.37}), (\ref{3.38}) and (\ref{3.51}),
(\ref{3.52}) that in the $S$-matrix element (\ref{3.48})
the $\delta(k,j)$ $(\delta(k',j'))$ terms cancel with the
$\eins$ terms in ${\cal M}^r_{kj}(G)\ ({{\cal M}'}^r_{k'j'}(G))$
in the limit (\ref{3.47}). This leads us to the
following simple rules for obtaining the $S$-matrix element in the high
energy limit:
For the right-moving quark $(1\to 3)$ we have to insert the
factor:
\bear\label{3.52b}
&&{\cal S}_{q+}(3,1)=\sqrt{p_{3+}p_{1+}}\cdot\delta_{s_3,s_1}
\int {\di}x_-{\di}^2x_T\nonumber\\
&&\exp\left[\frac{i}{2}(p_3-p_1)_+x_--i(\vec p_3-\vec p_1)_T\cdot
\vec x_T\right]\nonumber\\
&&V_-(\infty,x_-,\vec x_T)_{A_3,A_1}.\ear
For the right-moving antiquark $(1'\to3')$ we have to
insert the factor:
\bear\label{3.52c}
&&{\cal S}_{\bar q+}(3',1')=\sqrt{p_{3+}'p_{1+}'}
\cdot\delta_{s_3',s_1'}
\int {\di}x_-{\di}^2x_T\nonumber\\
&&\exp\left[\frac{i}{2}(p_3'-p_1')_+x_--i(\vec p_3'-\vec p_1')_T
\cdot\vec x_T\right]\nonumber\\
&&V^*_-(\infty,x_-,\vec x_T)_{A_3',A_1'}.\ear
For the left-moving quark $(2\to 4)$ and antiquark $(2'\to4')$
we have to exchange the $+$ and $-$ labels everywhere in (\ref{3.52b})
and (\ref{3.52c}). This gives:
\bear\label{3.52d}
&&{\cal S}_{q-}(4,2)=\sqrt{p_{4-}p_{2-}}\cdot\delta_{s_4,s_2}
\int {\di}y_+{\di}^2y_T\nonumber\\
&&\exp\left[\frac{i}{2}(p_4-p_2)_-y_+-i(\vec p_4-\vec p_2)_T
\cdot\vec y_T\right]\nonumber\\
&&V_+(y_+,\infty,\vec y_T)_{A_4,A_2},\ear
\bear\label{3.52e}
&&{\cal S}_{\bar q-}(4',2')=\sqrt{p_{4-}'p_{2-}'}
\cdot\delta_{s_4',s_2'}
\int {\di}y_+{\di}^2y_T\nonumber\\
&&\exp\left[\frac{i}{2}(p_4'-p_2')_-y_+-i(\vec p_4'-\vec p_2')_T
\cdot\vec y_T\right]\nonumber\\
&&V^*_+(y_+,\infty,\vec y_T)_{A_4',A_2'}.\ear

Finally we have to multiply together the factors
${\cal S}_{q^\pm},{\cal S}_{\bar q^{\pm}}$ and
integrate over all gluon potentials with the functional
integral measure (\ref{3.13}) to get
\be\label{3.52f}
\langle 3,3',4,4'|S|1,1',2,2'\rangle=
\langle{\cal S}_{q+}(3,1){\cal S}_{\bar q+}(3',1'){\cal S}
_{q-}(4,2){\cal S}_{\bar q-}(4',2')\rangle_G.\ee

Going from quarks to antiquarks corresponds, of course, just
to the change from the fundamental representation (3) of $SU(3)_c$
to the complex conjugate representation $(3^*)$ as we see by
comparing (\ref{3.52b}) with (\ref{3.52c}) and (\ref{3.52d})
with (\ref{3.52e}).

It is an easy exercise to show that these rules can be generalized
in an obvious way for the scattering of arbitrary systems of
quarks and antiquarks on each other. Here we always assume that
we have one distinguished collision axis and that one group
of partons moves with momenta approaching infinity to the
right, the other group to the left. The transverse
momenta are assumed to stay limited.

In appendix B we show that these rules can also be extended
to gluons participating in the scattering. We simply have
to change the colour representation in (\ref{3.52b}),
(\ref{3.52d}) from the fundamental to the adjoint one.
In detail we find that for a right-moving gluon
\be\label{3.52g}
G(p_1,j_1,a_1)\to G(p_3,j_3,a_3)\ee
the following factor has to be inserted in the
$S$-matrix element (cf. appendix B):
\bear\label{3.52h}
&&{\cal S}_{G+}(3,1)=\sqrt{p_{3+}p_{1+}}\cdot\delta_{j_3,j_1}
\nonumber\\
&&\int {\di}x_-{\di}^2x_T\exp\left[\frac{i}{2}
(p_3-p_1)_+x_--i(\vec p_3-\vec p_1)_T
\cdot\vec x_T\right]\nonumber\\
&&{\cal V}_-(\infty,x_-,\vec x_T)_{a_3,a_1}.\ear
Here $j_{1,3}$ are the spin indices which are
purely transverse, $1\leq j_{1,3}\leq 2$. The colour indices
are $a_1,a_3$ with $1\leq a_{1,3}\leq 8$ and ${\cal V}$ is the
connector for the adjoint representation of $SU(3)_c$ (cf.
(\ref{B.22})).

For a left-moving gluon we have again to exchange
$+$ and $-$ labels in (\ref{3.52h}).

\subsection{ The scattering of wave packets of partons 
\protect\\ representing mesons}

In this section we will go from the parton-parton to hadron-hadron
scattering. Our strategy will be to represent hadrons by wave
packets of partons, where we make simple ``Ans\"atze'' for the wave
functions. Then the partonic $S$-matrix element obtained by the
rules derived in section 3.5 will be folded with these wave
functions to give the hadronic $S$-matrix elements.
Of course, we always work in the limit of high energies and
small momentum transfers.

Let us start by considering meson-meson scattering:
\be\label{3.53}
M_1(P_1)+M_2(P_2)\to M_3(P_3)+M_4(P_4),\ee
where $M_{1,3}$ are again the right movers, $M_{2,4}$ the left
movers. We make simple ``Ans\"atze'' for the mesons
as $q\bar q$ wave packets as follows:
\bear\label{3.53a}
&&|M_j(P_j)\rangle=\int {\di}^2p_T\int^1_0
{\di}\zeta\frac{1}{(2\pi)^{3/2}}
h^j_{s_j,s_j'}(\zeta,\vec p_T)\nonumber\\
&&\frac{1}{\sqrt3}\delta_{A_j,A_j'}|q(p_j,s_j,A_j),
\bar q(P_j-p_j,s_j',A_j')\rangle\nonumber\\
&&(j=1,...,4),\ear
where for $j=1,3$:
\bear\label{3.54}
&&P_{j+}\to\infty\nonumber\\
&&p_{j+}=\zeta P_{j+},\nonumber\\
&&\vec p_{jT}=\frac{1}{2}\vec P_{jT}+\vec p_T\ear
and for $j=2,4$:
\bear\label{3.55}
&&P_{j-}\to\infty\nonumber\\
&&p_{j-}=\zeta P_{j-},\nonumber\\
&&\vec p_{jT}=\frac{1}{2}\vec P_{jT}+\vec p_T.\ear
Here $\zeta$ is the longitudinal momentum fraction of the
quark in the meson, $\vec p_T$ is the relative transverse
momentum of $q$ and $\bar q$.

We stress that we are not restricting ourselves to spin 0 mesons
only. With appropriate functions $h(\zeta, \vec p_T)$ in
(\ref{3.53a}) we can represent states of $q\bar q$-mesons of arbitrary
spin.

We choose for our states (\ref{3.53a})
the usual continuum normalization:
\be\label{3.56}
\langle M_j(P')|M_j(P)\rangle=(2\pi)^32P^0\delta^3
(\vec P'-\vec P).\ee
With (\ref{3.5a}) this requires:
\be\label{3.57}
\int {\di}^2p_T\int^1_0{\di}\zeta\ 2\zeta(1-\zeta)h^{*j}_{s,s'}(\zeta,
\vec p_T)h^j_{s,s'}(\zeta,\vec p_T)=1\ee
In (\ref{3.56}) and (\ref{3.57}) no summation over $j$
is to be taken.

For later use we define the wave functions in transverse position space
at fixed longitudinal momentum fraction $\zeta$:
\bear\label{3.58}
&&{\varphi}^j_{s,s'}(\zeta,\vec x_T):=\sqrt{2\zeta(1-\zeta)}
\frac{1}{2\pi}\int {\di}^2p_T\exp(i\vec p_T\cdot\vec x_T)h^j_{s,s'}
(\zeta,\vec p_T)\nonumber\\
&&(j=1,...,4).\ear
With this we define profile functions for the transitions
$M_j\to M_k$ for right and left movers as:
\be\label{3.59}
w_{k,j}(\vec x_T):=\int^1_0{\di}\zeta({\varphi}^k_{s,s'}(\zeta,
\vec x_T))^*\varphi^j_{s,s'}(\zeta,
\vec x_T),\ee
where $k,j$ are both odd or even.
Clearly we have (cf. (\ref{3.57}))
\bear\label{3.60}
&&w_{j,j}(\vec x_T)\geq0,\nonumber\\
&&\int {\di}^2x_Tw_{j,j}(\vec x_T)=1,\nonumber\\
&&({\rm no\ summation\ over}\ j).\ear

Let us first study the transition of the right movers
alone, i.e. the ``reaction'':
\be\label{3.61}
M_1(P_1)\to M_3(P_3).\ee
For stable mesons $M_{1,3}$ we should find that the
corresponding $S$-matrix elements are identical to the
matrix elements of the unit operator. Is this borne out
in our approach?
\par From the rules given in sect. 3.5 we find easily the
$S$-matrix element for the transition
\be\label{3.62}
q(1)+\bar q(1')\to q(3)+\bar q(3')\ee
in the form
\be\label{3.63}
\langle 3,3'|S|1,1'\rangle=\langle {\cal S}_{q+}(3,1){\cal S}
_{\bar q+}(3',1')\rangle_G.\ee
After folding (\ref{3.63}) with the mesonic wave functions
(\ref{3.53a}) we get:
\bear\label{3.64}
&&\langle M_3(P_3)|S|M_1(P_1)\rangle=(2\pi)^32P_1^0
\delta^3(\vec P_3-\vec P_1)\nonumber\\
&&\int {\di}^2z_T\int^1_0{\di}\zeta_3\int^1_0
{\di}\zeta_1\ \varphi^{3^*}_{s,s'}
(\zeta_3,\vec z_T)\varphi^1_{s,s'}
(\zeta_1,\vec z_T)\nonumber\\
&&\frac{1}{2}P_{1+}\int\frac{{\di}z_-}{2\pi}\exp\big[\frac{i}{2}
P_{1+}(\zeta_3-\zeta_1)z_-\big]\nonumber\\
&&\langle\frac{1}{3}{\Tr}[V_-(\infty,z_-,\vec z_T)V^\dagger_-
(\infty,0,0)]\rangle_G.\ear
Now we remember that we consider the limit $P_{1+}\to\infty$.
Therefore we perform a change of variables in the $z_-$ integral by setting:
\be\label{3.65}
z_-':=\frac{1}{2}P_{1+}z_-.\ee
This gives:
\bear\label{3.66}
&&\langle M_3(P_3)|S|M_1(P_1)\rangle=(2\pi)^32P_1^0
\delta^3(\vec P_3-\vec P_1)\nonumber\\
&&\int {\di}^2z_T\int^1_0{\di}\zeta_3\int^1_0
{\di}\zeta_1\ \varphi^{3^*}_{s,s'}
(\zeta_3,\vec z_T)\varphi^1_{s,s'}
(\zeta_1,\vec z_T)\nonumber\\
&&\int\frac{{\di}z_-'}{2\pi}\exp[i(\zeta_3-\zeta_1)z_-']\nonumber\\
&&\langle\frac{1}{3}{\Tr}[V_-(\infty,\frac{2}{P_{1+}}
z_-',\vec z_T)V^\dagger_-(\infty,0,0)]\rangle_G\nonumber\\
&&\longrightarrow
(2\pi)^32P_1^0\delta^3(\vec P_3-\vec P_1)\int {\di}^2z_T\ w_{3,1}(\vec  
z_T)\nonumber\\
&&\langle\frac{1}{3}{\Tr}[V_-(\infty,0,\vec  
z_T)V^\dagger_-(\infty,0,0)]\rangle_G\quad {\rm for}\quad
P_{1+}\to\infty.\ear

In (\ref{3.66}) $V_-(\infty,0,\vec z_T)(V_-^\dagger(\infty,0,0))$
is the quark (antiquark) connector taken along
the line $C_q (C_{\bar q})$, where:
\be\label{3.67}
C_q:\tau\to z_q(\tau)=\left(\begin{array}{c}
\tau\\ \vec z_T\\ \tau\end{array}\right),\quad
(-\infty<\tau<\infty),\ee
\be\label{3.68}C_{\bar q}:\tau\to z_{\bar q}(\tau)=
\left(\begin{array}{c}
\tau\\ 0\\ \tau\end{array}\right),\quad
(-\infty<\tau<\infty).\ee
But from sect. 2.1 we know that the connector
$V_-^\dagger$ along $C_{\bar q}$
is equal to the connector $V_-$ taken along
the oppositely oriented line $\bar C_{\bar q}$
(cf. (\ref{2.25})).
Now we will allow ourselves to join the lines $C_q$ and
$\bar C_{\bar q}$ at large positive and negative times.
We can imagine the gluon potentials to be turned
off adiabatically there. We obtain then from the
product of the connectors in (\ref{3.66}) a connector
taken along a closed lightlike Wegner-Wilson loop
\be\label{3.69}
\frac{1}{3}{\Tr}[V_-(\infty,0,\vec z_T)V^\dagger_-(\infty,
0,0)]\longrightarrow {\W}_+(\frac{1}{2}\vec z_T,\vec z_T).\ee
Here we define
\be\label{3.70}
{\W}_\pm(\vec y_T,\vec z_T)=\frac{1}{3}{\Tr}
{\rm P}\exp[-ig\int_{C_\pm}{\di}x_\mu G^\mu(x)]\ee
with $C_+(C_-)$ a lightlike Wegner-Wilson loop
in the plane $x_-=0 (x_+=0)$, where in the transverse
space the centre of the loop is at $\vec y_T$ and
the vector from the antiquark to the quark line is
$\vec z_T$ (Fig. 15).

Inserting now everything in (\ref{3.66}) we get the
simple answer:
\bear\label{3.71}
&&\langle M_3(P_3)|S|M_1(P_1)\rangle=(2\pi)^32P_1^0\delta^3
(\vec P_3-\vec P_1)\nonumber\\
&&\int {\rm d}^2z_T\  w_{3,1}(z_T)\langle{\W}_+
(\frac{1}{2}\vec z_T,\vec z_T)\rangle_G.\ear
In the next section we will evaluate the functional
integral in (\ref{3.71}) in the SVM. We will find
\be\label{3.72}
\left.\langle{\W}_+(\vec y_T,\vec z_T)\rangle_G
\right|_{SVM}=1.\ee
Inserting this in (\ref{3.71}) leads to the expected result (cf.
(\ref{3.53a})-(\ref{3.59})):
\be\label{3.73}
\langle M_3(P_3)|S|M_1(P_1)\rangle=\langle M_3(P_3)|M_1(P_1)\rangle.
\ee

In our approach the $q\bar q$ pair in the right-moving
meson $M_1\to M_3$ does not interact. Of course this is only
valid over our finite time interval (3.3)!

The techniques developed thus far are now easily employed
for the reaction (\ref{3.53}). After performing similar
steps as above we arrive at the following $S$-matrix
element:
\bear\label{3.74}
S_{fi}&=&\delta_{fi}+i(2\pi)^4\delta(P_3+P_4-P_1-P_2)T_{fi},
\nonumber\\[0.2cm]
T_{fi}&\equiv&\langle M_3(P_3),M_4(P_4)
|T|M_1(P_1),M_2(P_2)\rangle\nonumber\\
&=&-2is\int {\di}^2
b_T{\di}^2x_T{\di}^2y_Te^{i\vec q_T\cdot
\vec b_T}w_{3,1}(\vec x_T)w_{4,2}(\vec y_T)\nonumber\\
&\hphantom{=}&\langle {\W}_+(\frac{1}{2}\vec b_T,\vec x_T)
{\W}_-(-\frac{1}{2}\vec b_T,\vec y_T)-1\rangle_G.\ear
Here $s=(P_1+P_2)^2$ is the c.m. energy squared and $\vec q_T$
is the momentum transfer, which is purely transverse in the
high energy limit:
\be\label{3.75}
\vec q_T=(\vec P_1-\vec P_3)_T.\ee
\par From (\ref{3.74}) we see that the amplitude for soft
meson-meson scattering at high energies is governed by the
correlation function of two lightlike Wegner-Wilson loops, where
one is in the hyperplane $x_-=0$, the other in
$x_+=0$. The transverse separation between the centres
of the two loops is given by $\vec b_T$, the impact parameter.
The vectors $\vec x_T$ and $\vec y_T$ give the extensions
and orientations of the loops in transverse space (Fig. 16).
The loop-loop correlation function has to be integrated over
all orientations of the loops in transverse space
with the (transition) profile functions of the mesons,
$w_{3,1}$ and $w_{4,2}$. Finally a Fourier transform in
the impact parameter has to be done.

The methods presented here for meson-meson scattering
can of course also be employed for scattering reactions
involving baryons and antibaryons. This is sketched
in appendix C.

\subsection{The evaluation of scattering amplitudes in
the Minkowskian version of the stochastic vacuum model}

In the previous section we have derived expressions for the
amplitudes of soft meson-meson scattering at high energies
in terms of correlation functions of light-like
Wegner-Wilson loops. The task is now to evaluate the
corresponding functional integral
$\langle\rangle_G$ in (\ref{3.74}).
Surely we do not want to make a perturbative expansion there,
remembering our argument of sect. 1 (cf. (\ref{1.8})).
Instead, we will turn to the SVM which did quite well
in its applications in Euclidean QCD (sect. 2). Of
course, the generalization of the SVM to Minkowski
space-time is a bold step which was done in \cite{58},
\cite{44}. The authors of these refs. proposed to use in
Minkowski space-time just the assumptions 1-3 of the
SVM (cf. sect. 2.4), but after having made a suitable
analytic continuation. In this way we obtain for instance
the Minkowski version of Ass. 1 as (cf. (\ref{2.60}),
(\ref{2.61})):

\begin{itemize}
\item{\bf Ass. 1:}
The correlator of two field strengths, shifted to a
common reference point $y$, is independent of the
connecting path and given by:
\bear\label{3.76}
&&\langle\frac{g^2}{4\pi^2}\hat G^a_{\mu\nu}(y,x;C_x)\hat G
^b_{\rho\sigma}(y,x';C_{x'})\rangle_G
=\frac{1}{4}\delta^{ab}F_{\mu\nu\rho\sigma}(z),\nonumber\\[0.2cm]
&&F_{\mu\nu\rho\sigma}(z)=\frac{1}{24}G_2
\big\{(g_{\mu\rho}g_{\nu\sigma}-g_{\mu\sigma}
g_{\nu\rho})\kappa D(z^2)\nonumber\\
&&+\frac{1}{2}\big[\frac{\partial}{\partial z^\nu}
\left(z_\sigma g_{\mu\rho}-z_\rho g_{\mu\sigma}\right)
+\frac{\partial}
{\partial z^\mu}\left(z_\rho g_{\nu\sigma}
-z_\sigma g_{\nu\rho}\right)\big]
\cdot(1-\kappa)D_1(z^2)\big\}.\nonumber\\
&&\ear
\end{itemize}

Here $z=x-x'$ and $D(z^2)$ and $D_1(z^2)$ are defined
as in (\ref{2.64}), (\ref{2.66}) for
$z^2\leq0$ and by analytic continuation for $z^2>0$.

As a first application let us calculate the expectation
value of one lightlike Wegner-Wilson loop. We have from
(\ref{3.70}) using the non-abelian Stokes theorem:
\bear\label{3.77}
&&\big\langle{\W}_+(\vec y_T,\vec z_T)\big\rangle_G
=\big\langle\frac{1}{3}{\Tr}{\rm P}\exp\big[-ig\int_{{\cal C}_+}
{\di}x_\mu G^\mu(x)\big]\big\rangle_G\nonumber\\
&&=\big\langle\frac{1}{3}{\Tr}\ {\rm P}\exp\big[-i\frac{g}{2}
\int_{S_+}{\di}u{\di}v\frac{\partial(x^\mu,x^\nu)}{
\partial(u,v)}\hat G_{\mu\nu}(R,x;C_x)\big]\big\rangle_G.
\ear
Here $S_+$ is the (planar) surface spanned into
$C_+$ (Fig. 15) and parametrized by
\bear\label{3.78}
&&x^\mu(u,v)=u\ n^\mu_++y^\mu-(v-\frac{1}{2})z^\mu,
\nonumber\\[0.2cm]
&&-\infty<u<\infty,\ 0\leq v\leq1,\ear
where $n_+$ is as in (\ref{3.40}) and
\be\label{3.79}
y^\mu=\left(\begin{array}{c} 0\\ \vec y_T\\
0\end{array}\right),\qquad z^\mu=\left(\begin{array}{c}
0\\ \vec z_T\\ 0\end{array}\right).\ee
The reference point on the surface $S_+$ is denoted by
$R$ and $C_x$ are straight lines running from $x$
to $R$. From (\ref{3.78}) we find
\be\label{3.80}
\frac{\partial(x^\mu,x^\nu)}{\partial(u,v)}=
z^\mu n^\nu_+-n_+^\mu z^\nu.\ee

Now we apply the cumulant expansion formulae (cf. sect.
2.3) to (\ref{3.77}) and use assumptions 1-3 of the SVM.
This is completely analogous to the calculations
done in sect. 2.5. We get:
\bear\label{3.81}
\langle{\W}_+(\vec y_T,\vec z_T)\rangle_G&=&\exp
\big\{-\frac{\pi^2}{6}\int_{S_+}{\di}u{\di}v\int_{S_+} {\di}
u'{\di}v'K_2
(x-x')\nonumber\\
&&+\quad {\rm higher\ cumulant\ terms}\},
\ear
where
\bear\label{3.82}
&&K_2(x-x')=\frac{\partial(x^\mu,x^\nu)}{\partial(u,v)}
\frac{\partial({x'}^\rho,{x'}^\sigma)}{\partial(u',v')}
F_{\mu\nu\rho\sigma}(x-x'),\nonumber\\[0.1cm]
&&x\equiv x(u,v),\quad x'\equiv x'(u',v').\ear
It is an easy exercise to evaluate $K_2(x-x')$
using (\ref{3.80}) and $F_{\mu\nu\rho\sigma}$ from
(\ref{3.76}). The result is
\be\label{3.83}
K_2(x-x')=0\ee
for all $x,x'\in S_+$. With assumption 3 of the
SVM (cf. (\ref{2.71})-(\ref{2.73}))
all higher cumulants are related to the second one. Thus,
(\ref{3.83}) implies also the vanishing of all higher
cumulant terms in (\ref{3.81}) and we get
\be\label{3.84}
\langle{\W}_+(\vec y_T,\vec z_T)\rangle_G=1.\ee
A similar argument leads, of course, also to
\be\label{3.85}
\langle{\W}_-(\vec y_T,\vec z_T)\rangle_G=1.\ee

In sect. 3.6 we have already used the result (\ref{3.84})
in the discussion of the transition $M_1\to M_3$ to obtain
(\ref{3.73}). We can also use it for the calculation
of the wave function renormalization constant $Z_\psi$.
The expression we obtained for $Z_\psi$ in (\ref{3.45a})
can be interpreted as the expectation value of the non-abelian
phase factor picked up by a very fast right-moving
quark. Now, isolated quarks do not exist. The best
approximation for it we can think of is a fast,
right-moving quark-antiquark pair with the antiquark
being very far away from the quark in transverse
direction. In this way we obtain for $Z_\psi$ instead
of (\ref{3.45a}):
\bear\label{3.86}
Z_\psi&=&\lim_{|\vec z_T|\to\infty}\langle \frac{1}{3}
{\Tr}[V_-(\infty,0,0)V^\dagger_-(\infty,0,\vec z_T)]\rangle_G
\nonumber\\
&=&\lim_{|\vec z_T|\to\infty}\langle{\W}_+(-\frac{1}{2}
\vec z_T,-\vec z_T)\rangle_G\nonumber\\
&=&1.\ear
This result was already used in sect. 3.5, (\ref{3.52a})ff.

We come now to the evaluation of the loop-loop correlation
function of (\ref{3.74}):
\bear\label{3.87}
&&\langle{\W}_+(\frac{1}{2}{\vec b}_T,\vec x_T){\W}_-(-
\frac{1}{2}{\vec b}_T,\vec y_T)-1\rangle_G\nonumber\\
&&=\langle[{\W}_+(\frac{1}{2}\vec b_T,\vec x_T)-1][{\W}_-(-
\frac{1}{2}\vec b_T,\vec y_T)-1]\rangle_G.\ear
Here we used (\ref{3.84}), (\ref{3.85}). The strategy is as
before. We want to transform the line integrals of ${\W}_\pm$
(cf. (\ref{3.70})) into surface integrals using the non-abelian
Stokes theorem of sect. 2.2. Following the authors of
\cite{58}, \cite{44} we choose as surface with
boundary $C_+$ and $C_-$ a double pyramid with apex at
the mid-point of $C_+$ and $C_-$
which is the origin of our coordinate system (Fig. 17). The
mantle of this pyramid is ${\cal P}_++{\cal P}_-$ and we have
with suitable orientation
\be\label{3.88}
\partial({\cal P}_++{\cal P}_-)=C_++C_-.\ee
In the transverse projection of Fig. 16 the basis
surface $S_+(S_-)$ appears as the line $\bar q_+q_+$
$(\bar q_-q_-)$ and the mantle surface ${\cal P}_+({\cal P}_-)$
as the triangle $0\bar q_+q_+(0\bar q_-q_-)$. From
the non-abelian Stokes theorem we obtain now
for (\ref{3.87}):
\bear\label{3.89}
&&\big\langle\big[{\W}_+(\frac{1}{2}\vec b_T,
\vec x_T)-1\big]\big[{\W}_-(-\frac{1}{2}\vec b_T,
\vec y_T)-1\big]\big\rangle_G\nonumber\\
&&=\big\langle\big\{\frac{1}{3}{\Tr}{\rm P}{\exp}\big[
-i\frac{g}{2}\int_{{\cal P}_+}{\di}u{\di}v
\frac{\partial(x^\mu,x^\nu)}
{\partial(u,v)}\hat G_{\mu\nu}(0,x;C_x)\big]-1\big\}
\nonumber\\
&&\hphantom{=\big\langle}
\big\{\frac{1}{3}{\Tr}\ {\rm P}\exp\big[-i\frac{g}{2}
\int_{{\cal P}_-}{\di}u'{\di}v'\frac{\partial({x'}^\rho,{x'}^\sigma)}
{\partial(u',v')}\hat G_{\rho\sigma}(0,x';C_{x'})\big]-1\big\}
\big\rangle_G.\nonumber\\
&&\ear
So far, it has not been possible to use some version of
the cumulant
expansion for (\ref{3.89}). Thus in \cite{58}, \cite{44} the
path-ordered exponentials on the r.h.s. of (\ref{3.89})
were expanded directly. The structure of this expansion
is as follows:

\bear\label{3.90}
&&\langle[{\W}_+-1][{\W}_--1]\rangle_G\nonumber\\
&&\sim \langle\frac{1}{3}{\Tr}[\int_{\cal P_+}\hat G+
\int_{\cal P_+}\int_{\cal P_+}\hat G\hat G+...]\nonumber\\
&&\hphantom{\sim\langle}
\frac{1}{3}{\Tr}[\int_{\cal P_-}\hat G+\int_{\cal P_-}
\int_{\cal P_-}\hat G
\hat G+...]\rangle_G.\ear
The trace of a single shifted field strength vanishes. This
means that we cannot exchange a single coloured object in
meson-meson scattering (\ref{3.53}). The first non-trivial
contribution in (\ref{3.90}) comes from the term
with two shifted field strengths in each trace. The corresponding
correlation function
\[\langle\frac{1}{3}{\Tr}[\int_{{\cal P}_+}\int_{{\cal P}_+}
\hat G\hat G]
\quad\frac{1}{3}{\Tr}[\int_{{\cal P}_-}\int_{{\cal P}_-}\hat G\hat G]
\rangle_G\]
can now be evaluated using the assumptions 1-3 of the
Minkowskian version of the SVM. In doing so it is
advantageous to transform the surface integrals over
the mantles of the pyramids ${\cal P}_\pm$ into surface
integrals over $S_\pm$ and integrals over the volumes $V_\pm$
enclosed by ${\cal P}_+$ and $S_+$ and ${\cal P}_-$ and $S_-$,
respectively:
\be\label{3.91}
\partial(V_\pm)={\cal P}_\pm-S_\pm.\ee
Then one can use the ordinary Gauss theorem
to get simpler integrals.

We will not enter into the details of these calculations
here, but only note that the integrations along the
directions $x_+$ and $x_-$ can easily be done analytically
and that one ends up with integrals over the projections
of $S_\pm$ and $V_\pm$ into the transverse space. Thus,
\underbar{one finally needs the correlator
(\ref{3.76}) for}\\ \underbar{space-like
separations only:}
\bear\label{3.92}
&&(x-x')^2=z^2\leq0,\nonumber\\
&&x=\left(\begin{array}{c}
0\\ \vec x_T\\ 0\end{array}\right),\quad
x'=\left(\begin{array}{c}
0\\ \vec x'_T\\ 0\end{array}\right),\ear
where $\vec x_T$ runs over the triangle $\Delta_+=
0\bar q_+q_+$ and $\vec {x'}_T$ over $\Delta_-=0\bar q_-
q_-$ in Fig. 16. For space-like separations the correlator functions
$D(z^2)$ and $D_1(z^2)$ in (\ref{3.76}) are as in Euclidean
space time. The resulting expressions are then of the
following structure
\bear\label{3.93}
&&\langle[{\W}_+-1][{\W}_--1]\rangle_G\sim\nonumber\\
&&\big\{\int_{\Delta_+}{\di}^2z_T\int_{\Delta_-}{\di}^2z_T'\
G_2[..D(-(\vec z_T-\vec z_T')^2)+...D_1(-(\vec z_T-\vec z_T')
^2)]\big\}^2.\nonumber\\
&&\ear
These integrals have to be evaluated numerically.

With the methods outlined above, we have obtained an
(approximate) expression (\ref{3.93}) for the functional
integral $\langle\rangle_G$ governing the meson-meson
scattering amplitude (\ref{3.74}). Note that the
nonperturbative gluon condensate parameter $G_2$ sets
the scale in (\ref{3.93}) and in the integrals to be
performed there the vacuum correlation
length $a$ enters through the
$D$ and $D_1$ functions. Thus, on dimensional grounds, we
must have in our approximation:
\be\label{3.94}
\langle({\W}_+-1)({\W}_--1)\rangle_G=G^2_2a^8 f\left(
\frac{\vec b_T}{a},\ \frac{\vec x_T}{a},\
\frac{\vec y_T}{a}\right),\ee
where $f$ is a dimensionless function. To obtain the meson-meson
scattering amplitude (\ref{3.74})
we still have to integrate over the profile functions $w_{3,1}(\vec
x_T)$ and $w_{4,2}(\vec y_T)$. Here the \underbar{transverse
extensions} of the mesons - i.e. of the wave packets representing
them - enter in the results.

For a detailed exposition of the numerical results obtained
in the way sketched above we refer to \cite{44}.
Here we only discuss the outcome for proton-proton
scattering when treated in a similar way (cf. appendix C).
A fit to the numerical results gives for the total cross
section and the slope parameter at $t=0$ of elastic
proton-proton scattering the following
representation:
\be\label{3.95}
\sigma_{\rm tot}(pp)=0.00881\left(\frac{R_p}{a}\right)^{3.277}\cdot
(3\pi^2G_2)^2\cdot a^{10},\ee
\be\label{3.96}
b_{pp}:=\frac{\rm d}{{\rm d}t}\ln\frac{{\rm d}
\sigma_{el}}{{\rm d}t}(pp)|_{t=0}=1.558 a^2+0.454 R^2_p.
\ee
Here $R_p$ is the proton radius and the formulae (\ref{3.95}),
(\ref{3.96})
are valid for
\be\label{3.97}
1\leq R_p/a\leq 3.\ee

To compare (\ref{3.95}), (\ref{3.96}) with experimental results, we can,
for instance, consider the c.m. energy $\sqrt s=20$ GeV and take as
input the following measured values (cf. \cite{44}):
\bear\label{3.98}
\sigma_{\rm tot}(pp)|_{\rm Pomeron\ part}&=&35\ {\rm mb},\nonumber\\
b_{pp}&=&12.5\ {\GeV}^{-2},\nonumber\\
R_p&\equiv &R_{p,elm}=0.86\ {\rm fm}.\ear
We obtain then from (\ref{3.95}) and (\ref{3.96}):
\bear\label{3.99}
a&=&0.31\ {\rm fm},\nonumber\\
G_2&=&(507\ {\rm MeV})^4.\ear
The values for the correlation length $a$
and for the gluon condensate $G_2$ come out in surprising good
agreement with the determination of these quantities from the fit
to the lattice results (\ref{2.68}).

But perhaps we were lucky in picking out the right c.m. energy $\sqrt s$
and radius for our comparison of theory and experiment. What about
the $s$-dependence of the total cross section $\sigma_{tot}$ and slope
parameter $b$? The vacuum parameters $G_2$ and $a$ should be
independent of the energy $\sqrt s$. On the other hand,
from the discussion in sect. 3.1 leading to (\ref{3.2}),
it seems quite
plausible to us that the effective strong interaction radii
$R$ of hadrons may depend on $\sqrt s$. Let us consider again $pp$
(or $p\bar p$) elastic scattering. Once we have fixed $G_2$ and $a$
from the data at $\sqrt s=20\GeV$ (\ref{3.95}) and (\ref{3.96}) give us
$\sigma_{\rm tot}(pp)$ and $b_{pp}$ in terms of the single parameter
$R_p$, i.e. we obtain as prediction of the model a curve in the
plane $b_{pp}$ versus $\sigma_{\rm tot}(pp)$.
This is shown in Fig. 18. It
is quite remarkable that the data from $\sqrt s=20\GeV$
up to Tevatron
energies, $\sqrt s=1.8$ TeV follow this curve.

Summarizing this section we can say that explicit calculations for high
energy-elastic hadron-hadron scattering near the forward direction have
been performed combining the field-theoretic methods of \cite{52} and the
Minkowski version of the
stochastic vacuum model of \cite{58},\cite{44}.
The results are encouraging and
support the idea that the vacuum structure of QCD plays an essential
role in soft high-energy scattering. We want to point out that these
calculations also resolve a possible paradoxon of QCD: On the one
hand there are suggestions that the gluon propagator must be highly
singular (probably $\propto(Q^2)^{-2}$, cf. e.g. \cite{60})
for momentum transfers $Q^2\to0$ in order to produce confinement.
On the other hand high energy scattering amplitudes are
completely regular for $t\to0$. A singular gluon propagator
will lead in the 2-gluon exchange model
to a singularity for $t=0$ not only for quark-quark
scattering but also for hadron-hadron scattering if the latter
are considered e.g. as colour dipoles.
The resolution of the paradoxon
which we can present is intimately connected with the
\underbar{confinement} mechanism which we found in the SVM
in Sect. 2. The \underbar{short range} correlation of the
\underbar{gluon
field strengths} governs the $t$-dependence of the hadronic
scattering amplitudes and gives rise to their regularity for
$t=0$. The gluon propagator on the other hand can be singular
for $Q^2\to0$, since the \underbar{gluon potentials} have a
\underbar{long range} correlation as we have seen in Sect. 2.
The result (\ref{3.95}) for the total cross section depends
also on the proton radius. This radius dependence does not
saturate for large radii in the calculation with non-abelian
gluons but does saturate in an abelian model \cite{44}.
Thus with non-abelian gluons we do \underbar{not} get the
additive quark model result \cite{17}, and thus not the picture
of Donnachie and Landshoff \cite{22a}, where the ``soft
Pomeron'' couples to individual quarks in the hadrons. The
strong radius dependence in (\ref{3.95}) is due to the
$D$-term in the correlator (\ref{3.76}) which is related
to the effective chromomagnetic monopole condensate in the
QCD vacuum and which gives rise to
the linearly rising quark-antiquark potential, i.e. to string
formation (see sect. 2.5). The calculations
reported above suggest that in
high energy scattering this same term gives rise to a string-string
interaction which leads to the radius dependence in (\ref{3.95}).
Note that one does not have to put in the strings by hand. They
enter the picture automatically through our lightlike
Wegner-Wilson loops. The radius dependence of the cross section
occurs, of course, also for meson-baryon and meson-meson
scattering and gives a quantitative understanding of the
difference between the $Kp$ and $\pi p$ total cross sections
and slope parameters at high energies.
For this and for further results we refer to \cite{44},
\cite{61}. The success of the calculation for $pp(p\bar p)$
scattering describing correctly the relation of the total
cross section versus the slope parameter from $\sqrt s\simeq
20$ GeV up to $\sqrt s=1.8$ TeV suggests the following
simple interpretation: In soft elastic scattering the
hadrons act like effective ``colour dipoles'' with a
radius increasing with c.m. energy. The dipole-dipole interaction
is governed by the correlation function of two lightlike
Wegner-Wilson loops which receives the dominant contribution
from the same non-perturbative phenomenon - effective chromomagnetic
monopole condensation - which leads to string formation and
confinement.
\newpage
\section{``Synchrotron
Radiation'' from the \protect\\
Vacuum, Electromagnetic Form Factors \protect\\
of Hadrons, and Spin Correlations
in the \protect\\ Drell-Yan Reaction}
\setcounter{equation}{0}
Let us consider for definiteness again a proton-proton collision at high
c.m. energy $\sqrt s\gg m_p$. We look at this collision in the c.m. system
and choose as $x^3$-axis the collision axis (Fig. 19). According to
Feynman's parton dogma \cite{62} the hadrons look like jets of almost
non-interacting partons, i.e. quarks and gluons. Accepting our
previous views of the QCD vacuum (Sect. 2), these partons travel
in a background chromomagnetic field.

What sort of new effects might we expect to occur in this situation?
Consider for instance a quark-antiquark collision in a chromomagnetic
field. In our picture this is very similar to an electron-positron
collision in a storage ring (Fig. 20). We know that in a storage ring
$e^-$ and $e^+$ are deflected and emit synchrotron radiation. They also
get a transverse polarization due to emission of spin-flip synchrotron
radiation \cite{63}, \cite{64}. Quite similarly we can expect the quark
and antiquark to be deflected by the vacuum fields. Since quarks have
electric and colour charge, they should then emit both \underbar{photon}
and \underbar{gluon ``synchrotron radiation''}. Of course, as long as
we have quarks within a single, isolated
proton (or other hadron) travelling through the vacuum
no emission of photons can occur, and we should
consider such processes as contributing to the cloud of quasi-real photons
surrounding a fast-moving proton. (This is similar in spirit to the
well-known Weizs\"acker-Williams approximation.) But in a collision
process the parent quark or antiquark will be scattered away and the
photons of the cloud can become real, manifesting themselves as
\underbar{prompt photons in hadron-hadron collisions}.

In ref. \cite{14} we have given an estimate
for the rate and the spectrum
of such prompt photons using the classical formulae for synchrotron
radiation \cite{64}. A more detailed study of soft photon production
in hadronic collisions was made in \cite{65}. A sketch of our
arguments and calculations is as follows.

In Sect. 2 we discussed the domain picture of the QCD vacuum.
In Euclidean space time we have domains in the vacuum of
linear size $\simeq a$. Inside one domain the colour fields
are highly correlated. The colour field orientations and domain
sizes fluctuate, i.e. have a certain distribution. If we
translate this picture naively to Minkowski space, we arrive
at colour correlations there being characterized by
\underbar{invariant} distances of order $a$. Then the colour fields
at the origin of Minkowski space, for instance,
should be highly correlated with the fields in the region
\be\label{4.1} |x^2|\klgl a^2\ee
(cf. Fig. 21). Consider now a fast hadron passing by with one
of its quarks going right through $x=0$ on a nearly lightlike
world line. It is clear that in such a situation the quark
will, from the point of view of the observer, spend a long
time in a correlated colour background field. An easy exercise
shows furthermore that two quarks of the same hadron will have
a very good chance to travel in \underbar{two different} colour
domains. The argument is in essence as follows. The quarks have
a transverse separation of the order of the hadron radius $R$
whereas the transverse size of a domain is of order $a$ and
we have $a^2/R^2\ll1$ (cf. (\ref{2.70})). Each quark
will then wiggle around due to the deflection by the background
colour fields in an uncorrelated fashion. This gives us
a justification for adding the synchrotron gluon and photon
emission of the quarks \underbar{incoherently}.
The result we found can be summarized as follows:
In the overall c.m. system of the hadron-hadron
collision ``synchrotron''
photons should appear with energies
$\omega<300-500$ MeV, i.e. in the
very central region of the rapidity space.
The number of photons per
collision and their spectrum are --- apart
from logarithms --- independent
of the c.m. energy $\sqrt s$. The dependence of the number of
photons on the  energy $\omega$
and on the emission angle $\vartheta^*$ with respect to the
beam axis is obtained  as follows for $pN$ collisions:
\be\label{4.2}
\frac{{\rm d}n_\gamma}{{\rm d}\omega {\rm d}
\cos\vartheta^*}=\frac{2\pi\alpha}
{\omega^{1/3}}(l_{\rm eff})^{2/3}\cdot\Sigma(\cos\vartheta^*)\ee
Here $\alpha$ is the fine structure constant and $l_{\rm eff}$
is the length or time over which the fast quark travelling
in the background chromomagnetic field $B_c$ obtains
by its deflection a transverse momentum of order $\bar p_T
\approx 300$ MeV, the mean transverse momentum of quarks
in a hadron  (Fig. 22):
\be\label{4.3}
l_{\rm eff}=\frac{\bar p_T}{gB_c}.\ee
The quantity $\Sigma$ in (\ref{4.2}) sums up the contributions from
all quarks of the initial and final state hadrons. It involves
an integration over
the quark distribution functions of these hadrons.
In \cite{65} we found
\be\label{4.3a}
\Sigma(\cos\vartheta^*)\simeq\frac{0.21}{(\sin\vartheta^*)^{2/3}}\ee
for $pN$ collisions at $\sqrt s=29$ GeV.

Our result (\ref{4.2}) for synchrotron photons
should be compared to the inner-bremsstrahlungs spectrum
\be\label{4.4}
\left.\frac{{\rm d}n_\gamma}{{\rm d}\omega {\rm d}\cos\vartheta^*}
\right|_{\rm
bremsstr.}\propto\frac{1}{\omega\sin^2\vartheta^*}\ee
The ``synchrotron'' radiation from the quarks (\ref{4.2}) is
thus harder than the hadronic bremsstrahlung
spectrum. This is welcome, since for $\omega
\to 0$ brems\-strahlung should dominate according to Low's theorem
\cite{66}.

It is amusing to note that in several experiments an excess of soft
prompt photons over the bremsstrahlung calculation has been observed
\cite{67}-\cite{71}. The gross features and the order of magnitude of
this signal make it a candidate for our ``synchrotron'' process. A
detailed comparison with our formulae has
been made in \cite{65}
for the results from one experiment \cite{71} with encouraging results.
This is shown in Fig. 23 for the $k_T$ spectrum of photons
at $y=0$, where
\bear\label{4.4a}
&&k_T=\omega\sin\vartheta^*,\nonumber\\
&&y=-\ln\tan(\vartheta^*/2).\ear
We see that the addition of synchrotron photons to the bremsstrahlung
ones improves the agreement of theory with the data considerably.
We deduce from Fig. 23 $l_{\rm eff}\simeq 20-40\ {\rm fm}$. Taking
$l_{\rm eff}=20\ {\rm fm}$ and $\bar p_T=300$ MeV, we find for the effective
chromomagnetic deflection field from (\ref{4.3})
\be\label{4.5}
gB_c=\frac{\bar p_T}{l_{\rm eff}}=(55\ {\rm MeV})^2.\ee
This is much smaller than the vacuum field strength (\ref{2.7}).
Our interpretation of this puzzle is as follows: The colour
fields in a fast moving hadron must be shielded. Indeed, a
chromomagnetic field of the strength (\ref{2.7}) would lead
to a ridiculously small value for the radius of cyclotron
motion of a fast quark. The necessary shielding could be done
by gluons in a fast hadron. We know from the deep inelastic
lepton-nucleon scattering results that a fast nucleon contains
many gluons. We may even be brave and turn the argument around:
in order for a fast hadron to be able to move through the
QCD vacuum, the very strong vacuum chromomagnetic fields must
be shielded, making gluons in a fast hadron a \underbar{necessity}.
Thus soft photon production in $pp$ collisions may give us a quite
unexpected insight into the quark and gluon structure of
\underbar{fast} hadrons.

The next topic we want to discuss briefly concerns electromagnetic
form factors of hadrons.

We have argued above that the colour fields in  the vacuum should
give a contribution to the virtual photon cloud of hadrons and
we made an estimate of the distribution of these photons using the
synchrotron radiation formulae. Consider now any reaction where a
quasi-real photon is emitted from a hadron with the hadron staying
intact and the photon interacting subsequently. In Fig.~24 we draw the
corresponding diagram for a nucleon $N$:
\be\label{4.6}
N(p)\to N(p')+\gamma(q).\ee
The flux of these quasi-real photons is well known. The first calculations
in this context are due to Fermi, Weizs\"acker, and Williams \cite{72}.
For us the relevant formula is given in eq.~(D.4) of \cite{73}.
Let $E$ be the energy of the initial nucleon, $G^N_E(Q^2)$ its
electric Sachs formfactor, and let $\omega$ and $q^2=-Q^2$ be the
energy and mass of the virtual photon. Then the distribution of
quasi-real photons in the fast-moving nucleon is given by
\be\label{4.7}
{\rm d}n_\gamma^{\rm (excl)}=\frac{\alpha}{\pi}\frac{{\rm  
d}\omega}{\omega}\frac{{\rm d}Q^2}{Q^2}
\left[G^N_E(Q^2)\right]^2\ee
where we neglect terms of order $\omega/E$ and $Q^2/m^2_N$ and assume
\be\label{4.8}
Q^2\gg Q^2_{\rm min}\simeq\frac{m^2_N\omega^2}{E^2}.\ee
We call (\ref{4.7}) the exclusive flux since the nucleon stays intact.
Now we want to translate (\ref{4.7}) into a distribution in
$\omega$ and the angle $\vartheta^*$ of emission of the $\gamma$
(cf. Fig.~24). A simple calculation gives
\be\label{4.9}
Q^2\simeq\omega^2\sin^2\vartheta^*,\ee
\be\label{4.10}
{\rm d}n_\gamma^{\rm(excl)}\simeq\frac{2\alpha}{\pi}\frac{{\rm d}\omega}{\omega}
\frac{{\rm d}\vartheta^*}{\sin\vartheta^*}\left[G^N_E(\omega^2\sin^2
\vartheta^*)\right]^2.\ee
Now we made an ``exclusive-inclusive connection'' argument in \cite{65}:
We require ${\rm d}n_\gamma^{\rm(excl)}/{\rm d}
\omega {\rm d}\cos\vartheta^*$ to
behave as $\omega^{-1/3}$ for fixed $\vartheta^*$ as we found
in (\ref{4.2}). This implies for the form factor $G_E^N(Q^2)$ a
behaviour as $(Q^2)^{1/6}$.

Thus we arrive at the following conclusion: The proton form factor $G_E^p$
should contain in addition to a ``normal'' piece connected with the total
charge and the hadronic bremsstrahlung in inelastic collisions a
piece $\propto (Q^2)^{1/6}$ connected with ``synchrotron'' radiation from
the QCD vacuum.  For the neutron
which has total charge zero we would expect the ``normal'' piece in
$G_E^n$ to be quite small and the ``anomalous'' piece to be quite
important for not too large $Q^2$. Thus the neutron electric formfactor
should be an interesting quantity to look for ``anomalous''
effects $\propto(Q^2)^{1/6}$.

In Fig.~25 we show the data on the electric formfactor of the neutron from
\cite{74,75}. We superimpose the curve
\be\label{4.11}
G^n_{(\rm syn)}(Q^2)=3.6\cdot 10^{-2}\left(\frac{Q^2}{5\,
{\rm fm}^{-2}}\right)
^{1/6}\ee
which is normalized to the data at $Q^2=5\, {\rm fm}^{-2}$.
We see that except in
the very low $Q^2$ region we get a decent description of the data. For
$Q^2\to0$ (\ref{4.11}) has to break down since $G_E^n(Q^2)$ is regular
at $Q^2=0$. Indeed one knows the slope of $G_E^n(Q^2)$ for $Q^2=0$ from
the scattering of thermal neutrons on electrons (\cite{76} and references
cited therein):
\be\label{4.12}
\frac{{\rm d}G_E^n(Q^2)}{{\rm d}Q^2}
\Biggl|_{Q^2=0}=0.019\ {\rm fm}^2.\ee
We see from Fig.~25 that the behaviour of $G_E^n(Q^2)$ has to change
rather quickly as we go away from $Q^2=0$.  We will now make a simple
ansatz which takes into account that $G_E^n(Q^2)$ can have
singularities in the complex $Q^2$-plane only for
\[-\infty<Q^2\leq-4m_\pi^2,\]
where $m_\pi$ is the pion mass.
We require a $(Q^2)^{1/6}$ behaviour for positive $Q^2$
and take the slope of $G_E^n(Q^2)$ at $Q^2=0$ from experiment
(\ref{4.12}). This leads us to the following functional form
for $G_E^n(Q^2)$:
\be\label{4.13}
G_E^n(Q^2)=0.019\ {\rm fm}^2\cdot Q^2\left[1+\frac{Q^2}{4m^2_\pi}\right]
^{-5/6}.\ee
It is amusing to see that this gives a decent description of the
data (Fig. 25).

What about the electric form factor of the proton
$G_E^p(Q^2)$? Here, clearly, we have a dominant ``normal''
piece connected with the total charge. We will assume
that this normal contribution is represented by the usual dipole
formula
\bear\label{4.14}
G_D(Q^2)&=&\left(1+\frac{Q^2}{m_D^2}\right)^{-2},\nonumber\\
m^2_D&=&18.23\ {\rm fm}^{-2}\hat=0.710\ {\rm GeV}^2\ear
which gives a good representation of the data for $Q^2=2-4\ {\rm GeV}^2$
\cite{77}. Let us add to this an anomalous piece for smaller
$Q^2$, connected with synchrotron radiation, and let us
assume that this is a purely isovector contribution,
consistent with the singularity at $Q^2=-4m^2_\pi$ in
(\ref{4.13}). We obtain then from (\ref{4.13}) the
following ansatz for $G_E^p(Q^2)$:
\be\label{4.15}
G_E^p(Q^2)=G_D(Q^2)-G_E^n(Q^2)=G_D(Q^2)(1-\Delta(Q^2)),\ee
where
\be\label{4.16}
\Delta(Q^2)=0.019\ {\rm fm}^2\cdot Q^2(1+\frac{Q^2}{4m^2_\pi})^{-5/6}
\cdot(1+\frac{Q^2}{m^2_D})^2.\ee
We predict  a deviation of the ratio $G_E^p(Q^2)
/G_D(Q^2)$ from unity for small $Q^2$. It is again amusing to
note that such a deviation is indeed observed experimentally
\cite{46,47}. Our ansatz does even quite well quantitatively
(Fig. 26). For the electromagnetic radius of the proton we
predict from (\ref{4.15})
\bear\label{4.17}
\langle r^2_E\rangle:&=&-6\frac{{\rm d}G^p_E(Q^2)}
{{\rm d}Q^2}\bigr|_{Q^2=0}
\nonumber\\
&=&\frac{12}{m^2_D}+6\cdot0.019\ {\rm fm}^2\nonumber\\
&=&(0.88\ {\rm fm})^2.\ear
This checks well with the experimental values quoted in \cite{46}:
\bear\label{4.18}
\langle r^2_E\rangle^{1/2}&=&0.88\pm0.03\ {\rm fm}, \ {\rm or}
\nonumber\\
&&0.92\pm0.03\ {\rm fm},\ear
depending on the fit used for $G^p_E(Q^2)$ at low $Q^2$.

To conclude this brief discussion of nucleon form factors,
we can summarize our picture as follows: The quarks in the
nucleon make a cyclotron-type motion in the chromomagnetic
vacuum field. This leads to a spreading out of the charge
distribution of the neutron
\be\label{4.19}
\rho_n(\vec x)\propto |{\bf x}|^{-10/3}\ee
corresponding to $G_E^n(Q^2)\propto (Q^2)^{1/6}$.
The same effect should lead to a deviation of the proton
form factor $G_E^p(Q^2)$ from the dipole formula for small
$Q^2$. Concerning the sign of $G_E^n(Q^2)$ we can in
essence follow the arguments put forward in \cite{78}.

One might think --- maybe rightly --- that these ideas are a little
crazy. But we have also worked out some consequences of them for
the Drell-Yan reaction (\ref{1.3}), which make us optimistic. In the
lowest order parton process contributing there,  we have a quark-antiquark
annihilation giving a virtual photon $\gamma^*$, which decays then into
a lepton pair (Fig. 1):
\be\label{4.22}
q+\bar q\to \gamma^*\to \ell^+\ell^-.\ee
In the usual theoretical framework $q$ and $\bar q$ are assumed
to be uncorrelated and
unpolarized in spin and colour if the original hadrons are
unpolarized. From our point of view we expect a different situation.
Let us go back to Fig. 21 where we sketched the world line
of a quark of one fast hadron in a colour domain of extension
$|x^2|\klgl a^2$. Let the quark $q$ and antiquark $\bar q$
in (\ref{4.22}) annihilate at the point $x=0$. Here $q$
and $\bar q$ come from two different hadrons $h_1,h_2$ moving
with nearly light-like velocity in opposite directions. It is
clear that in this situation $q$ and $\bar q$ will spend a
\underbar{long} time in a highly correlated colour background
field (Fig. 27). In \cite{14} we speculated that this may lead
to a correlated transverse spin and colour spin polarization of $q$
and $\bar q$ due to the chromomagnetic
Sokolov-Ternov effect \cite{63,64}.
In \cite{79} we worked out this idea in more detail and found that
a transverse $q\bar q$ spin correlation
influences the $\ell^+\ell^-$ angular distribution
in a profound way. Then
our colleague H. J. Pirner pointed out to us that data which may be
relevant in this connection existed already \cite{80}.
And very obligingly
these data show a large deviation from the standard perturbative QCD
prediction. On the other hand, we can nicely understand the
data in terms of our spin correlations and
thus vacuum effects in high energy collisions.
For more details we refer
to \cite{79}. If such spin correlations are
confirmed by experiments
at higher energies, we would presumably have to reconsider the
fundamental factorization hypothesis for
hard reactions which we sketched
in Sect.~1 and which is discussed in detail in \cite{11}.

\section{Conclusions}
In these lectures we have discussed various ideas connected with
non-pertur\-bative QCD and in particular with
the QCD vacuum structure. In Sect. 2
we introduced connectors, the non-abelian Stokes theorem and the
cumulant expansion. Then we
presented the domain picture of the QCD vacuum and
the stochastic vacuum model (SVM). The latter is consistent
with the view of the QCD vacuum acting like a dual superconductor:
We found that in the SVM the vacuum contains an effective
chromomagnetic monopole
condensate, whose effect is parametrized by the $D$-term
in the gluon field strength correlator (\ref{2.61}). With
these tools we could calculate the expectation value
of the Wegner-Wilson loop in the SVM. We found
a linearly rising potential between a heavy quark-antiquark
pair, $Q\bar Q$. This is related to the formation
of a ``string'', a chromoelectric flux tube between $Q\bar Q$ as
can be seen explicitly in the SVM \cite{51}.
Thus in this framework confinement is an effect of the nontrivial QCD
vacuum structure. All calculations in sect. 2 were done in Euclidean
space time.

In appendix A we discuss some problems which arise when one
considers higher cumulant terms in the SVM. We propose
as remedy for these problems to do the calculation of the
expectation value of the Wegner-Wilson loop in an
iterative way, summing step by step over the contributions
of various plaquettes. This also leads us to a
proposal for including dynamical fermions in the
calculation with the expected result that
the linearly rising potential levels off and goes
to a constant at large $Q\bar Q$ separation, where,
of course, we have then mesons $Q\bar q, q\bar Q$
formed of a heavy quark $Q$ (antiquark $\bar Q$) and a
light antiquark $\bar q$ (quark $q$).

In sections 3, 4 we looked at various possible effects of the
nontrivial vacuum structure
in the Minkowski world. In section 3 we gave a detailed account of a
field-theoretic method for the calculation of
scattering amplitudes of high-energy soft hadronic
reactions. We started from the functional integral
and made high energy approximations in the integrand. This
led us to give general rules for writing down scattering amplitudes
for high energy soft reactions at the parton level. Then
we considered as representation for hadrons wave packets of partons.
The corresponding scattering amplitudes for (quasi-)elastic
hadron-hadron scattering were found to be governed
by the correlation functions of lightlike Wegner-Wilson
loops. The evaluation of these correlation functions was
possible with the help of the Minkowski version of
the stochastic vacuum model. The comparison with experiment gave very
good consistency, supporting the view that the QCD vacuum
structure plays an essential role in high energy soft reactions.
The framework developed in these lectures can be applied
directly to elastic and diffractive hadron-hadron scattering
at high energies. In principle we should also be able
to apply it to non-diffractive reactions, fragmentation
processes etc., but this remains to be worked out.

In sect. 4 we argued that some more
startling QCD vacuum effects in high
energy collisions may be
the appearance of anomalous
soft photons in hadron-hadron collisions
due to ``synchrotron radiation'' and spin correlations in the Drell-Yan
reaction due to the chromodynamic Sokolov-Ternov effect. Furthermore,
we gave an argument that electromagnetic form factors at small $Q^2$
should reflect the vacuum structure. Finally we would like to
mention that in \cite{81} the rapidity gap events observed at
HERA are quantitatively described in terms of the parton model
but invoking again nonperturbative QCD effects, possibly
connected with the vacuum structure.
Another place where the QCD vacuum structure may show
up is in certain correlations of hadrons in $Z^0$ decays
to two jets \cite{14,82} for which there is also some
experimental evidence \cite{83}.

We hope to have convinced the reader that the non-perturbative
structure of the QCD vacuum is useful in order to understand
confinement of heavy quarks. In our view this vacuum structure
manifests itself also in high energy soft and hard reactions.
We think it is very worth-while to study such effects
both from the theoretical and the experimental point of view.

\section*{Acknowledgements}
The author is grateful to the organizers of the 1996 Schladming
winter school headed by H. Latal for the invitation to lecture
there. As basis for the present article the author could
use lectures - partly collected by U. Grandel - given at
the Workshop on Topics in Field Theory, Graduiertenkolleg
Erlangen-Regensburg, Banz (1993). The author also thanks
the organizers of that meeting, especially F. Lenz,
for the invitation to lecture
there. The present article is the contribution of the author
to the proceedings of both meetings.
For many fruitful discussions on topics of these
lecture notes the author extends his gratitude to
A. Brandenburg, W. Buchm\"uller, H. G. Dosch, U. Ellwanger,
D. Gromes, P. Haberl,
P. V. Landshoff, P. Lepage, H. Leutwyler, Th. Mannel,
E. Mirkes, H. J. Pirner,
M. Rueter, Yu. A. Simonov, G. Sterman, and W. Wetzel. Special
thanks are due to H. G. Dosch for a critical
reading of the manuscript and to E. Berger and P. Haberl
for their help in the
preparation of the manuscript. Finally the author thanks
Mrs. U. Einecke for her excellent typing of the article.

\section*{Appendix A:\protect\\
Higher cumulant \hskip-.2pt terms and dynamic 
\hskip-.2pt fermions in the calculation
of the Wegner-Wilson loop in the stochastic vacuum model}
\renewcommand{\theequation}{A.\arabic{equation}}
\setcounter{equation}{0}
In this appendix we discuss first some problems arising
in the calculation of the Wegner-Wilson loop in the SVM (cf.
sect. 2.5) if higher cumulants are taken into account. Then we
outline a possible remedy which may also point to a way of including
the effects of dynamical fermions in the SVM.
We start with the replacements (\ref{2.77}) which allow us to use the
cumulant expansion (\ref{2.41}) for calculating $W(C)$. The
second cumulant is given in (\ref{2.78}):
\be\label{A.1}
K_2(1,2)=\frac{4}{3}\frac{\pi^2}{g^2}{\F}(1,2),\ee
where we set with $F_{\mu\nu\rho\sigma}$ as defined in
(\ref{2.61}):
\be\label{A.2}
{\F}(i,j)\equiv F_{1414}(X^{(i)}-X^{(j)}),\ee
\bear\label{A.2a}
&&F_{1414}(Z)=\frac{G_2}{24}[\kappa D(-Z^2)\nonumber\\
&&+\frac{1}{2}\left(\frac{\partial}{\partial Z_1}Z_1+\frac{\partial}
{\partial Z_4}Z_4\right)(1-\kappa)D_1(-Z^2)],\nonumber\\
&&Z^2=Z^2_1+Z^2_4.\ear
With the assumptions 1-3 of the SVM (sect. 2.4), the
next nonvanishing cumulant is $K_4$, for
which we obtain from (\ref{2.47}), (\ref{2.60}),
(\ref{2.71}), (\ref{2.73}):
\be\label{A.3}
K_4(1,2,3,4)=-2\left(\frac{\pi^2}{g^2}\right)^2
[{\F}(1,3){\F}(2,4)\Theta(1,2,3,4)\ +\ perm.].\ee
Here $\Theta(1,2,3,4)=1$ if $X^{(1)}>
X^{(2)}>X^{(3)}>X^{(4)}$ in the sense of the path-ordering
on the surface $S$ (cf. Figs. 8, 10) and $\Theta(1,2,3,4)=0$
otherwise.

We start now again from the expression (\ref{2.76})
for the expectation
value of the Wegner-Wilson loop and use the cumulant expansion
(\ref{2.41}) with the identifications (\ref{2.77}).
We get then:
\be\label{A.4}
W(C)=\exp\left\{\sum^\infty_{n=1}\frac{(-ig)^n}{n!}
\int_{S_1}...\int_{S_n}K_n(1,...,n)\right\},\ee
where
\be\label{A.5}
\int_{S_i}\equiv\int_S{\di}X^{(i)}_1{\di}X^{(i)}_4.\ee

In the SVM as formulated in sect. 2.4
the cumulants for odd $n$ vanish (cf. Ass. 3). The lowest
contribution in (\ref{A.4}) is then from $n=2$, the next from
$n=4$. Cutting off the infinite sum in (\ref{A.4})
at $n=4$, we get with (\ref{A.1})
and (\ref{A.3}):
\bear\label{A.6}
W(C)&=&\exp\left\{-\frac{g^2}{2!}\int_{S_1}\int_{S_2}K_2(1,2)\right.
\nonumber\\
&&\left.+\frac{(g^2)^2}{4!}\int_{S_1}...\int_{S_4}K_4(1,2,3,4)\right\}
\nonumber\\
&=&\exp\left\{-\frac{1}{2!}\frac{4\pi^2}{3}\int_{S_1}
\int_{S_2}{\F}(1,2)
\right.\nonumber\\
&&\left.-\frac{1}{4!}2\pi^4\int_{S_1}...\int_{S_4}[{\F}(1,3)
{\F}(2,4)\Theta(1,2,3,4)\ +\ perm.]\right\}\ear

Consider now the contribution of the second cumulant in (\ref{A.6})
for a large Wegner-Wilson loop (Fig. A1 with $R,T\gg a$):
\be\label{A.7}
I_2=\int_{S_1}\int_{S_2}{\F}(1,2).\ee
For fixed integration point (1) on $S$ the
integration over (2) will give significant contributions
only for a region of radius $\sim a$ around (1) since the
functions $D(-Z^2)$ and $D_1(-Z^2)$, where $Z=(X^{(1)}-
X^{(2)})$, are assumed to fall off rapidly with increasing
$Z^2$. From (\ref{A.2a}) we see that the $D_1$-term contributes
as a total divergence in ${\F}(1,2)$. Thus we can apply Gauss'
law in 2 dimensions for it to transform it to an integral
over the boundary $\partial S=C$. In this way we find a
contribution from the $D_1$ term of order $a^4/R^2,
\ a^4/RT,\ a^4/T^2$. From the $D$-term in (\ref{A.2a}) we get
a factor $\propto a^2$ from the integration over (2)
in (\ref{A.7}). Then the integration over (1) is unconstrained and
gives a factor proportional to the whole area of $S$:
\be\label{A.8}
I_2\propto RTa^2+O\left(a^4,a^4\frac{T}{R},a^4\frac{R}{T}
\right).\ee
Putting in all factors from (\ref{A.2a}) and using the explicit
form (\ref{2.64}) for the function $D(-Z^2)$ gives for $T\to\infty$
and $R\gg a$ the result:
\be\label{A.9}
\frac{1}{2!}\frac{4\pi^2}{3}I_2=RT\sigma\ee
with $\sigma$ as given in (\ref{2.81}).

We turn next to the contribution of the 4th
cumulant in (\ref{A.6}) and
study the integral
\be\label{A.10}
I_4=\int_{S_1}...\int_{S_4}{\F}(1,3){\F}(2,4)\Theta(1,2,3,4).
\ee
Consider again $R,T\gg a$ and fix the integration point (1)
(Fig. A2). Then the short-range correlation of the field
strengths requires (3) and (1) as well as (4) and (2)
to be near to each other, i.e. at a distance $\stackrel{
<}{\sim} a$. The function $\Theta(1,2,3,4)$ requires $(1)>(2)
>(3)>(4)$ in the sense of the path ordering on the surface $S$.
Using the fan-type net as indicated in Fig. 8 for the application
of the non-abelian Stokes theorem with straight line handles from
the points $X^{(i)}$ in the surface to the reference point $Y$
we see the following. The path-ordering function
$\Theta(1,2,3,4)$ restricts (2) to be in the hatched
sector of $S$ in Fig. A2. In general this does
\underbar{not} restrict (2) to a region close to (1).
On the contrary, (2) can vary freely over a triangle of area
 $\stackrel{
>}{\sim} a.L$, where $L$ is of order $R$ or $T$ or in between.
Thus, the integration over (2) in (\ref{A.10})
will give at least a factor $\propto (aT)$ if the hatched
triangle has its long sides in $X_4$ direction (cf. Fig. A2).
The integrations over (3) and (4) in (\ref{A.10}) should give
factors of $a^2$ each. The integration over (1) will give a factor
RT. Thus we estimate
\be\label{A.11}
I_4\propto RT.a^4.aT\propto T^2.\ee
This is very unpleasant. The 4th cumulant gives a
contribution which \underbar{domi-}\\
\underbar{nates} over the one
from the second cumulant for $T\to\infty$. There is no finite
limit for $T\to\infty$ in the expression (\ref{2.75}). The
quark-antiquark potential comes out infinite.

This problem was recognized in ref. \cite{51} and eliminated
by hand making an additional assumption: that all but
the leading powers in the quotient of the correlation
length to the extension parameters $R, T$ of the loop could
be neglected. Another simple cure of the problem would be
to postulate instead of Ass. 3 (cf. (2.73)-(2.75)) a behaviour
of the higher point correlation functions of the shifted
gluon field strengths which gives precisely zero for the
cumulants $K_n$ with $n>2$ in (A.5). In our opinion this
would be an unsatisfactory solution, since the model would then
only work for a particular fine-tuned set of correlation
functions whereas generically the above problem would remain.

We think that the origin of these difficulties in the SVM is
the assumption 1, which states that the correlator should be
independent of the reference point $Y$ for arbitrary $Y$.
But why should the field strength correlation (\ref{2.60})
(cf. Fig. 5) be the same if we use a straight line path
$C_X+\bar C_{X'}$ to connect $X$ and $X'$ or a path which runs
on a loop behind the moon? We will replace Ass. 1 by a milder
one:

\begin{itemize}
\item{\bf Ass. 1':} $F_{\mu\nu\rho\sigma}$ in (\ref{2.60}) is
independent of the reference point $Y$ and of the curves $C_X$
and $C_{X'}$ if the reference point $Y$ is close to $X$ and
$X'$:
\[|X-Y| \stackrel{\scriptstyle
<}{\sim} a',\ |X'-Y| \stackrel{\scriptstyle
<}{\sim} a',\]
where $a'$ is of order of $a$.
\end{itemize}

Now we try to calculate the expectation value of the
Wegner-Wilson loop (\ref{2.74}) using only this weaker
hypothesis. We start from the rectangle $S$ of area $RT$
and insert a smaller rectangular loop $C_1$ with
sides $R-2a', T-2a'$. On $C_1$ we choose $N_1$ reference
points $Y_1,...,Y_{N_1}$ and we partition the area between
$C$ and $C_1$ in $N_1$ plaquettes $P_1,...,P_{N_1}$ of
size $\sim {a'}^2$ (Fig. A3).
We can now apply the non-abelian Stokes theorem. We start
from $Y_0$ on $C$ and construct a path equivalent to
$C$ in the following way: From $Y_0$ to $Y_1$, then in a
fan-type net over the plaquette $P_1$ with $Y_1$ as reference
point, etc., until we arrive at $Y_{N_1}\equiv Y_1$
from where the path runs back to $Y_0$. According to
(\ref{2.36}) we get then:
\be\label{A.12}
W(C)=\frac{1}{3} {\Tr}\langle V_{0,N_1}V(P_{N_1})
V_{N_1,N_1-1...}V_{2,1}V(P_1)V_{1,0}\rangle,\ee
where
\be\label{A.13}
V(P_j)={\rm P}\ \exp\big[-ig\int_{P_j}\hat G_{14}(Y_j,X^{(j)};
C_{X^{(j)}})\big]\ee
and $V_{i,j}$ are  the connectors from $Y_j$ to $Y_i$ on
straight lines.

\begin{itemize}
\item{\bf Ass. 4:}
Now we will make a \underbar{mean field-type
approximation} and replace the path-ordered
integrals over the plaquettes $P_j$ by the corresponding vacuum
expectation values:
\be\label{A.14}
V(P_j)\to\langle V(P_j)\rangle\cdot\eins.\ee
\end{itemize}

This is similar in spirit to the
``block spin'' transformations considered in \cite{300}.
For the r.h.s. of (\ref{A.14}) we can use the cumulant
expansion and assumptions 1', 2 and 3 of the SVM. Here
the reference point $Y_j$ is never too far away from $X^{(j)}$.
We get with $\eins$ the unit matrix in colour space
\bear\label{A.15}
&&\langle V(P_j)\rangle=\eins\cdot\exp\big\{
\sum^\infty_{n=1}\frac{(-ig)^n}{n!} \int_{P_{j,1}}...
\int_{P_{j,n}}K_n(1,...,n)\big\}\nonumber\\
&&=\eins\cdot\exp\big\{-|P_j|\sigma\big[
1+O\big(\frac{1}{4!}\sigma\cdot a\cdot a'\big)\big]\big\}.
\ear
Here we cut off the cumulant expansion at $n=4$ and use the
results and estimates (\ref{A.6})-(\ref{A.11}),
but now for each plaquette $P_j$ instead of the whole
surface $S$. We see that the correction terms from the 4th
cumulant are now of manageable size since (cf. (\ref{2.68}),
(\ref{2.83}))
\be\label{A.16}
\sigma aa'=\left(\frac{a'}{a}\right)\sigma a^2=
0.56\left(\frac{a'}{a}\right)=O(1).\ee
Then (hopefully) the factorials $1/n!$ in the
cumulant expansion will lead to small corrections
from higher cumulants.

We can now insert (\ref{A.14}), (\ref{A.15}) in (\ref{A.12})
and get
\bear\label{A.17}
&&W(C)\simeq\frac{1}{3}{\Tr}\langle V_{0,N_1}\langle
V(P_{N_1})\rangle V_{N_1,N_1-1}...V_{2,1}\langle V(P_1)\rangle
V_{1,0}\rangle\nonumber\\
&&=\frac{1}{3} {\Tr}\langle V_{0,N_1}V_{N_1,N_1-1}...
V_{2,1}V_{1,0}\rangle\cdot \exp\big\{-\sum^{N_1}_{j=1}|P_j|\sigma
\big[1+O\big(\frac{\sigma aa'}{4!}\big)\big]\big\}
\nonumber\\
&&=\frac{1}{3}{\Tr}\langle V_{N_1,N_1-1}...V_{2,1}\rangle\exp
\big\{-(|S|-|S_1|)\sigma\big[1+O\big(\frac{\sigma
aa'}{4!}\big)\big]\big\}.\ear
Here we used also the cyclicity of the trace and
$V_{1,0}V_{0,N_1}=\eins$ (cf. (\ref{2.24})). The
product of the remaining connectors in (\ref{A.17}) gives just
the Wegner-Wilson loop of the curve $C_1=\partial S_1$ in
Fig. A.3 and we obtain:
\be\label{A.18}
W(C)=W(C_1)\exp\left\{-(|S|-|S_1|)\sigma\left[1+O\left(
\frac{\sigma aa'}{4!}\right)\right]\right\}.\ee
The procedure can easily be repeated by inscribing a rectangle
$S_2$ with boundary $C_2=\partial S_2$ in $S_1$ etc. The final
result of this iterative procedure obviously is:
\be\label{A.19}
W(C)=\exp\left\{-|S|\sigma\left[1+O\left(\frac{\sigma
aa'}{4!}\right)\right]\right\}.
\ee From
this we deduce for the ``true'' string tension in the SVM:
\be\label{A.20}
\sigma_{\rm true}=\sigma\left[1+O\left(
\frac{\sigma aa'}{4!}\right)\right]
=\sigma\left[1+\ {\rm terms\ of\ order}\ 5\%\right],\ee
where $\sigma$ is given in (\ref{2.81}).
Thus we have justified the result for the string
tension in section 2.5, where we used the second cumulant
only. We have now relied on the assumption 1' and avoided
transportation of field strengths to far away reference
points $Y$. The fourth cumulant is estimated to give only
a correction of a few percent and the contribution of the
6th, 8th, etc. cumulants can be estimated in an analogous way
to be even smaller\footnote{Numerical
studies suggest that $a'\approx 3a$
should be large enough for obtaining the
area law for the plaquettes $P_j$ in
(\ref{A.15}) (H. G. Dosch, private communication).
We obtain then $\sigma aa'\simeq 1.7$ and still correction
terms $\stackrel{\scriptstyle<}{\sim}$ 10 \% in (\ref{A.20}).}
Most important, we have avoided the
unpleasant result (\ref{A.11}).

We will now discuss another question which can be raised
in connection with the calculation of $W(C)$ in the SVM.
Why should we span a \underbar{minimal} surface $S$ into
the loop $C$ and not use some other, more
complicated surface $S'$ with the same boundary, $\partial
S'=C$? From the point of view of applying the non-abelian
Stokes theorem (\ref{2.37})
any wiggly surface $S'$ would be as good as the flat
rectangle $S$ which is the surface of minimal area. However,
from the point of view of the iterative calculation with the
plaquettes, as explained in this appendix, an
arbitrary surface $S'$ is clearly \underbar{not}
equivalent to $S$. In (\ref{A.14}) we made the approximation
of replacing $V(P_j)\ (j=1,...,N_1)$
by its vacuum expectation value. This should be a good
approximation if the various plaquettes are well separated.
Indeed, we would expect that then
we can perform the functional
integral over the variables related to the regions in
Euclidean space time where the various plaquettes are located
in a separate and independent way. If,
however, two plaquettes overlap or are very near to each
other, the above approximation will break down. The
point is now that on the minimal surface $S$ neighbouring
plaquettes will be at maximal distance from each other.
For some arbitrary surface $S'$ with wiggles we will
inevitably find plaquettes closer together which will make
our approximation worse. This gives us some rationale
to use a minimal surface $S$ in the applications of
the non-abelian Stokes theorem in the framework
of the SVM.

So far our calculations should apply to QCD with dynamical
gluons and static quarks only. The quark-antiquark potential
$V(R)$ in (\ref{2.82}) rises linearly for $R\to\infty$.
In real life we have, of course, dynamic quarks. If we
separate a heavy quark-antiquark pair $Q\bar Q$ starting
from small $R$ we will first see a linear potential as in
(\ref{2.82}) but at some point the heavy quark $Q$ (antiquark
$\bar Q$) will pick up a light antiquark $\bar q$ (quark $q$)
from the vacuum and form a meson $Q\bar q$ $(\bar Qq)$.
The two mesons can escape to infinity, i.e. the force between
them vanishes, the potential $V(R)$ must go to a constant
as $R\to\infty$:
\be\label{A.21}
V(R)\to V_\infty\quad {\rm for}\quad R\to\infty.\ee
Can we understand also this feature of nature in an
extension of the SVM?

Let us go back to the approximation (\ref{A.14}),
where we have replaced the integral over the plaquette $P_j$ by its
vacuum expectation value. More generally we can argue that
$V(P_j)$ as defined in (\ref{A.13}) is an object transforming
under a gauge transformation (\ref{2.17}) as follows (cf.
(\ref{2.20})):
\be\label{A.22}
V(P_j)\to U(Y_j)V(P_j)U^{-1}(Y_j).\ee
Any approximation we make should respect this gauge property
and indeed the replacement (\ref{A.14}) does. Now we can ask
for a generalisation of (\ref{A.14}) in the presence of
dynamic light quarks. We have then the quark and antiquark
variables at the point $Y_j$ at our
disposal and can construct from them the object
\be\label{A.23}
q(Y_j)\bar q(Y_j)\ee
which has the gauge transformation property (\ref{A.22})
and is also rotationally invariant. Let us consider only
$u$ and $d$ quarks. Then we suggest as generalization of
(\ref{A.14}) the following replacement:
\be\label{A.24}
V_{AB}(P_j)\to\langle V(P_j)\rangle_0 \delta_{AB}-
w(P_j)q_{A,f,\alpha}(Y_j)\bar q_{B,f,\alpha}(Y_j),\ee
where $A,B$ are the colour indices, $f=u,d$ is the flavour index
and $\alpha$ the Dirac index. Furthermore, $\langle
V(P_j)\rangle_0$
is as in (A.15), (A.16), and $w(P_j)$ is a coefficient
depending on the size of the plaquette $P_j$.
It can be thought of as representing the chance of producing
a $q\bar q$ pair from the vacuum gluon fields over the
plaquette $P_j$. (This is inspired by the discussions
of particle production in the LUND model \cite{53}). On dimensional
grounds $w(P_j)$ must be proportional to a volume, thus we will
set
\be\label{A.25}
w(P_j)=c|P_j|\cdot a,\ee
where  $c$ is a constant. The idea is that $q\bar q$
``production'' should feel the gluon fields in a disc
of area $|P_j|$ and thickness $a$.

We can now insert the ansatz (\ref{A.24}) in (\ref{A.12}).
The resulting expression for $W(C)$ is of the form:
\bear\label{A.26}
W(C)&\cong&\frac{1}{3}{\Tr}\langle V_{0,N_1}[\langle
V(P_{N_1})\rangle\nonumber\\
&&-w(P_{N_1})q(Y_{N_1})\bar q(Y_{N_1})]V_{N_1,N_1-1}
\nonumber\\
&&...V_{2,1}[\langle V(P_1)\rangle-w(P_1)q(Y_1)\bar q(Y_1)]
\cdot V_{1,0}
\rangle.\ear
If we mutliply out these brackets we get terms where we have
again the Wegner-Wilson loop along $C_1$ (Fig. A.3) but then also
terms where quarks and antiquarks at different points $Y_k,
Y_l$ are connected. For two neighbouring points,
for instance, $k=l+1,l$ this would read:
\[...q(Y_{l+1})[-w(l)\bar q(Y_{l+1})V_{l+1,l}q(Y_l)]\bar q(Y_l)...
\]
The importance of these terms will increase with increasing
$w(P_j)$.  Starting from (\ref{A.26}) we can now inscribe
plaquettes into the
rectangle $S_1$ and transport everything, including the
quark variables $q(Y_k)\bar q(Y_k)(k=1,...,N_1)$ to a curve
$C_2$ etc.

To get an orientation we will now make a drastic approximation:
For a loop with $R\ll R_c$, where $R_c$ is some critical
value to be determined, we neglect the dynamical
quarks, saying that the $w(P_j)$ factors will still be
too small. The result for $W(C)$ and $V(R)$ in the range
$a\stackrel{\scriptstyle <}{\sim} R\ll R_c$ will then be
as in (\ref{A.19}):
\bear\label{A.27}
&&W(C)=\exp(-RT\sigma),\nonumber\\
&&V(R)=\sigma R,\nonumber\\
&&(a\stackrel{\scriptscriptstyle <}{\sim}R\ll R_c),\ear
where we set $\sigma_{\rm true}\cong\sigma$.

In the other limiting case, $R\gg R_c$, we can expect the
$q\bar q$ terms in (\ref{A.26}) to dominate. Indeed, let us
start with the loop $C$ with sides $R$ and $T$ of Fig. A.3
and inscribe first plaquettes $P_{1,...}, P_{N_1}$ of size
${a'}^2$. With the replacement (\ref{A.24}) this gives for
$W(C)$ the expression (\ref{A.26}). Now we inscribe plaquettes
of size ${a'}^2$ in the loop $C_1$ and parallel transport
everything to a smaller loop $C_2$. The new element is that
we will now have to parallel-transport also the quark variables,
but this poses no problem. We will again obtain an expression
like the one shown in (\ref{A.26}) but now for the
curve $C_2$ and with increased weight factors $w$ in front
of the $q\bar q$ terms. We assume that we continue this
procedure. Finally, we will obtain an expression like
(\ref{A.26}) but with the terms involving the quark
variables at the reference points $Y_j$ dominating the
expression. Thus we set:
\be\label{A.28}
W(C)\cong\frac{1}{3}{\Tr}\langle[-w(N)q(N)\bar q(N)
]V_{N,N-1}...[-w(1)q(1)\bar q(1)]V_{1,N}\rangle,\ee
where the points $1,...,N$ are on a curve $C'$ of sides
$R',T'$.
This summing up of quark contributions from various plaquettes
of size ${a'}^2$ should be reasonable as long as the quarks
are inside an area corresponding to their own correlation
length, $a_\chi$, the chiral correlation length. This can
be estimated from the behaviour of the non-local $\bar qq$
condensate [86-88] for which one typically makes an ansatz
of the form $(q=u,d)$:
\be\label{A.29}
\langle\bar q(X_2)V_{2,1}\ q(X_1)\rangle=\langle\bar qq\rangle
e^{-|X_2-X_1|^2/a^2_\chi}.\ee
Here $\langle\bar qq\rangle$ is the local quark condensate
for which we set, neglecting isospin breaking (cf. \cite{38}):
\be\label{A.30}
\langle\bar qq\rangle=\frac{1}{2}\langle\bar uu+\bar dd\rangle
=-(0.23\ {\GeV})^3.\ee
For the correlation length $a_\chi$ one estimates (cf. \cite{303}):
\be\label{A.31}
a_\chi\cong 1\ {\rm fm}.\ee
It will be reasonable to choose the reference points $1,...,N$ on
$C'$ also such that the distance between neighbouring points
is $a_\chi$. Then
\bear\label{A.32}
R'&=&R-2a_\chi,\nonumber\\
T'&=&T-2a_\chi,\nonumber\\
N&=&\frac{2T'+2R'}{a_\chi}=\frac{2(T+R)}{a_\chi}-8.\ear

We will now estimate $W(C)$ from (\ref{A.28}) as product of
the expectation values of the nonlocal $\bar qq$ condensate.
We set
\be\label{A.33}
w(1)=w(2)...=w(N)=c\cdot aa_\chi^2,\ee
where $c$ should be a numerical constant of order 1. Furthermore we
have from (\ref{A.29})
\bear\label{A.34}
&&\langle \bar q_{A,f,\alpha}(j)(V_{j,j-1})_{AB}\ q_{B,f',\alpha'}
(j-1)\rangle\nonumber\\
&&=\frac{1}{4}\delta_{f,f'}\delta_{\alpha,\alpha'}\langle
\bar qq\rangle\exp\left[-\frac{|X_j-X_{j-1}|^2}{a_\chi^2}\right]
\nonumber\\
&&=\frac{1}{4e}\delta_{f,f'}\delta_{\alpha,\alpha'}
\langle \bar qq\rangle\ear
for $|X_j-X_{j-1}|=a_\chi$. This gives:
\bear\label{A.35}
W(C)&\cong&-\frac{1}{3}.\left\{
\left[-w(N)\langle\bar q_{f_1,\alpha_1}(1)V_{1,N}\ q_{f_N,
\alpha_N}(N)\rangle\right]\right.
\nonumber\\
&\hphantom{=}
&\left....\left[-w(1)\langle\bar q_{f_2,\alpha_2}(2)V_{2,1}\
q_{f_1,
\alpha_1}(1)\rangle\right]\right\}\nonumber\\
&=&-\frac{8}{3}\left[-\frac{w(1)\langle\bar qq\rangle}
{4e}\right]^N\nonumber\\
&=&-\frac{8}{3}\left[-\frac{w(1)\langle\bar qq\rangle}
{4e}\right]^{[\frac{2}{a_\chi}(T+R)-8]}.\ear
For the potential this leads to
\be\label{A.36}
V(R)=-\lim_{T\to\infty}\frac{1}{T}\ln W(C)=\frac{2}{a_\chi}
\ln\frac{4e}{[-w(1)\langle\bar qq\rangle]}
\equiv V_\infty.\ee
Thus we find indeed a constant potential for $R\to\infty$.

For intermediate values of $R$ the potential should change
smoothly from the linear rise to the constant behaviour. As a
crude approximation we will assume the potential to be
continuous but turn abruptly from one to the other behaviour
at a critical length $R=R_c$ which we define as
\be\label{A.37}
R_c=V_\infty/\sigma.\ee
Our simple ansatz for $V(R)$ is then as follows (cf. Fig. A.4):
\be\label{A.38}
V(R)=\left\{\begin{array}{ccc}
\sigma R&{\rm for}& R\leq R_c,\\
V_\infty=\sigma R_c&{\rm for}& R\geq R_c.\end{array}\right..\ee

How does the numerics work out taking (\ref{A.36}) and (\ref{A.37})
seriously? We can estimate the constant $V_\infty$
crudely from the mass difference of two $B$-mesons and the
$\Upsilon$ resonance
\be\label{A.39}
V_\infty\cong 2m_B-m_\Upsilon\cong 1.1\ {\GeV}.\ee
With $\sigma$ from (\ref{2.83}) we find for $R_c$
(\ref{A.37}):
\be\label{A.40}
R_c\cong 1,2\ {\rm fm}.\ee
A much more elaborate estimate of the string-breaking
distance gives \cite{304}
$R_c\leq 1,6$ to $2,1 {\rm fm}$. The lattice calculations
of \cite{305} suggest an even larger value for
$R_c$. From (\ref{A.39}) and (\ref{A.36}), (\ref{A.31}) we get
for $w(1)$ and $c$:
\bear\label{A.41}
w(1)&=&-\frac{4\exp[1-\frac{1}{2}a_\chi V_\infty]}
{\langle\bar qq\rangle}=0.42\ {\rm fm}^3,\nonumber\\
c&=&\frac{w(1)}{a^2_\chi a}=1.2.\ear
Thus the probability factor $w(1)$ comes out as we
estimated it on geometrical grounds in (\ref{A.25}).

With these remarks we close this appendix. Of course, much
more work is needed in order to decide if our simple ansatz
for incorporating dynmical fermions in the stochastic
vacuum model is viable or not.

\section*{Appendix B: The scattering of gluons}
\renewcommand{\theequation}{B.\arabic{equation}}
\setcounter{equation}{0}

In this appendix we will discuss the contribution of gluons to
the scattering amplitude of a general parton reaction,
generalizing (\ref{3.46}):
\be\label{B.1}
G(1)+G(2)+...+q(k)+\bar q(k')+...\to G(3)+G(4)+...
+q(l)+\bar q(l')+... .\ee
Here, as in all of sect. 3, we use the convention that partons
with odd (even) number are moving fast in positive (negative)
$x^3$-direction. Consider thus the transition of a right-moving
gluon:
\be\label{B.2}
G(1)\equiv G(P_1,\vec\epsilon_1,a_1)\to G(3)\equiv
G(P_3,\vec\epsilon_3,a_3),\ee
where $a_{1,3}\ (1\leq a_{1,3}\leq8)$ are the
colour indices,  $P_1,P_3$ are the momenta
with $P_{1+},P_{3+}\to\infty$, and $\vec\epsilon_{1,3}$ are
the polarization vectors which satisfy:
\bear\label{B.3}
&&\vec P_1\vec\epsilon_1=0,\nonumber\\
&&\vec P_3\vec\epsilon_3=0,\nonumber\\
&&|\vec\epsilon_1|=|\vec\epsilon_3|=1.\ear
In the high energy limit the vectors $\vec\epsilon_{1,3}$ are
transverse with corrections of order $|\vec p_{T1,3}|/P_{+1,3}$
which we can neglect. We will argue that such a gluon in a
high energy soft reaction is equivalent to a quark-antiquark
pair with the same quantum numbers in the limit of the $q$
and $\bar q$ being so close to each other in position space that
their separation cannot be resolved in the collision. From
the contribution of such a $q\bar q$ pair to the
scattering amplitude we will get in the above limit the
gluon contribution.

We start by constructing wave packets similar to the mesonic
wave packets of (\ref{3.53a})
\bear\label{B.4}
&&|q\bar q; P_j,\vec\epsilon_j,a_j\rangle=
\int {\di}^2p_T\int^1_0{\di}\zeta\frac{1}{(2\pi)^{3/2}}
h^j(\zeta,|\vec p_T|)\cdot
\nonumber\\
&&\frac{1}{2} (\lambda_{a_j})_{A_j{A'}_j}
(\vec\epsilon_j\cdot\vec\sigma\epsilon)_{s_js_j'}
|q(p_j,s_j,A_j)\bar q(P_j-p_j,s_j',A_j')\rangle\nonumber\\
&&(j=1,2,3,4),\ear
where $p_j$ is as in (\ref{3.54}), (\ref{3.55}) and
\be\label{B.5}
\epsilon=\left(\begin{array}{cc}
0&1\\
-1& 0\end{array}\right).\ee
We require the functions $h^j$ to satisfy:
\bear\label{B.6}
&&(h^j(\zeta,|\vec p_T|))^*=h^j(\zeta,|\vec p_T|),\nonumber\\
&&h^j(\zeta,|\vec p_T|)=h^j(1-\zeta,|\vec p_T|),\nonumber\\
&&\int {\di}^2p_T\int_0^1{\di}\zeta\ 2\zeta
(1-\zeta)|h^j(\zeta,|\vec p_T|)
|^2=1.\ear
This gives us the normalization of the states (\ref{B.4}) as
\bear\label{B.7}
&&\langle q\bar q;P_j',\vec\epsilon_j',
a_j'|q\bar q,P_j,\vec\epsilon_j,a_j\rangle\nonumber\\
&&=\vec\epsilon_j^{\prime*}\cdot\vec\epsilon_j\delta_{a_{j'},a_j}
(2\pi)^32P^0_j\delta^3(\vec P_j'-\vec P_j).\ear

The $q\bar q$ states (\ref{B.4}) have the same transformation
properties under a parity (P) charge conjugation (C)
and time reversal (T) transformation as a gluon state:
\bear
&&U({\rm P})|q\bar q; P^0,\vec P,\vec\epsilon,a\rangle
=-|q\bar q;P^0,-\vec P,\vec\epsilon,a\rangle,\label{B.8}\\[0.2cm]
&&U({\rm C})|q\bar q; P^0,\vec P,\vec\epsilon,a\rangle=-|q\bar q;P^0,\vec  
P,\vec\epsilon,b\rangle
\cdot\frac{1}{2}{\rm Tr}(\lambda_b\lambda_a^T),\label{B.9}\\
&&V({\rm T})|q\bar q; P^0,\vec P,\vec\epsilon,a\rangle
=-|q\bar q;P^0,-\vec P,\vec\epsilon^*,b\rangle
\frac{1}{2} {\rm Tr}(\lambda_b\lambda_a^T).\label{B.9a}\ear
Here $U({\rm P}),U({\rm C})$ are the unitary operators, $V$(T)
is the antiunitary operator representing the P, C
and T transformations, respectively, in the Fock space
of parton states.

As in (\ref{3.58}) we define the wave functions in transverse
position space and longitudinal momentum fraction:
\be\label{B.10}
\varphi^j(\zeta,\vec y_T):=\sqrt{2\zeta(1-\zeta)}
\frac{1}{2\pi}\int {\di}^2p_T\exp(i\vec p_T\cdot\vec y_T)h^j
(\zeta,|\vec p_T|).\ee
Here we also define for the right (left) movers the wave
functions in $y_-,\vec y_T$ space ($y_+,\vec y_T$ space)
as:
\bear\label{B.11}
&&\tilde\varphi^j(y_-,\vec y_T):=\frac{1}{\sqrt{4\pi}}
\sqrt{P_{j+}}\int^1_0{\di}\zeta\ \varphi^j(\zeta,\vec y_T)
\exp[-\frac{i}{2}P_{j+}(\zeta-\frac{1}{2})y_-],\nonumber\\
&&(j\ {\rm odd}),\ear
\bear\label{B.12}
&&\tilde\varphi^j(y_+,\vec y_T):=\frac{1}{\sqrt{4\pi}}
\sqrt{P_{j-}}\int^1_0{\di}\zeta\ \varphi^j(\zeta,\vec y_T)
\exp[-\frac{i}{2}P_{j-}(\zeta-\frac{1}{2})y_+],\nonumber\\
&&(j\ {\rm even}),\ear
The normalization condition (\ref{B.6}) implies:
\bear\label{B.13}
&&\int {\di}^2y_T\int^1_0{\di}\zeta|\varphi^j
(\zeta,\vec y_T)|^2=1,
\nonumber\\
&&\int {\di}y_-\int {\di}^2y_T|\tilde\varphi^j(y_-,\vec y_T)
|^2=1\qquad(j\ {\rm odd}).\nonumber\\
&&\int {\di}y_+\int {\di}^2y_T|\tilde\varphi^j(y_+,\vec y_T)
|^2=1\qquad (j\ {\rm even}).\ear

To realize the condition that the $q\bar q$ pair acts like a gluon,
we require for right (left) movers that they have similar
longitudinal momenta and that their wave function in
the relative $q$-$\bar q$ coordinates,
$y_-,\vec y_T (y_+,\vec y_T)$
is nearly a $\delta$ function. To be concrete, we require:
\be\label{B.14}
\varphi^j(\zeta,\vec y_T)\not=0\quad {\rm only\ for}
\quad |\zeta-\frac{1}{2}|\leq\xi_0,\ee
where
\be\label{B.15}
0<\xi_0\ll\frac{1}{2},\ee
and for $j$ odd:
\be\label{B.16}
\tilde\varphi^j(y_-,\vec y_T)\not=0\quad{\rm only\ if}\quad
|y_-|\stackrel{\scriptstyle<}{\sim}\frac{1}{P_+\xi_0}
\quad {\rm and}\quad |\vec y_T|\ll a.\ee
For right movers we replace plus by minus signs in
(\ref{B.16}). Any $q\bar q$ wave packets with these
properties should then look identical to a gluon for an
observer in the femto universe with regard to
``soft'' scatterings.

Now we replace $G(1)$ and $G(3)$ in reaction (\ref{B.1})
by the $q\bar q$ wave packets (\ref{B.4}). According
to the rules derived in sect. 3.5 the scattering of the
$q\bar q$ system
\be\label{B.17}
q(1)\bar q(1')\to q(3)\bar q(3')\ee
gives the following factor in the $S$-matrix (cf. (\ref{3.52b})
(\ref{3.52c})):
\[{\cal S}_{q+}(3,1){\cal S}_{\bar q+}(3',1')\]
which still has to be integrated over the wave functions
(\ref{B.4}). In this way we obtain for the $q\bar q$-pair:

\newpage

\bear\label{B.18}
&&{\cal S}_{q\bar q}(3,1)=\int {\di}^2p_T'\int^1_0{\di}
\zeta'\int {\di}^2p_T
\int^1_0{\di}\zeta\nonumber\\
&&\hphantom{{\cal S}_{q\bar q}(3,1)=}
\frac{1}{(2\pi)^3}h^3(\zeta',|\vec p_T'|)h^1(\zeta,
|\vec p_T|)\nonumber\\
&&\hphantom{{\cal S}_{q\bar q}(3,1)=}\frac{1}{2}(\lambda_{a_3})
_{A_3',A_3}({\vec\epsilon_3}^*
\cdot \epsilon^T\vec\sigma)_{s_3',s_3}\nonumber\\
&&\hphantom{{\cal S}_{q\bar  
q}(3,1)=}\frac{1}{2}(\lambda_{a_1})_{A_1,A_1'}(\vec\epsilon_1
\cdot \vec\sigma\epsilon)_{s_1,s_1'}\nonumber\\
&&\hphantom{{\cal S}_{q\bar q}(3,1)=}{\cal S}_
{q+}(3,1)\ {\cal S}_{\bar q+}(3',1'),\ear
\bear\label{B.19}
{\cal S}_{q\bar q}(3,1)&=&\int {\di}z_-\int {\di}y_-
\int {\di}^2z_T\int {\di}^2y_T
\sqrt{P_{3+}P_{1+}}\vec \epsilon_3^*\cdot\vec\epsilon_1
\nonumber\\
&&\exp[\frac{i}{2}(P_{3+}-P_{1+})z_--i(\vec P_3-\vec P_1)_T
\vec z_T]\nonumber\\
&&{\tilde\varphi^{3*}}(y_-,\vec y_T)\tilde\varphi^1
(y_-,\vec y_T)\nonumber\\
&&\frac{1}{2}{\Tr}[\lambda_{a_3}V_-(\infty,z_-+\frac{1}{2}y_-,
\vec z_T+\frac{1}{2}\vec y_T)\nonumber\\
&&\lambda_{a_1}V^\dagger_-(\infty,z_--\frac{1}{2}y_-,\vec z
_T-\frac{1}{2}\vec y_T)].\ear
With our assumptions (\ref{B.16})
${\tilde\varphi^{3*}}(y_-,\vec y_T)\tilde\varphi^1
(y_-,\vec y_T)$ acts like a $\delta$-function at $y_-=0,\
\vec y_T=0$ and we get:
\bear\label{B.20}
{\cal S}_{q\bar q}(3,1)&=&\sqrt{P_{3+}P_{1+}}\
\vec\epsilon_3^*\cdot\vec\epsilon_1\nonumber\\
&&\int {\di}y_-\int {\di}^2y_T\ \tilde\varphi^{3*}(y_-,\vec y_T)
\tilde\varphi^1
(y_-,\vec y_T)\nonumber\\
&&\int {\di}z_-\int {\di}^2z_T\ \exp[\frac{i}{2}(P_{3+}-P_{1+})z_-
-i(\vec P_3-\vec P_1)_T\cdot
\vec z_T]\nonumber\\
&&\frac{1}{2}{\Tr}[\lambda_{a_3}V_-(\infty,z_-,
\vec z_T)\lambda_{a_1}V^\dagger_-(\infty,z_-,\vec z
_T)].\ear
An easy exercise shows that
\be\label{B.21}
\frac{1}{2}{\Tr}[\lambda_{a_3}V_-(\infty,z_-,\vec z_T)\lambda
_{a_1}V^\dagger_-(\infty,z_-,\vec z_T)]={\cal V}
_-(\infty,z_-,\vec z_T)_{a_3,a_1},\ee
where ${\cal V}_-$ is the connector analogous to (\ref{3.33}) but
for the adjoint representation (cf. (\ref{2.22})):
\bear\label{B.22}
&&{\cal V}_-(\infty,z_-,\vec z_T)={\rm P}\big\{\exp\big[
-\frac{i}{2}g\int^\infty_{-\infty}{\di}z_+G^a_-(z_+,z_-,\vec z_T)
T_a\big]\big\},\nonumber\\
&&(T_a)_{bc}=\frac{1}{i}f_{abc}.\ear
The $f_{abc}$ are the structure constants of $SU(3)_c$.
Inserting (\ref{B.21}) in (\ref{B.20}) gives
\be\label{B.23}
{\cal S}_{q\bar q}(3,1)=\int {\di}y_-\int {\di}^2y_T
\ {\tilde\varphi^{3*}}
(y_-,\vec y_T)\tilde\varphi^1(y_-,\vec  
y_T)\epsilon^*_{3j_3}\epsilon_{1j_1}{\cal S}_{G+}(3,1),\ee
where
\bear\label{B.24}
&&{\cal S}_{G+}(3,1)=\sqrt{P_{3+}P_{1+}}\
\delta_{j_3,j_1}\int {\di}z_-\int {\di}^2z_T\nonumber\\
&&\exp\big[\frac{i}{2}(P_{3+}-P_{1+})z_--i(
\vec P_{3T}-\vec P_{1T})\cdot\vec z_T\big]\nonumber\\
&&{\cal V}_-(\infty,z_-,\vec z_T)_{a_3,a_1}.\ear

In a general scattering reaction (\ref{B.1}) the factor ${\cal S}
_{q\bar q}(3,1)$ (\ref{B.23}) has to be inserted with other factors
${\cal S}_q,{\cal S}_{\bar q},...$ and then integrated over all
gluon potentials as explained in sect. 3.5. We note that
${\cal S}_{q\bar q}(3,1)$ in (\ref{B.23}) factorizes
into ${\cal S}_{G+}(3,1)$ (\ref{B.24}) times the overlap
of the internal wave functions of the incoming
and outgoing $q\bar q$ pairs:
\be\label{B.25}
\int {\di}y_-\int {\di}^2y_T\ \tilde\varphi^{3*}(y_-,\vec y_T)
\tilde\varphi^1(y_-,\vec y_T).\ee
This means that the $q\bar q$ pair will come out with some
distribution in total momentum $P_3$ and polarization
vector $\vec\epsilon_3=\vec\epsilon_1$ but
always with internal wave function $\tilde\varphi^1(y_-,\vec y
_T)$. With the conditions (\ref{B.14})-(\ref{B.16})
$\tilde\varphi^1(y_-,\vec y_T)$ leads to a ``permissible''
internal wave function for a $q\bar q$ pair of momentum
$P_3$, to be regarded as a \underbar{gluon} of momentum
$P_3$ by our observer in the
femto universe. Indeed we have from (\ref{B.11}):
\bear\label{B.26}
\tilde\varphi^1(y_-,\vec y_T)&=&\sqrt{\frac{P_{1+}}
{4\pi}}\int^1_0{\di}\zeta\ \varphi^1(\zeta,\vec  
y_T)\exp[-\frac{i}{2}P_{1+}(\zeta-\frac{1}{2})y_-]\nonumber\\
&=&\sqrt{\frac{P_{1+}}
{4\pi}}\int^{1/2}_{-1/2}{\di}\xi\ \varphi^1(\frac{1}{2}+\xi,
\vec y_T)\exp[-\frac{i}{2}P_{1+}\xi\cdot y_-]\nonumber\\
&=&\sqrt{\frac{P_{3+}}
{4\pi}}\int^{\xi+}_{\xi-}{\di}\xi'\ \varphi^{\prime 3}(
\frac{1}{2}+\xi',\vec y_T)\exp[-\frac{i}{2}P_{3+}\xi'\cdot y_-]\ear
Here we define the internal wave function of the outgoing $q\bar q$
pair as
\be\label{B.27}
\varphi^{\prime 3}\left(\frac{1}{2}+\xi',\vec y_T\right)=
\sqrt{\frac{P_{3+}}{P_{1+}}}\varphi^1\left(
\frac{1}{2}+\frac{P_{3+}}{P_{1+}}\xi',\vec y_T\right)\ee
and
\be\label{B.28}
\xi_\pm=\pm\frac{P_{1+}}{P_{3+}}\cdot\frac{1}{2}.\ee
The condition (\ref{B.14}) for $\varphi^1$ guarantees a
similar condition for $\varphi^{\prime 3}$ if $P_{3+}/P_{1+}$ is
of order 1, as we will always assume. Thus the integration
limits $\xi_\pm$ in (\ref{B.26}) can be replaced by
$\pm\frac{1}{2}$ and also the normalization conditions
(\ref{B.13}) can easily be checked for $\varphi^{\prime 3}$.

To summarize: In this appendix we have shown that suitable
$q\bar q$ pairs which are indistinguishable from gluons
for our observer in the femto universe scatter as entities
in a soft reaction. Their internal wave function in momentum
space is modified, but in a way not observable in a soft reaction.
Thus we can consider ${\cal S}_{G+}(3,1)$ in (\ref{B.24})
as the scattering contribution of a right-moving gluon in
the transition (\ref{B.2}). We quoted this result already
in (\ref{3.52h}) in sect. 3.5. For left moving gluons
we just have to
exchange everywhere $+$ and $-$ components.

\section*{Appendix C: The scattering of baryons}
\renewcommand{\theequation}{C.\arabic{equation}}
\setcounter{equation}{0}
In this appendix we discuss high energy soft reactions,
in particular elastic reactions involving baryons and antibaryons.
We represent baryons by $qqq$ wave packets:
\bear\label{C.1}
&&|B_j(P_j)\rangle=\frac{1}{6\cdot(2\pi)^3}\int {\di}\mu \ h^j
(f^i,s^i,\zeta^i,\vec p^i_T)\cdot\nonumber\\
&&\epsilon_{A^1A^2A^3}|q(p^1_j,f^1,s^1,A^1),q(p_j^2,f^2,s^2,A^2),
q(p_j^3,f^3,s^3,A^3)\rangle.\ear
Here $f^i,s^i,A^i\ (i=1,2,3)$ are the flavour, spin, and colour
indices of the quarks and $\epsilon_{ABC}\ (\epsilon_{123}=1)$
is the totally antisymmetric tensor.  For the momenta
$p^i_j$ we set for right-moving baryons ($j$ odd, $P_{j+}
\to\infty$):
\bear\label{C.2}
&&p_{j+}^i=\zeta^iP_{j+},\nonumber\\
&&\vec p^i_{jT}=\frac{1}{3}\vec P_{jT}+
\vec p^i_T,\nonumber\\
&&(i=1,2,3)\ear
and for left-moving baryons ($j$ even, $P_{j-}\to\infty$):
\bear\label{C.3}
&&p^i_{j-}=\zeta^iP_{j-},\nonumber\\
&&\vec p_{jT}^i=\frac{1}{3}\vec P_{jT}+\vec p^i_T,\nonumber\\
&&(i=1,2,3).\ear
The integral with the measure ${\di}\mu$ stands for
\bear\label{C.4}
\int{\di}\mu\equiv&&\int\int\int\prod^3_{i=1}{\di}^2p^i_T\delta^2
(\sum^3_{i=1}\vec p_T^i)\cdot
\nonumber\\
&&\int^1_0\int^1_0\int^1_0
\prod^3_{i=1}{\di}\zeta^i
\delta(1-\sum^3_{i=1}\zeta^i).\ear
The flavour and spin of the baryon states (\ref{C.1})
is, of course, determined by the functions $h^j$ which
must be totally symmetric under simultaneous exchange
of the arguments
$(f^i,s^i,\zeta^i,\vec p^i_T)$ for $i=1,2,3$. In the
following we will collectively set $(f,s,\zeta,\vec p_T)\equiv\alpha$
and $q(\alpha,A)\equiv q(p_j,f,s,A)$. We have then
\bear
&&h^j(\alpha,\beta,\gamma)=h^j(\beta,\alpha\gamma)
=h^j(\alpha,\gamma,\beta).\label{C.5}\\[0.2cm]
&&|B_j(P_j)\rangle=\frac{1}{6.(2\pi)^3}\int {\di}\mu\  h^j(\alpha,\beta,
\gamma)\epsilon_{ABC}|q(\alpha,A)q(\beta,B)q(\gamma,C)\rangle.
\quad\label{C.6}\ear
The normalization condition
\be\label{C.7}
\langle B_j(P_j')|B_j(P_j)\rangle=
(2\pi)^32P_j^0\delta^3(\vec P_j'-\vec P_j)\ee
requires $h^j$ to satisfy:
\bear\label{C.8}
&&\int {\di}\mu\ 4\zeta^1\zeta^2\zeta^3\ {h^j}^*(\alpha,\beta,\gamma)h^j
(\alpha,\beta,\gamma)=1\nonumber\\
&&({\rm no\ summation\ over} j).\ear
We define the wave functions $\varphi^j$ in transverse position and
longitudinal momentum space and the transition profile functions
$w_{kj}^B$ for $B_j\to B_k$ as:
\newpage
\bear
&&\varphi^j(f^i,s^i,\zeta^i,\vec x^i_T):=
\frac{1}{(2\pi)^2}\int\prod
^3_{i=1}{\di}^2p^i_T\cdot\delta^2(\sum_{i=1}^3\vec p^i_T)\cdot\nonumber\\
&&6\cdot
(\zeta^1\zeta^2\zeta^3)^{1/2}
\exp(i\sum^3_{i=1}\vec p_T^i\cdot\vec x^i_T)h^j
(\alpha^1,\alpha^2,\alpha
^3),\label{C.9}\\[0.2cm]
&&w^B_{k,j}(\vec x^1_T,\vec x_T^2,\vec x_T^3):=
\int^1_0\int^1_0\int^1_0\prod^3_{i=1}
{\di}\zeta^i\ \delta(1-\sum^3_{i=1}\zeta^i)\nonumber\\
&&\sum_{f^i,s^i}{\varphi^3}^*(f^1,s^1,\zeta^1,\vec x_T^1;
f^2,s^2,\zeta^2,\vec x^2_T;f^3,s^3,\zeta^3,\vec x^3_T)\nonumber\\
&&\varphi^1(f^1,s^1,\zeta^1,\vec x_T^1;
f^2,s^2,\zeta^2,\vec x^2_T;f^3,s^3,\zeta^3,\vec x^3_T).
\label{C.10}\ear
The symmetry relations for $h$ (\ref{C.5}) and the
normalization condition (\ref{C.8}) imply:
\be\label{C.10a}
w^B_{k,j}(\vec x_T^1,\vec x_T^2,\vec x_T^3)=w_{k,j}^B(\vec x_T^2,
\vec x_T^1,\vec x_T^3)=w_{k,j}^B(\vec x_T^1,\vec x_T^3,\vec x_T^2),
\ee
\bear\label{C.10b}
&&\int\prod^3_{i=1}{\di}^2x^i_T\
\delta^2(\vec x_T^1+\vec x_T^2+\vec x_T^3)
w_{jj}^B(\vec x_T^1,\vec x_T^2,\vec x_T^3)=1\nonumber\\
&&({\rm no\ summation\ over}\ j).\ear

As a concrete scattering reaction let us consider meson-baryon
scattering:
\be\label{C.11}
B_1(P_1)+M_2(P_2)\to B_3(P_3)+M_4(P_4).
\ee From the rules of sect. 3.5 we get
\bear\label{C.12}
&&S_{fi}\equiv \langle
B_3(P_3),M_4(P_4)|S|B_1(P_1)M_2(P_2)\rangle=\nonumber\\
&&\frac{1}{6(2\pi)^6}\int {\di}\mu'\ {h^3}^*(\alpha',\beta',\gamma')
\epsilon_{A'B'C'}\int {\di}\mu\ h^1(\alpha,\beta,\gamma)\epsilon_{ABC}
\nonumber\\
&&\frac{1}{3(2\pi)^3}\int {\di}^2p_T'\int^1_0{\di}\zeta'
\int {\di}^2p_T\int^1_0{\di}\zeta\
{h^4}^*_{s_4s_4'}(\zeta',\vec p_T')h^2_{s_2s_2'}(\zeta,
\vec p_T)\nonumber\\
&&\big\langle{\cal S}_{q+}(\alpha',\alpha)_{A'A}{\cal S}_{q+}
(\beta',\beta)_{B'B}{\cal S}_{q+}(\gamma',\gamma)_{C'C}\nonumber\\
&&\hphantom{\big\langle}
\delta_{A_2'A_2}\delta_{A_4'A_4}
{\cal S}_{q-}(4,2)_{A_4A_2}{\cal S}_{\bar q-}(4',2')_
{A_4'A_2'}\big\rangle_G.\ear
After some straightforward algebra we get:
\bear\label{C.13}
S_{fi}&=&\delta_{fi}+i(2\pi)^4\delta(P_3+P_4-P_1-P_2){\T}_{fi}
,\nonumber\\[0.2cm]
{\T}_{fi}&=&-\frac{i}{(2\pi)^6}\int {\di}\mu'\ {h^3}^*(\alpha',\beta',
\gamma')\nonumber\\
&&\int {\di}\mu\  h^1(\alpha,\beta,\gamma)\prod^3_{i=1}
(\delta_{f^{\prime i},f^i}\delta_{s^{\prime i},s^i})
(P_{1+})^3\left[\prod^3_{i=1}\zeta^{\prime i}\zeta^i\right]^{1/2}
\nonumber\\
&&\frac{1}{(2\pi)^3}\int {\di}^2 p'_T
\int^1_0 {\di}\zeta'\int {\di}^2 p_T\int^1_0 {\di}\zeta\
{h^4}^*_{s,r}(\zeta',\vec p'_T)h^2_{s,r}(\zeta,\vec p_T)\nonumber\\
&&2(P_{2-})^2\left[\zeta'(1-\zeta')
\zeta(1-\zeta)\right]^{1/2}\nonumber\\
&&3^3\cdot\int {\di}^2 b_T\exp(i\vec q_T\cdot\vec b_T)
\int\prod^3_{i=1}({\di}x^i_- {\di}^2 x^i_T)\delta \left(
\sum^3_{i=1} x^i_-\right)\ \delta^2
\left(\sum^3_{i=1}\vec x^i_T\right)\nonumber\\
&&\int {\di}y_+{\di}^2y_T\exp \big\{ i\sum^3_{i=1}
\left[\frac{1}{2} P_{1+}
(\zeta^{\prime i}-\zeta^i) x^i_--(\vec p_T^{\prime i}-
\vec p^i_T)\cdot\vec x_T^i\right]\nonumber\\
&&+i\frac{1}{2}P_{2-}(\zeta'-\zeta)y_+-i(\vec p'_T-\vec p_T)\cdot
\vec y_T\big\}\nonumber\\
&&\big\langle\big\{ V_-(\infty,x^1_-,\frac{1}{2}
\vec b_T+\vec x_T^1)_{A'A}V_-(\infty,x^2_-,\frac{1}{2}\vec b_T+\vec x^2_T)_{B'B}
\nonumber\\
&&\hphantom{\langle\{}
V_-(\infty,x^3_-,\frac{1}{2}\vec b_T+\vec x^3_T)_{C'C}\
\frac{1}{6}\varepsilon_{A'B'C'}\ \varepsilon_{ABC}\nonumber\\
&&\hphantom{\langle\{}
\frac{1}{3}{\Tr}\big[ V_+\big(y_+,\infty,-\frac{1}{2}
\vec b_T+\frac{1}{2}\vec y_T\big)V_+^\dagger(0,\infty,-\frac{1}{2}\vec  
b_T-\frac{1}{2}\vec y_T\big)
\big]-1\big\}\big\rangle_G,\nonumber\\
&&\ear
where
\be\label{C.14}
\vec q_T=(\vec P_1-\vec P_3)_T.\ee
Now we make the transformation of variables
\bear\label{C.15}
x^i_-&\to&\frac{2}{P_{1+}}x_-^i,\nonumber\\
y_+&\to&\frac{2}{P_{2-}}y_+\ear
and use $P_{1+}\to\infty,\ P_{2-}\to\infty$. With the same arguments
which led us from (\ref{3.64}) to (\ref{3.66}) we get
\bear\label{C.16}
{\T}_{fi}&=&-2is\int {\di}^2 b_T\exp(i\vec q_T\cdot\vec b_T)
\int\prod^3_{i=1}{\di}^2 x^i_T\ \delta^2(\vec x^1_T+\vec x^2_T
+\vec x^3_T)\nonumber\\
&&w^B_{3,1}(\vec x^1_T,\vec x^2_T,\vec x^3_T)
\int {\di}^2 y_T w^M_{4,2}(\vec y_T)\nonumber\\
&&\big\langle\big\{ V_-(\infty,0,\frac{1}{2}
\vec b_T+\vec x^1_T)_{A'A}V_-(\infty,0,\frac{1}{2}\vec b_T+\vec  
x_T^2)_{B'B}\nonumber\\
&&\hphantom{\langle\{}
V_-(\infty,0,\frac{1}{2}\vec b_T+\vec  
x_T^3)_{C'C}\frac{1}{6}\epsilon_{A'B'C'}\epsilon_{ABC}\nonumber\\
&&\hphantom{\langle\{}
\frac{1}{3}{\Tr}\big[V_+(0,\infty,-\frac{1}{2}\vec b_T
+\frac{1}{2}\vec y_T)V^\dagger
_+(0,\infty,-\frac{1}{2}\vec b_T-\frac{1}
{2}\vec y_T\big]-1\big\}\big\rangle_G.\nonumber\\
&&\ear
Here $w_{4,2}^M(y_T)$ is the transition profile function
for the mesons as defined in (\ref{3.59}) and
$w^B_{3,1}$ is the corresponding function
for the baryons (cf. (\ref{C.10})).

In the next step we follow \cite{58}, \cite{44} and use
relations which are valid for any $3\times3$ matrix: $H=(H_{AB})$:
\be\label{C.16a}
H_{A'A''}H_{B'B''}H_{C'C''}\cdot
\epsilon_{A''B''C''}=\det H\cdot\epsilon_{A'B'C'},\ee
\bear\label{C.17}
&&\det H\cdot\epsilon_{A'B'C'}\epsilon_{ABC}=\nonumber\\
&&H_{A'A}H_{B'B}H_{C'C}
+H_{A'B}H_{B'C}H_{C'A}+H_{A'C}
H_{B'A}H_{C'B}\nonumber\\
&&-H_{A'B}H_{B'A}H_{C'C}-H_{A'A}H_{B'C}H_{C'B}-
H_{A'C}H_{B'B}H_{C'A}.
\ear
We take $H$ equal to the antiquark line integral at the
central point of the baryon in transverse space:
\be\label{C.18}
H=V_-^*(\infty,0,\frac{1}{2}\vec b_T).\ee
As a $SU(3)$-connector $V_-^*$ satisfies
\be\label{C.19}
\det V_-^*(\infty,0,\frac{1}{2}\vec b_T)=1.\ee
This is easy to prove. From (\ref{C.16a})
we see that $\det V_-^*$ is the connector in
the singlet part of the product of three $SU(3)$ antiquark
representations: $\bar 3\times\bar 3\times \bar 3$. But for
the singlet representation the connector equals 1
since we have to set $T_a=0$ in (\ref{2.22}).

In the following we will use as shorthand notation
\bear\label{C.20}
&&V(i)\equiv V_-(\infty,0,\frac{1}{2}\vec b_T+\vec x^i_T),\nonumber\\
&&(i=1,2,3),\nonumber\\
&&V(0)\equiv V_-(\infty,0,\frac{1}{2}\vec b_T).
\ear
With this we get for the $qqq$-contribution to the integrand
in the functional integral in (\ref{C.16}) using
(\ref{C.17}) to (\ref{C.19}):
\bear\label{C.21}
&&{\W}_+^B(\frac{1}{2}\vec b_T,\vec x^1_T,\vec x^2_T,\vec x^3_T):
=\frac{1}{6}V(1)_{A'A}V(2)_{B'B}V(3)_{C'C}\ \epsilon_{A'B'C'}
\epsilon_{ABC}\nonumber\\
&&=\frac{1}{6}\left\{{\Tr}[V(1)V^\dagger(0)]
\cdot{\Tr}[V(2)V^\dagger(0)]
\cdot{\Tr}[V(3)V^\dagger(0)]\right.\nonumber\\
&&\hphantom{=\frac{1}{6}\{}+
{\Tr}[V(1)V^\dagger(0)V(2)V^\dagger(0)V(3)V^\dagger(0)]
\nonumber\\
&&\hphantom{=\frac{1}{6}\{}+{\Tr}[V(1)V^\dagger(0)
V(3)V^\dagger(0)V(2)V^\dagger(0)]
\nonumber\\
&&\hphantom{=\frac{1}{6}\{}-{\Tr}[V(1)V^\dagger(0)V(2)
V^\dagger(0)]\cdot{\Tr}[V(3)V^\dagger(0)]
\nonumber\\
&&\hphantom{=\frac{1}{6}\{}-{\Tr}[V(2)V^\dagger(0)V(3)
V^\dagger(0)]\cdot{\Tr}[V(1)V^\dagger(0)]
\nonumber\\
&&\hphantom{=\frac{1}{6}\{}\left.-{\Tr}[V(1)V^\dagger
(0)V(3)V^\dagger(0)]\cdot
{\Tr}[V(2)V^\dagger(0)]
\right\}.\ear
As for the meson case (cf. (\ref{3.69}))
we will now add suitable connectors at $\pm\infty$.
In this way all the traces in (\ref{C.21})
become closed light-like Wegner-Wilson loops.
To give an example: the term
\[{\Tr}[V(1)V^\dagger(0)V(2)V^\dagger(0)]\]
should then be read as the loop in the hyperplane $x_-=0$
in the limit $T\to\infty$ which connects the following points
$(x_+,x_-,\vec x_T)$:
\bear\label{C.22}
&&(T,0,\frac{1}{2}\vec b_T),\nonumber\\
&&(-T,0,\frac{1}{2}\vec b_T),\nonumber\\
&&(-T,0,\frac{1}{2}\vec b_T+\vec x_T^2),\nonumber\\
&&(T,0,\frac{1}{2}\vec b_T+\vec x_T^2),\nonumber\\
&&(T,0,\frac{1}{2}\vec b_T),\nonumber\\
&&(-T,0,\frac{1}{2}\vec b_T),\nonumber\\
&&(-T,0,\frac{1}{2}\vec b_T+\vec x_T^1),\nonumber\\
&&(T,0,\frac{1}{2}\vec b_T+\vec x_T^1),\nonumber\\
&&(T,0,\frac{1}{2}\vec b_T)\ear
on straight lines in the order indicated.

Inserting (\ref{C.21}) in (\ref{C.16}) and denoting
the mesonic Wegner-Wilson loop as defined in (\ref{3.70})
by ${\cal W}_-^M$, we get finally for the $T$-matrix element
of baryon-meson scattering:
\bear\label{C.23}
&&{\T}_{fi}=-2is\int {\di}^2b_T\exp(i\vec q_T\cdot
\vec b_T)\nonumber\\
&&\int\prod^3_{i=1}{\di}^2x_T^i\ \delta^2
(\vec x_T^1+\vec x_T^2+\vec x_T^3)
w_{3,1}^B(\vec x_T^1,\vec x_T^2,\vec x_T^3)\nonumber\\
&&\int {\di}^2y_T\ w^M_{4,2}(\vec y_T)\nonumber\\
&&\left\langle{\cal W}^B_+\left(\frac{1}{2}
\vec b_T,\vec x_T^1,\vec x_T^2,\vec x^3_T\right)
{\cal W}^M_-\left(-\frac{1}{2}\vec b_T,\vec y_T
\right)-1\right\rangle_G.
\ear
This formula is the starting point for the evaluation
of the baryon-meson elastic scattering amplitude:
One can now apply the Minkowskian version of the SVM
to calculate the functional integral $\langle\quad\rangle_G$
in an appropriate way. Then one has to fold the result with the
profile functions of the mesonic and baryonic transitions.
At the present state one has to make a suitable ansatz for
these profile functions.

For the case of right-moving antibaryons in an
elastic reaction we just have to substitute the loop
factor ${\cal W}^B_+$ by ${\cal W}_+^{\bar B}$
which is obtained by replacing the quark connectors
$V_-$ by the antiquark connectors $V_-^*$ and vice
versa in (\ref{C.20}), (\ref{C.21}). In an equivalent way we
can get $W_+^{\bar B}$ from ${\cal W}_+^B$ by reversing
the directional arrows on all Wegner-Wilson loops
obtained in the way discussed above from (\ref{C.21}).
For left-moving baryons and antibaryons we have to exchange
$+$ and $-$ components.

For the further treatment of scattering amplitudes involving
mesons, bary\-ons and antibaryons, for many results and a
comparison with experiments we refer to \cite{58}, \cite{44},
\cite{61}.

\bigskip

\section*{Figure Captions}
\begin{description}
\item[Fig. 1:]   The lowest order diagram for the Drell-Yan reaction
(\ref{1.3})
in the QCD improved parton model.
\item[Fig. 2:]  The schematic behaviour of the vacuum energy density
$\varepsilon (B)$ as function of a constant chromomagnetic
field $B$ according to
Savvidy's calculation (eq. (\ref{2.2})).
\item[Fig. 3:]  A ``snapshot'' of the QCD vacuum
showing a domain structure of
spontaneously created chromomagnetic fields.
\item[Fig. 4:]  The QCD-vacuum according to Ambj\o rn and Olesen \cite{32} (a).
The ether according to Maxwell \cite{37} (b).
\item[Fig. 5:] Points $X,X',Y$ in Euclidean space time and curves
$C_X, C_{X'}$ running from $X$ to $Y$ and $X'$ to $Y$, respectively.
\item[Fig. 6:] Curves in Euclidean space time: $C_1$ going
from $X_1$ to $X_2$, $C_2$ from $X_2$ to $X_3$, $C$ from
$X$ to $Y$, and $\bar C$ from $Y$ to $X$.
\item[Fig. 7:] A surface $S$ with boundary $C=\partial S$.
A plaquette with corner points $Z_1...,Z_4$ and boundary
formed by lines $u=const,\ v=const.$.
\item[Fig. 8:] A reference point $Y$ on $S$ and the
fan-type net with center $Y$ spanned over $S$.
\item[Fig. 9:] The correlator function $4\pi^2\kappa G_2D(-Z^2)$
(cf. (\ref{2.61})) as function of $|Z|$. The dashed line is the
lattice result of \cite{43} with the arrows indicating the range
where these results are considered reliable. The solid line corresponds
to the ansatz (\ref{2.64}) with $a=0.35\ {\rm fm}$ (cf. Fig. 5.1 of
\cite{44}).
\item[Fig. 10:] Rectangular Wegner-Wilson loop in Euclidean space-time
in the $X_1-X_4$ plane. The linear extensions are $R$ in $X_1$
direction and $T$ in $X_4$ direction. $Y$ is the reference
point used in the application of the non-abelian Stokes theorem.
\item[Fig. 11:] The $t$-channel (a) and $u$-channel (b)
exchange topologies for the diagrams describing quark-quark
scattering.
\item[Fig. 12:] Projection of the world lines of the quarks
1(2) moving at high velocity in positive (negative) $x^3$
direction onto the $x^0-x^3$ plane in Minkowski space. The
non-abelian phase factors $V_-$ in (\ref{3.33})
$(V_+$ in (\ref{3.35})) are the connectors taken along
lines $x_-,\vec x_T=const.\ (x_+,\vec x_T=const.).$
\item[Fig. 13:] Two light-like lines on which the associated string
operators $V_\pm$ in (\ref{3.43}) are evaluated. Their correlation
function governs quark-quark scattering at high energies.
\item[Fig. 14:] Sketch of reaction (\ref{3.46}) in the overall
c.m. system
\item[Fig. 15:] The light-like Wegner-Wilson loop
in Minkowski space-time, $C_+$ consisting of two
light-like lines in the hyperplane
$x_-=0$ and connecting pieces at infinity. In transverse space
the centre of the loop is at $\vec y_T$, the vector from
the antiquark to the quark line is $\vec z_T$ (from $A$ to $B$).
\item[Fig. 16:] The projections of the two lightlike Wegner-Wilson
loops $C_+,C_-$ occurring in the definition of $W_+(\frac{1}
{2}\vec b_T,\vec x_T),W_-(-\frac{1}{2}\vec b_T,
\vec y_T)$ (cf. (\ref{3.74}), (\ref{3.70})) into
transverse space. The points marked $q_+,\bar q_+\ (q_-,\bar q_-)$
correspond to the projections of the quark and antiquark
lines of $C_+\ (C_-)$.
\item[Fig. 17:] The curves $C_+$ and $C_-$ along which
the path integrals ${\cal W}_\pm$ in (\ref{3.87}) are taken. The
mantle of the pyramid with apex at the origin of the coordinate
system and boundary $C_+$ $(C_-)$ is ${\cal P}_+\ ({\cal P}_-)$,
the basis surface $S_+\ (S_-)$.
\item[Fig. 18:]  The relation between the total cross
section $\sigma_{\rm
tot}$ and the slope parameter $b$ for proton-proton
and proton-antiproton
scattering. The dotted line is the prediction from Regge theory. The
prediction of the calculation for soft high energy scattering
in the stochastic
vacuum model is that the data points should lie in the
area between the full lines.
In essence this is given by (\ref{3.95}), (\ref{3.96}) with an
uncertainty estimate from different assumptions for
the proton wave functions (cf. \cite{44}).
\item[Fig. 19:]   A proton-proton collision at high energies in the
parton picture.
\item[Fig. 20:]   A quark and antiquark traversing a
region of chromomagnetic
field (a). An electron and a positron in a storage  ring (b).
In both cases we
expect the emission of synchrotron radiation to occur.
\item[Fig. 21:]
Sketch of a ``colour domain'' in Minkowski space and of
the worldline of a quark
from a fast hadron moving through it.
\item[Fig. 22:]
A quark moving in 3-direction in a transverse (in 1-direction)
chromomagnetic field of strength $gB_c$ and picking up a transverse
momentum (in 2-direction) of magnitude $\bar p_T$ over a length
$l_{\rm eff}$. Here $\bar p_T$ is the mean transverse momentum of
quarks in the hadron.
\item[Fig. 23:] The $|{\bf k}_T|$ distribution for direct photons
emitted at c.m. rapidity $y=0$ in $p-Be$ collisions at 450 GeV
incident proton momentum from \cite{71}. The normalization is
according to a private communication by H. J. Specht. The background
of decay photons is subtracted. The dash-dotted line gives the
expected yield of photons from hadronic bremsstrahlung, the dashed
lines show the upper and lower limits including the systematic errors
in the shape of the decay background and the bremsstrahlung calculation
(cf. \cite{71}). The lower (upper) solid line is the result of the
calculation for synchrotron photons ((\ref{4.2}ff.) with $l_{\rm eff}
=20\ {\rm fm}\ (l_{\rm eff}=40\ {\rm fm})$ added to
the spectrum of hadronic
bremsstrahlung (cf. \cite{65}).
\item[Fig. 24:] A nucleon interacting by emission of a quasi-real
photon.
\item[Fig. 25:] The data for the electric form factor of the neutron
$G_E^n(Q^2)$ from refs. \cite{74,75}. Dash-dotted line: our naive
``synchrotron'' prediction $\propto (Q^2)^{1/6}$ (\ref{4.11})
normalized to the data at $Q^2=5\ {\rm fm}^{-2}$.
Dashed line: the slope of
$G^n_E(Q^2)$ at $Q^2=0$ as deduced from thermal neutron-electron
scattering \cite{76}. Full line: the ansatz (\ref{4.13}).
\item[Fig. 26:] The ratio $G_E^p/G_D$ of the electric fom factor
of the proton to the dipole fit versus $Q^2$. The
data points are from various experiments as summarized in \cite{47}.
The solid line corresponds to the ansatz (\ref{4.15}), (\ref{4.16}).
\item[Fig. 27:] Annihilation of a $q\bar q$ pair with production of
a virtual photon $\gamma^*$  in a colour domain. Here $q$ and $\bar q$
come from two different hadrons $h_1$ and $h_2$, respectively.
\item[Fig. A1:] The relevant integration region for the integral
(\ref{A.7}). The point (1) runs freely over the surface $S$,
the point (2) is constrained to a distance $\stackrel{\scriptstyle <}
{\sim} a$ from (1).
\item[Fig. A2:] The relevant integration region for the integral
(\ref{A.10}). The point (1) runs freely over $S$. The points
(1), (2), (3), (4) are ordered in the angle as seen from the
reference point $Y$ due to the path ordering function
$\Theta(1,2,3,4)$. The points (1), (3) and (2), (4) must be at a distance
$\stackrel{\scriptstyle <}
{\sim} a$ to each other.
\item[Fig. A3:] The rectangle $S$, $C=\partial S$ and $N_1$ reference
points $Y_1,...,Y_{N_1}$ on the curve $C_1$. The area between
$C_1$ and $C$ is partitioned in $N_1$ plaquettes
$P_1,...,P_{N_1}$.
\item[Fig. A4:] The potential $V(R)$ as defined in (\ref{A.38})
with $R_c$ the ``string breaking'' radius.

\end{description}
\end{document}